\documentclass[%
 reprint,
 superscriptaddress,
 amsmath,amssymb,
 aps,
 pra,
]{revtex4-1}

\usepackage{lipsum} 
\usepackage{graphicx}
\usepackage{dcolumn}
\usepackage{bm}
\usepackage[colorlinks=true,
linkcolor=blue,
anchorcolor=blue,
citecolor=blue,
urlcolor=blue]{hyperref}

\usepackage{mathtools}
\usepackage{color}
\usepackage{multirow}

\begin{document}

\preprint{APS/123-QED}

\title{Circularly polarized RABBIT applied to a Rabi-cycling atom}
\author{Yijie Liao}
\affiliation{School of Physics and Wuhan National Laboratory for Optoelectronics, \\Huazhong University of Science and Technology, Wuhan, 430074,China }
\author{Edvin Olofsson}
\affiliation{Department of Physics, Lund University, Box 118, SE-221 00 Lund, Sweden}
\author{Jan Marcus Dahlstr\"om}
\email{marcus.dahlstrom@matfys.lth.se}
\affiliation{Department of Physics, Lund University, Box 118, SE-221 00 Lund, Sweden}
\author{Liang-Wen Pi}
\affiliation{Center for Attosecond Science and Technology, \\Xi'an Institute of Optics and Precision Mechanics of the Chinese Academy of Sciences, Xi'an 710119, China}
\author{Yueming Zhou}
\email{zhouymhust@hust.edu.cn}
\affiliation{School of Physics and Wuhan National Laboratory for Optoelectronics, \\Huazhong University of Science and Technology, Wuhan, 430074,China }

\author{Peixiang Lu}
\affiliation{School of Physics and Wuhan National Laboratory for Optoelectronics, \\Huazhong University of Science and Technology, Wuhan, 430074,China }
\affiliation{Optics Valley Laboratory, Hubei 430074, China}

\date{\today}

\begin{abstract} 
We utilize the reconstruction of attosecond beating by interference of two-photon transitions (RABBIT) technique to study the phase of a Rabi-cycling atom using circularly polarized extreme ultraviolet and infrared (IR) fields, where the IR field induces Rabi oscillations between the 2s and 2p states of lithium. Autler-Townes splittings are observed in sidebands of the photoelectron spectra and the relative phases of outgoing electron wave packets are retrieved from the azimuthal angle. In this RABBIT scheme, more ionization pathways beyond the usual two-photon pathways are required. Our results show that the polar-angle-integrated and polar-angle-resolved RABBIT phases have different behaviors when the XUV and IR fields have co- and counter-rotating circular polarizations, which are traced back to the different ionization channels according to the selection rules in these two cases and their competition relying on the propensity rule in laser-assisted photoionization.
\end{abstract}

\maketitle


\section{\label{sec:level1}INTRODUCTION}
Rabi oscillations \cite{PhysRev.51.652} are ubiquitous in nuclear physics \cite{PhysRevLett.103.083002}, atomic physics \cite{Dudin2012}, condensed-matter physics \cite{PhysRevLett.100.077401} and quantum information science \cite{PhysRevLett.87.133603}. In the semiclassical interpretation, when a single atom is radiated by a laser field that is nearly resonant with the transition from the ground state to an excited bound state \cite{10.1119/10.0001897}, the population is periodically transferred via absorption and stimulated emission of a photon \cite{KNIGHT198021}. Furthermore, when Rabi oscillations are measured by light-induced perturbative or non-perturbative transitions to \textit{another} bound or continuum state, Autler-Townes (AT) splittings \cite{PhysRev.100.703} are manifested in the energy domain \cite{PhysRevResearch.3.L032052,Nandi2022,PhysRevA.105.063110,PhysRevA.106.063114,PhysRevResearch.5.043017,PhysRevA.87.013415,PhysRevA.88.043416,Ott2014,PhysRevLett.127.023202}, in analogy with Ramsey interference \cite{1956PhT.....9k..37R,PhysRevA.59.R922} from energy domain to time domain. The formation of the AT splitting separated by the Rabi frequency $\Omega_{\rm R}$ can be understood by the quasi-eigenenergies in Floquet theory \cite{ASENS_1883_2_12__47_0,CHU20041}. In the Jaynes-Cummings model, the spectral AT doublet is observed when the atom-field interaction splits a pair of near-degenerated “unperturbed” states \cite{1443594,doi:10.1080/09500349314551321}.

With the advent of attosecond extreme ultraviolet (XUV) pulses \cite{Hentschel2001,doi:10.1126/science.1059413}, attosecond pump-probe spectroscopy enables monitoring and controlling the electronic motion on its natural time scale \cite{RevModPhys.87.765}. For instance, the RABBIT technique is widely employed to investigate the electronic dynamics in laser-assisted photoionization of atoms \cite{doi:10.1126/science.aao7043}, molecules \cite{doi:10.1126/science.aao5624}, solids \cite{PhysRevLett.115.137401} and liquids \cite{doi:10.1126/science.abb0979}. In conventional RABBIT measurements, an XUV attosecond pulse train (APT) ionizes a target synchronized with a weak IR probe field \cite{2012,DAHLSTROM201353}. Owing to the interference between the outgoing electron wave packets, the photoelectron yield of sidebands (SBs) located between the main peaks oscillates with the time delay between the XUV and IR fields \cite{doi:10.1126/science.1059413,E_S_Toma_2002,PhysRevA.54.721}. From the modulations of the SB signals, the relative phase of the ionized electron wave packets can be retrieved, which contains two contributions, the phase of the laser fields \cite{doi:10.1126/science.1059413,Géneaux2016} and the phase related to the electronic dynamics in laser-matter interaction \cite{DAHLSTROM201353,Gruson734,Zhong2020,Biswas2020,PhysRevX.12.011002}. 

In the conventional RABBIT scheme with resonant two-photon ionization of atoms, photoelectrons are ionized from the solely populated ground state and an abrupt phase variation around the resonance is observed in SB modulations \cite{PhysRevLett.104.103003,PhysRevA.104.L021103,doi:10.1126/sciadv.abl7594}. Recently, the RABBIT simulations on a lithium atom have revealed AT splittings in the SB signals of the photoelectron spectra when the IR field induces several cycles of Rabi oscillations in the atom \cite{PhysRevA.105.063110}. During each Rabi cycle on femtosecond timescale, electrons are periodically transferred between the ground (2s) and excited (2p) states of lithium \cite{PhysRevA.105.063110}. Meanwhile, a train of attosecond XUV bursts assisted with the IR field samples this Rabi process by simultaneously releasing the electrons from \textit{both} Rabi states every half IR period. Hence, the \textit{out-of-phase} feature of the temporal Rabi flopping is imprinted in the spectral interference pattern of the emitted electron wave packets in this self-referenced interferometry. Most interestingly, it is revealed that this Rabi process leads to a near $\pi$ phase difference in SB modulations between the AT doublet in both angle-integrated and angle-resolved \cite{PhysRevA.94.063409} spectroscopies \cite{PhysRevA.105.063110}. This real-time evolution of the structured Rabi wave packets below the ionization threshold is out of reach in the previous pump-probe schemes where only \textit{one} of the two Rabi states is measured \cite{PhysRevResearch.3.L032052,PhysRevA.87.013415,PhysRevA.88.043416}. In addition, the influence of the ongoing Rabi oscillations on the \textit{phase} of the outgoing electron wave packets can not be uncovered by a one-color scheme where only an asymmetric AT doublet is observed at a resonant laser frequency \cite{Nandi2022}.

During Rabi oscillations, the atom is polarized by the resonant IR field and its dipole moment oscillates at the Rabi frequency $\Omega_{\rm R}$ and beats at the energy spacing of the two Rabi states. In the previous streaking \cite{PhysRevLett.104.043602,PhysRevLett.108.163001,Ossiander2017} and RABBIT measurements \cite{PhysRevA.102.033112}, it was revealed that a permanent dipole moment of a polar target (either in its initial or ionic states) leads to an additional term in the phase of the outgoing electron wave packet. Meanwhile, the induced dipole moment related to the target's polarizability only distorts the shape of the modulation in the streaking spectrum \cite{PhysRevLett.104.043602}. However, because there are no conceptual counterparts like permanent dipole moment nor polarizability in the case of Rabi oscillations \cite{10.1119/1.3553018}, the influence of the Rabi process on the phase of the ejected electron wave packets is investigated by comparing the two peaks in the AT doublet. 

In this work, we revisit the RABBIT process of a Rabi-cycling lithium atom and adopt circularly polarized XUV \cite{Kfir2015} and IR fields to avoid scanning the time delay between the two laser fields \cite{PhysRevLett.123.133203,Sorngard_2020,Han2023}. In this single time-delay RABBIT measurement, the relative phases of outgoing electron wave packets are retrieved from the photoemission anisotropy along the azimuthal direction. This relative phase encodes the information about the temporal Rabi dynamics in spite of the contamination related to the bound-free and free-free transitions in RABBIT measurements \cite{DAHLSTROM201353}. In accordance with our previous work \cite{PhysRevA.105.063110}, a near $\pi$ phase difference \textit{related to the Rabi dynamics} is observed between each pair of splitting SBs in AT doublets for both polar-angle-integrated and polar-angle-resolved photoelectron spectroscopies. Furthermore, the ionization channels are controlled by using co- or counter-rotating XUV and IR fields according to the dipole selection rules, as an advantage of utilizing circularly polarized fields. Correspondingly, the polar-angle-integrated and polar-angle-resolved RABBIT phases exhibit different behaviors due to different competition among partial waves. To give a quantitative interpretation of this complex RABBIT process, more ionization pathways with important roles are considered in perturbation theory than the previous work \cite{PhysRevA.105.063110}, which elaborates the modification on the Rabi-related phase by the measurement process.

This paper is structured as follows. In Sec.\,\ref{sec:level2}, we describe the numerical methods in our calculations, including the atomic parameters [Sec.\,\ref{atomPara}] and the numerical details [Sec.\,\ref{subsec:TDSE}]. In Sec.\,\ref{theo}, we introduce the theoretical model, including the perturbative treatment of the ionization amplitudes on the top of the Rabi model [Sec.\,\ref{subsec:pert}], the description of the ionization pathways [Sec.\,\ref{subsec:path}], and the review of the RABBIT scheme using circularly polarized laser fields [Sec.\,\ref{subsec:sche}]. In Sec.\,\ref{res}, we present the numerical results, including the co-rotating case in Sec.\,\ref{circular} and the counter-rotating case in Sec.\,\ref{circular_coun}. We finish with a summary in Sec.\,\ref{sec:con}. The paper is completed by Appendix\,\ref{appB:RabiWF}, which details the derivation of the wavefunction for the Rabi process; Appendix\,\ref{appA:DysonSer}, which gives the derivation of the Dyson series for the RABBIT measurement on the Rabi process; Appendix\,\ref{appC:OtherIonAmp}, which supplements the formulas of the amplitudes for less contributing ionization pathways; and Appendix\,\ref{appD:DME}, which selects some numerical results of the dipole transition matrix elements for readers' reference. Atomic units are used throughout this paper unless otherwise stated.

\section{\label{sec:level2}Methods}

\subsection{\label{atomPara}Atomic parameters}
In this work, we consider a lithium atom to study the \textit{Rabi-RABBIT} scheme, where the IR field is resonantly tuned to the transition between the 2s ($m=0$) ($\vert \psi_{2s}\rangle$) and 2p ($m=1$ or $m=-1$)($\vert\psi_{2p}\rangle$) states of lithium and thus induces the Rabi oscillations between these two states. The lithium atom has a single electron outside a closed shell and thus the single-active-electron approximation is reasonable for describing the photoionization of the lithium atom. In this study, we adopt the one-electron effective potential of the lithium atom from Ref.\,\cite{SARSA2004163} for the time-dependent Schr\"odinger equation [Sec.\,\ref{subsec:TDSE}] and for perturbative calculations [Sec.\,\ref{subsec:pert}]. Using this model potential, the calculated energies of the 2s, 2p, 3s, and 3d states of the lithium atom are given in Tab.\,\ref{EnergyLevel}. Note that in this model, the energy spacing between the 2s and 2p levels is 1.6707\,eV. It deviates from the experimental value given in NIST by 0.1771\,eV \cite{NIST_ASD}.
\begin{table}[htb]
  \centering
  \begin{tabular}{c c}
    \hline\hline
    \,\,Bound states\,\, & \,\,Energies (eV)\,\, \\
    \hline
    2s & -5.3821\\
    \hline
    2p & -3.7114\\
    \hline
    3s & -2.1822\\
    \hline
    3d & -1.7960\\
    \hline\hline    
  \end{tabular}
  \caption{The energies of the 2s, 2p, 3s, and 3d states of the lithium atom calculated using the effective potential from Ref.\,\cite{SARSA2004163}.}
  \label{EnergyLevel}
\end{table}  

\begin{figure}[htb]
  \centering
  \includegraphics[width=0.4\textwidth]{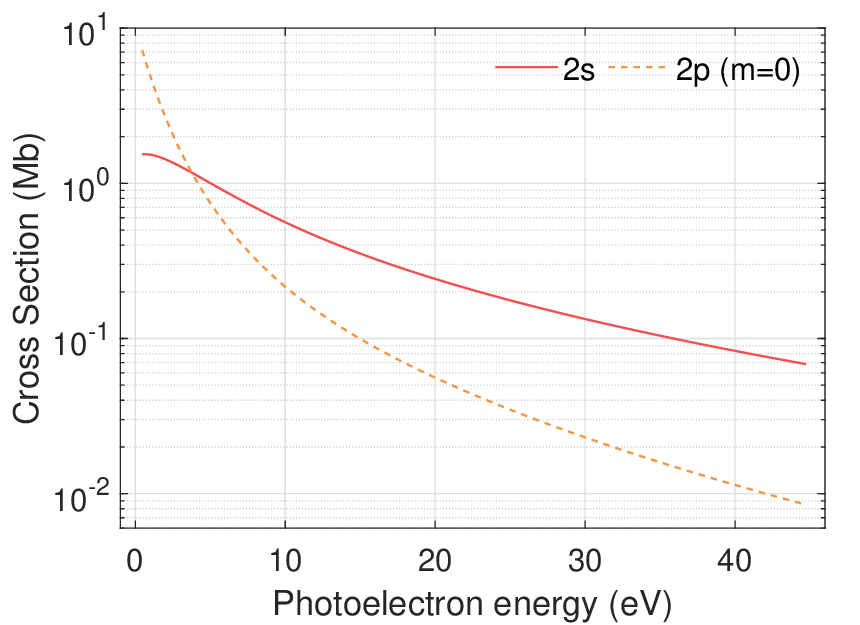}
  \caption{Photoionization cross sections from the 2s and 2p ($m=0$) states of the lithium atom as the functions of the photoelectron energy. }
  \label{fig:CrossSec}
\end{figure}

Figure \ref{fig:CrossSec} shows the photoionization cross sections as a function of the photoelectron energy for the 2s and 2p ($m=0$) states of the lithium atom ionized by a linearly polarized laser field. As shown, the photoionization cross sections both decrease monotonously with the photoelectron energy, but the cross-section of the 2p state decreases faster than that of the 2s state. The photoionization cross-section of the 2p state is larger (smaller) than that of the 2s state below (above) the photoelectron energy of around 4\,eV. For the laser parameters in our Rabi-RABBIT scheme, we will focus on the SBs of the photoelectron energies ranging from 10\,eV to 30\,eV, where single-photon ionization from the 2s state dominates over that of the 2p state by less than an order of magnitude.

\subsection{\label{subsec:TDSE}Time-dependent Schr\"odinger equation}

To uncover the effects of Rabi oscillations on the phase of the ejected wave packet, we solve the time-dependent Schr\"odinger equation (TDSE) for the lithium atom. The TDSE is written as 
\begin{equation}
    i \frac{\partial \Psi(\textbf{r},t)}{\partial t} =  [H_0+H^{\rm V}_{\rm int}(t)]\Psi(\textbf{r},t),
  \label{eq:TDSE}
\end{equation}
where the atomic Hamiltonian is $H_0=\textbf{p}^2/2 + V(\textbf{r})$. Here the canonical momentum operator of the electron is $\textbf{p}=-i\nabla$ and $V(\textbf{r})$ is the effective one-electron potential of the lithium atom \cite{SARSA2004163}. The atom-field interaction term is written in velocity gauge as $H^{\rm V}_{\rm int}(t;\tau)=\textbf{A}(t;\tau) \cdot \textbf{p}$, where the external laser field is described as
\begin{equation}
  \textbf{A}(t;\tau) =  \textbf{A}_{\textrm{IR}}(t;\tau) + \textbf{A}_{\textrm{XUV}}(t).
\end{equation}
Here $\textbf{A}_{\textrm{IR}}(t;\tau)$ and $\textbf{A}_{\textrm{XUV}}(t)$ are the vector potentials of the IR and XUV fields, respectively, and $\tau$ is the time delay between the two fields. In our coordinates, the polarizations of both fields are within the $zOx$ plane and they propagate in the $y$ direction. The vector potential of the time-delayed IR field is expressed as
\begin{subequations}
  \begin{equation}
  \begin{aligned}
  \textbf{A}_{\textrm{IR}}(t;\tau) = &\frac{A_\omega \Lambda^{\rm A}_{\textrm{IR}}(t;\tau)}{\sqrt{1+\xi ^2}}\times\\
  &\{\sin[\omega (t-\tau)]\hat{e}_z-\xi\cos[\omega (t-\tau)]\hat{e}_x\},
  \end{aligned}
  \label{IR_A}
\end{equation}
with the envelope function
\begin{equation}
  \begin{aligned}
  &\Lambda^{\rm A}_{\textrm{IR}}(t;\tau) = \\
  &\begin{cases}
    \frac{1}{2}-\frac{1}{2}\cos[\frac{\pi (t-\tau+\tau_{\textrm{IR}})}{2T}],&-\tau_{\rm IR}<t-\tau<  -\tau_{\rm IR}+2T\\
    1,&-\tau_{\rm IR}+2T\leqslant t-\tau\leqslant \tau_{\rm IR}-2T\\
    \frac{1}{2}-\frac{1}{2}\cos[\frac{\pi (t-\tau-\tau_{\textrm{IR}})}{2T}],&\tau_{\rm IR}-2T<t-\tau\leqslant \tau_{\rm IR}\\
    0,&\rm otherwise.
  \end{cases}\\
\end{aligned}
  \label{IR_Aenv}
\end{equation}
\end{subequations}
Here $A_\omega$ is the amplitude of the field, and $\hat{e}_x$ and $\hat{e}_z$ are the unit vectors along the $x$ and $z$ axes, respectively. In our numerical simulations, the central frequency $\omega=1.6707$\,eV is used for the IR field, which corresponds to a period of $T=2\pi/\omega=2.4963$\,fs. The duration of the IR field is $2\tau_{\textrm{IR}} = 160T = 399.4080$\,fs, corresponding to a spectral width of 0.023\,eV. The vector potential of the XUV field is modeled as an APT \cite{PhysRevA.104.L021103}
\begin{subequations}
\begin{equation}
  \begin{aligned}
     &\textbf{A}_{\textrm{XUV}}(t)  = \sum_{n=-120}^{120} \frac{A_{0}\Lambda^{\rm A, (n)}_{\rm APT}\Lambda^{\rm A, (n)}_{\rm XUV}(t) }{\sqrt{1+\xi ^2}}\times\\
    &\{\sin[\omega_{\textrm{XUV}}(t-nT/2)]\hat{e}_z-\xi\cos[\omega_{\textrm{XUV}}(t-nT/2)]\hat{e}_x\},
    \label{xuv}
  \end{aligned}
  \end{equation}
where the relative amplitude of the $n$th XUV pulse in the pulse train is 
 \begin{equation}
  \Lambda^{\rm A, (n)}_{\rm APT} = \exp\left[-2 \ln 2 \frac{(nT/2)^2}{\tau_{\textrm{APT}}^2}\right],
  \end{equation}
and the envelope function of the $n$th XUV pulse is
  \begin{equation}
  \Lambda^{\rm A, (n)}_{\rm XUV}(t) = (-1)^n \exp\left[-2\ln 2 \frac{(t-nT/2)^2}{\tau_{\textrm{XUV}}^2}\right].
  \end{equation}
\end{subequations}
Here $A_{0}$ is the amplitude of the field. In our simulations, the central frequency $\omega_{\textrm{XUV}} = 15\omega$ is used for the XUV field. $\tau_{\textrm{XUV}} = 0.08T = 0.1997$\,fs and $\tau_{\textrm{APT}} = 50T = 124.8150$\,fs are the durations of the XUV attosecond pulses and the APT, respectively. The spectral width of the XUV field is 0.021\,eV. In our calculations, both the IR and XUV fields are circularly polarized, with $\xi=-1$ and $\xi=+1$ corresponding to right- and left-hand circularly polarized laser fields, respectively.

The wavefunction of the TDSE is expanded as a partial wave series 
\begin{equation}
    \Psi(\textbf{r},t) =\sum_{l=0}^{l_{\rm max}} \sum_{m=-m_{\rm max}}^{m_{\rm max}}\frac{R_{l,m}(r,t)}{r} Y_{l,m}(\theta,\varphi),
\end{equation}
where $R_{l,m}(r,t)$ is the radial part of the wavefunction and $Y_{l,m}(\theta,\varphi)$ are spherical harmonics with polar angle $\theta$ and azimuthal angle $\varphi$. The angular momentum quantum number and the magnetic quantum number are denoted as $l$ and $m$, respectively.

In our calculations, $R_{l,m}(r,t)$ is discretized by the finite-element discrete variable representation method \cite{PhysRevA.62.032706}, where the box size is $r_{\rm max} = 400$\,a.u.. The numerical convergence is guaranteed with $l_{\rm max}=15$ and $m_{\rm max}=5$. The initial state of the lithium atom is obtained by imaginary-time propagation. The time propagation of the wavefunction $\Psi(\textbf{r},t)$ is implemented by the split-Lanczos method \cite{Jiang:17,doi:10.34133/2022/9842716} with the time step $\Delta t = 0.02$\,a.u.. In each propagation step, we apply an absorbing mask function, $F(r) = 1-1/[1+e^{(200.0-r)/2.0}]$, which splits the wavefunction $\Psi(\textbf{r},t)$ into the inner part $\Psi_{\rm in}(\textbf{r},t) = F(r) \Psi(\textbf{r},t)$ and the outer part $\Psi_{\rm out}(\textbf{r},t) = [1-F(r)]\Psi(\textbf{r},t)$. The inner part $\Psi_{\rm in}(\textbf{r},t)$ is kept in the propagation governed by the full Hamiltonian $H_0+H^{V}_{\rm int}(t)$ and the outer part $\Psi_{\rm out}(\textbf{r},t)$ is approximately propagated by Coulomb-Volkov propagator \cite{PhysRevA.77.013401}. Specifically, the ionization amplitude of the photoelectrons with the momentum $\textbf{k}$ at time $t_i$ is obtained by projecting the outer part $\Psi_{\rm out}(\textbf{r},t_i)$ on the set of Volkov states as
\begin{subequations}
\begin{equation}
  f(\textbf{k},t_i)=\langle \psi_{\textbf{k}}^{\rm V}(\textbf{r},t_i)\vert \Psi_{\rm out}(\textbf{r},t_i)\rangle.
\end{equation}  
Then the total ionization amplitude at the final time $t_f$ is expressed as 
\begin{equation}
  f(\textbf{k})=\sum_{i=1}^{N_{\rm step}} U_{\textbf{k}}(t_i,t_f)f(\textbf{k},t_i),
\end{equation}
where $N_{\rm step}$ is the number of propagation steps and the time evolution factor,
\begin{equation}
  U_{\textbf{k}}(t_i,t_f)=e^{-i\int_{t_i}^{t_f}[\frac{\textbf{k}^2}{2}+\textbf{A}(\tau)\cdot \textbf{k}]d\tau },
\end{equation}
is expressed in terms of the Volkov phase of the photoelectrons accumulated from $t_i$ to $t_f$. Finally, the ionization probability distributions are obtained as  
\begin{equation}
  P(\textbf{k})=\vert f(\textbf{k})\vert^2.
\end{equation}
\end{subequations}

\section{\label{theo}Theory}
\subsection{\label{subsec:pert}Perturbative treatment of the ionization amplitudes on the top of the Rabi model} 

In our perturbative calculations, we employ the length gauge to describe the atom-field interaction term, 
\begin{equation}
  H^{\rm L}_{\rm int}(t;\tau)=\textbf{E}(t;\tau) \cdot \textbf{r},
\end{equation}
where the external electric field $\textbf{E}(t;\tau)\coloneqq -d \textbf{A}(t;\tau)/d t$ contains the two contributions from the IR and XUV fields, i.e., $\textbf{E}(t;\tau)= \textbf{E}_{\textrm{IR}}(t;\tau)+\textbf{E}_{\textrm{XUV}}(t)$. To describe the physical process in the Rabi-RABBIT scheme, we preferentially deal with the Rabi oscillations within the two-level Rabi subspace $\mathcal{R}=\{\vert \psi_{2s}\rangle,\vert\psi_{2p}\rangle\}$, followed by a perturbative treatment of the transitions from Rabi subspace $\mathcal{R}$ to its orthogonal complement $\mathcal{S}$ in Hilbert space $\mathcal{H}=\mathcal{R}\bigoplus \mathcal{S}$. In doing so, the full Hamiltonian in $\mathcal{H}$ space is repartitioned as 
\begin{equation}
  H\coloneqq H_0+H^{\rm L}_{\rm int}(t;\tau)=H_{\mathcal{R}}(t)+H_{\rm int}^{\perp \mathcal{R}}(t;\tau).
  \label{Hdiv}
\end{equation}
Here the unperturbed Hamiltonian $H_{\mathcal{R}}$ governs both Rabi dynamics within $\mathcal{R}$ subspace and the field-free evolution dynamics within $\mathcal{S}$ subspace and it is written as
\begin{equation}
  H_{\mathcal{R}}\coloneqq RH_0R+H_{\rm int}^{\mathcal{R}}(t;\tau)+SH_0S,
  \label{RabiH}
\end{equation}
where $R\coloneqq \vert \psi_{2s}\rangle\langle\psi_{2s}\vert+\vert \psi_{2p} \rangle\langle\psi_{2p}\vert$ and $S\coloneqq 1-R$ are the two projectors corresponding to $\mathcal{R}$ and $\mathcal{S}$ subspaces, respectively. The interaction term $H_{\rm int}^{\mathcal{R}}(t;\tau)$ in the Rabi Hamiltonian $H_{\mathcal{R}}$ only contains the two rotating-wave terms related to the excitation and stimulated emission processes induced by the IR field within Rabi subspace $\mathcal{R}$ and it reads
\begin{equation}
  \begin{aligned}
  H_{\rm int}^{\mathcal{R}}(t;\tau)=&\frac{1}{2}E_{\omega}\Lambda^{\rm E}_{\rm IR}(t;\tau)\times\\
  [&e^{-i\omega (t-\tau)} \vert \psi_{2p} \rangle \langle \psi_{2p}\vert (\hat{\epsilon}_{\rm IR}\cdot \textbf{r})\vert \psi_{2s}\rangle\langle \psi_{2s}\vert\\
   + &e^{i\omega (t-\tau)} \vert \psi_{2s} \rangle \langle \psi_{2s}\vert (\hat{\epsilon}_{\rm IR}\cdot \textbf{r})\vert \psi_{2p}\rangle\langle \psi_{2p}\vert],
   \label{Hr_int}
  \end{aligned}
\end{equation}
where $E_{\omega}$ and $\Lambda^{\rm E}_{\rm IR}(t;\tau)$ are the amplitude and the envelope of the IR electric field, respectively. The polarization vector of the IR field is denoted as $\hat{\epsilon}_{\rm IR}$. The interaction term $H_{\rm int}^{\perp \mathcal{R}}(t;\tau)$ of the full Hamiltonian $H$ in Eq.\,(\ref{Hdiv}) is treated as a perturbation to the ongoing Rabi oscillations within $\mathcal{R}$ subspace. Using the projection operators $R$ and $S$, it can be further separated as 
\begin{equation}
  \begin{aligned}
  H_{\rm int}^{\perp \mathcal{R}}(t;\tau) &\coloneqq  H^{\rm L}_{\rm int}(t;\tau) - H_{\rm int}^{\mathcal{R}}(t;\tau)\\
  &=[RH^{\rm L}_{\rm int}(t;\tau)R-H_{\rm int}^{\mathcal{R}}(t;\tau)]\\
  &+SH^{\rm L}_{\rm int}(t;\tau)S\\
  &+[RH^{\rm L}_{\rm int}(t;\tau)S+SH^{\rm L}_{\rm int}(t;\tau)R],
  \end{aligned}
  \label{Hq}
\end{equation}
where the role of $H_{\rm int}^{\perp \mathcal{R}}$ has three aspects: within $\mathcal{R}$, $H_{\rm int}^{\perp \mathcal{R}}$ is responsible for the two counter-rotating-wave transitions induced by the IR field, and all possible non-resonant transitions induced by the XUV field; within $\mathcal{S}$, $H_{\rm int}^{\perp \mathcal{R}}$ is responsible for all possible transitions induced by both the IR and XUV fields; crossing between $\mathcal{R}$ and $\mathcal{S}$, $H_{\rm int}^{\perp \mathcal{R}}$ is responsible for all the possible transitions induced by both the IR and XUV fields.

According to the Rabi model \cite{PhysRev.51.652}, the Rabi oscillations within $\mathcal{R}$ subspace is described by the two-level wavefunction as 
\begin{equation}
  \vert\Psi_{\mathcal{R}}(t)\rangle=C_{2s}(t;\tau)e^{-i\omega_{2s}t}\vert \psi_{2s}\rangle+C_{2p}(t;\tau)e^{-i\omega_{2p}t}\vert \psi_{2p}\rangle,
  \label{rabi_wfn}
\end{equation}
where $H_0\vert\psi_{2s,2p}\rangle=\omega_{2s,2p}\vert\psi_{2s,2p}\rangle$ with $\omega_{2s}$ and $\omega_{2p}$ the energies of the 2s and 2p states. Within the subspace $\mathcal{R}$, the wavefunction $\Psi_{\mathcal{R}}$ obeys the Schr\"odinger equation governed by $H_{\mathcal{R}}$, i.e., $i \frac{d}{dt} \vert\Psi_{\mathcal{R}}(t)\rangle = H_{\mathcal{R}} \vert\Psi_{\mathcal{R}}(t)\rangle$. Symbolically, the propagator $U_{\mathcal{R}}$ corresponding to $H_{\mathcal{R}}$ is defined as
\begin{equation}
  \vert\Psi_{\mathcal{R}}(t)\rangle\coloneqq  U_{\mathcal{R}}(t,t_0)\vert\Psi_{\mathcal{R}}(t_0)\rangle.
\end{equation}
Considering that the system is initially in the 2s state at time $t_0$, i.e., $\vert\Psi_{\mathcal{R}}(t_0)\rangle=e^{-i\omega_{2s}t_0}\vert \psi_{2s}\rangle$, the time-dependent coefficients in Eq.\,(\ref{rabi_wfn}) are solved as (see Appendix \ref{appB:RabiWF})
\begin{equation}
  \begin{cases}
    &C_{2s}(t;\tau)=\cos\left[\frac{1}{2}\int_{t_0}^t\Omega_{\rm R}(t^\prime;\tau) dt^\prime\right]\\
    &C_{2p}(t;\tau)=- i e^{i\omega\tau}\sin\left[\frac{1}{2}\int_{t_0}^t\Omega_{\rm R}(t^\prime;\tau) dt^\prime\right],
  \end{cases}
  \label{coef}
\end{equation}
where $\Omega_{\textrm{R}}(t;\tau) =  \Omega_{\textrm{R}}^0\Lambda^{\rm E}_{\rm IR}(t;\tau)$ is the instantaneous Rabi frequency with its maximum value of $\Omega_{\textrm{R}}^0 =  E_{\omega} \langle \psi_{2p}\vert (\hat{\epsilon}_{\rm IR}\cdot \textbf{r}) \vert \psi_{2s}\rangle$. The factor $e^{i\omega\tau}$ in Eq.\,(\ref{coef}) comes from the time-delayed IR field. Here and hereafter the initial time for the evolution of the system is taken from minus infinity $t_0\rightarrow -\infty$. 
\begin{figure*}[htb]
  \centering
  \includegraphics[width=0.9\textwidth]{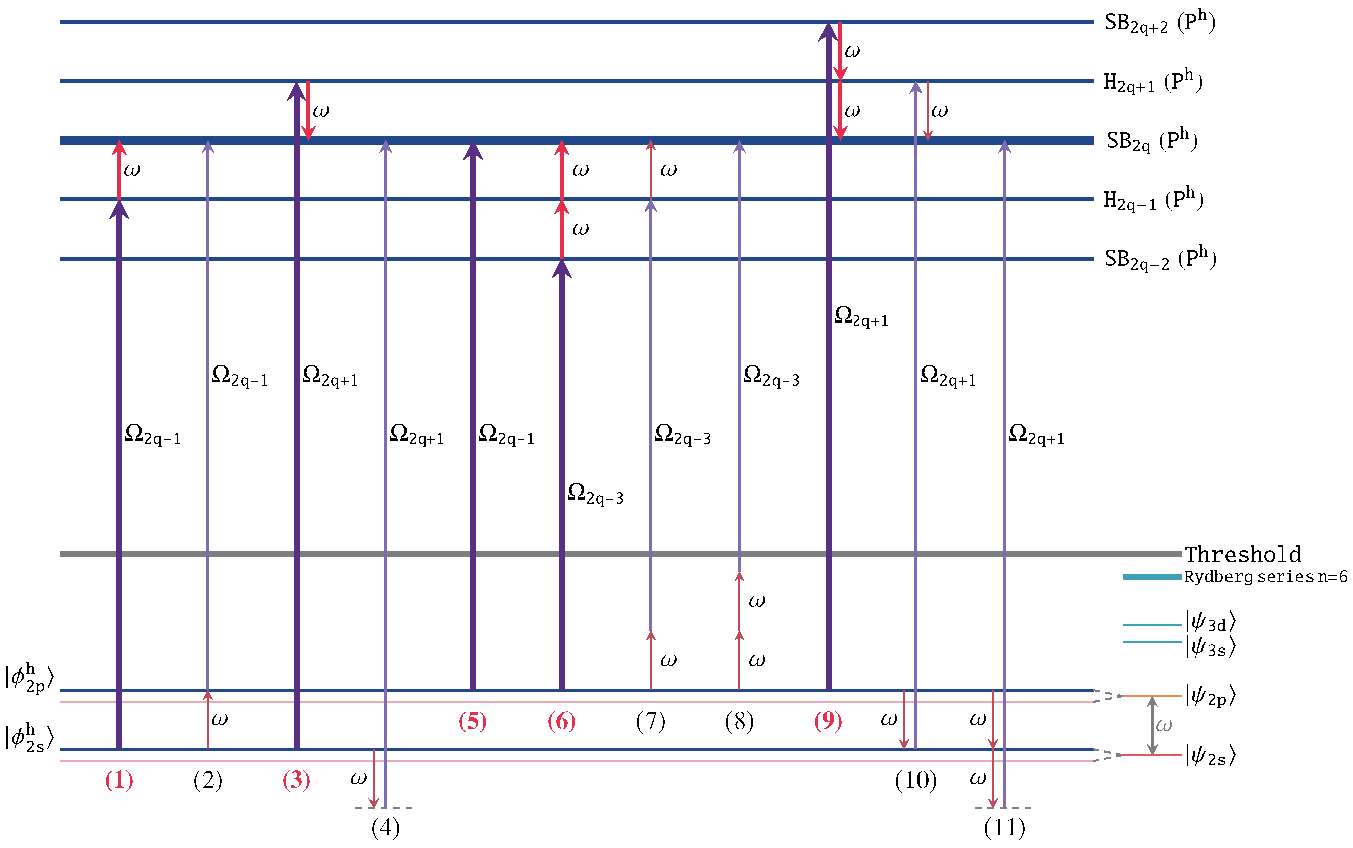}
  \caption{The schematic of ionization pathways for $\rm P^h$ with the photoelectron energy of $2q\omega-I^{2s}_p+\omega_{\rm R}$ ($\omega_{\rm R}\approx \frac{\Omega_{\rm R}^0}{2}$). The purple and red arrows indicate the transition by exchanging the XUV and IR photons, respectively. The essential pathways are drawn in thick-line arrows. The less-contributed pathways are represented by thin-line arrows.}
  \label{fig:schem}
\end{figure*}

In the Rabi-RABBIT scheme, assuming $H_{\rm int}^{\mathcal{R}}(t_0;\tau)=0$ and $H_{\rm int}^{\rm L}(t_0;\tau)=0$, the system is initially in the 2s state within $\mathcal{R}$ subspace at time $t_0$ and thus the initial wavefunction of Hilbert space $\mathcal{H}$ is $\Psi(t_0)=\Psi_{\mathcal{R}}(t_0)=e^{-i\omega_{2s}t_0}\vert \psi_{2s}\rangle$. Then the wavefunction $\Psi(t)$ describing the dynamics of the system at time $t$ in $\mathcal{H}$ space can be written in the form of the Dyson series as \cite{Nandi2022} (see Appendix \ref{appA:DysonSer})
\begin{equation}
  \begin{aligned}
&\vert \Psi(t)\rangle\coloneqq  U(t,t_0) \vert \Psi(t_0)\rangle\\
&= U_{\mathcal{R}}(t,t_0)\vert\Psi(t_0)\rangle\\
&-i\int_{t_0}^t dt_1 U(t,t_1) H_{\rm int}^{\perp \mathcal{R}}(t_1)U_{\mathcal{R}}(t_1,t_0)\vert \Psi(t_0)\rangle\\
&=\vert\Psi_{\mathcal{R}}(t)\rangle -i\int_{t_0}^t dt_1 U(t,t_1) H_{\rm int}^{\perp \mathcal{R}}(t_1)\vert\Psi_{\mathcal{R}}(t_1)\rangle,
  \end{aligned}
\label{dyson1}
\end{equation}
where $U$ is the propagator related to the full Hamiltonian $H$. In the last line of Eq.\,(\ref{dyson1}), the first term describes the unperturbed Rabi oscillations within the two-level subspace $\mathcal{R}$; and the second term describes the transition from the Rabi wavefunction $\Psi_{\mathcal{R}}$ of the subspace $\mathcal{R}$ outwards to its orthogonal complement $\mathcal{S}$ through the interaction $H_{\rm int}^{\perp \mathcal{R}}$ followed by its full propagation under $H$, which contains the ionization part of the wavefunction.

According to Eq.\,(\ref{Hdiv}), the full propagator $U(t,t_1)$ can be alternatively written as
\begin{equation}
  U(t,t_1)=U_0(t,t_1)-i\int_{t_1}^t dt_2U(t,t_2)H^{\rm L}_{\rm int}(t_2)U_0(t_2,t_1)
  \label{dyson2}
\end{equation}
where $U_0$ is the propagator related to the field-free Hamiltonian $H_0$. Here $U_0(t,t^\prime)$ can be expanded on the eigenstates of $H_0$ as $U_0(t,t^\prime)=\sum_{\nu}\kern-1.5em \int\,\,\,\vert \psi_\nu\rangle\langle \psi_\nu\vert \exp[-i\omega_\nu(t-t^\prime)]$ with $H_0\vert \psi_n\rangle=\omega_n\vert \psi_n\rangle$. Substituting Eq.\,(\ref{dyson2}) into Eq.\,(\ref{dyson1}), the wavefunction can be expanded iteratively to arbitrary order in the interaction term as
\begin{equation}
  \vert \Psi(t)\rangle = \sum_{N=0}^{\infty}\vert \bar{\Psi}^{(N)}(t)\rangle,
\end{equation}
where the zeroth-order wave packet is the Rabi wavefunction $\vert \bar{\Psi}^{(0)}(t)\rangle=\vert \Psi_{\mathcal{R}}(t)\rangle$ in Eq.\,(\ref{rabi_wfn}) and where the first-order wave packet is $\vert \bar{\Psi}^{(1)}(t)\rangle=-i\int_{t_0}^t dt_1 U_0(t,t_1) H_{\rm int}^{\perp \mathcal{R}}(t_1)\vert\Psi_{\mathcal{R}}(t_1)\rangle$, which describes the transition from the Rabi wavefunction $\Psi_{\mathcal{R}}$ through the first-order interaction $H_{\rm int}^{\perp \mathcal{R}}$. Likewise, for the wave packets to higher orders $N>1$, $\vert \Psi^{(N)}(t)\rangle$ originates from the $N$th-order interaction and it is derived from Eq.\,(\ref{dyson1}) as 
\begin{equation}
  \begin{aligned}
  &\vert \bar{\Psi}^{(N>1)}(t)\rangle=(-i)^N\int_{t_0}^tdt_1 \int_{t_1}^t dt_2\cdots \int_{t_{N-1}}^tdt_N\times\\
  &U_0(t,t_N)H^{\rm L}_{\rm int}(t_N)U_0(t_N,t_{N-1})H^{\rm L}_{\rm int}(t_{N-1})\cdots\\
  & H^{\rm L}_{\rm int}(t_2)U_0(t_2,t_1) H_{\rm int}^{\perp \mathcal{R}}(t_1)\vert\Psi_{\mathcal{R}}(t_1)\rangle,
  \end{aligned}
  \label{psin}
\end{equation}
with the time order $t>t_{N}>t_{N-1}>\cdots>t_2>t_1>t_0$. Here in our perturbative treatment of the ionization process on the top of the Rabi model, we assume the ongoing Rabi oscillations with $\mathcal{R}$ are perfect: the energy shifts of the 2s and 2p states, the population leakage to $\mathcal{S}$ subspace, and the dampening in Rabi oscillations are all negligible. Moreover, we assume the time scale of the process we are investigating is much shorter than the lifetime of the 2s and 2p states.

The $N$th-order transition amplitude $a_f^{(N)}$ is obtained by projecting the $N$th-order wavefunction $\vert \Psi^{(N)}(t)\rangle$ to the final state $\vert \phi_f\rangle$ of a specific energy $\omega_f$ as \cite{joachain_kylstra_potvliege_2011}
\begin{equation}
  a_f^{(N)}=\lim_{t\rightarrow\infty}\langle \phi_f(t)\vert \Psi^{(N)}(t)\rangle,
  \label{ionAmp}
\end{equation}
where $\vert \phi_f(t)\rangle =e^{-i\omega_f t}\vert \psi_f\rangle$, with $\vert \psi_f\rangle$ satisfying $H_0\vert\psi_f\rangle=\omega_f\vert\psi_f\rangle$. Here we take $t\rightarrow\infty$ because the photoelectrons are measured long after the interaction with the field is over. In the case of $\omega_f>0$, $a_f^{(N>0)}$ represents the $N$-photon ionization amplitude from the Rabi wavefunction $\Psi_{\mathcal{R}}$ with all possible combinations of photons in different time orders. To further specify different ionization processes in the total $N$th-order transition amplitude $a_f^{(N)}$, the interaction terms $H^{\rm L}_{\rm int}$ and $H_{\rm int}^{\perp \mathcal{R}}$ are separated into different laser components in Eq.\,(\ref{psin}). For example, the interaction term is $E_{\rm IR}(t;\tau)\hat{O}_{\omega}$ for the IR photon and it is $E_{\Omega_{2q+1}}(t)\hat{O}_{\Omega}$ for the XUV harmonics of the frequency $\Omega_{2q+1}=(2q+1)\omega$, which contain both absorption and emission terms. Here the transition operators are written in the length gauge as $\hat{O}_{\omega}\coloneqq \hat{\epsilon}_{\rm IR}\cdot\textbf{r}$ and $\hat{O}_{\Omega}\coloneqq \hat{\epsilon}_{\rm XUV}\cdot\textbf{r}$, with $\hat{\epsilon}_{\rm XUV}$ the polarization vector of the XUV field assuming all XUV harmonics have the same polarization. In particular, when considering the XUV, XUV+IR, and XUV+IR+IR absorption processes, the corresponding ionization amplitudes are derived as
\begin{widetext}
\begin{subequations}
  \begin{equation}
    \begin{aligned}
    a_{\rm XUV}^{(1)}=&-i\int_{-\infty}^\infty d\omega_{\rm R}\tilde{C}_{2s}(\omega_{\rm R};\tau) \tilde{E}_{\Omega_{2q+1}}(\omega_f-\omega_{2s}-\omega_{\rm R})\langle \psi_f\vert \hat{O}_\Omega\vert\psi_{2s}\rangle\\
    &-i\int_{-\infty}^\infty d\omega_{\rm R}\tilde{C}_{2p}(\omega_{\rm R};\tau) \tilde{E}_{\Omega_{2q+1}}(\omega_f-\omega_{2p}-\omega_{\rm R})\langle \psi_f\vert \hat{O}_\Omega\vert\psi_{2p}\rangle,
    \end{aligned}
    \label{1st_order}
  \end{equation}
  \begin{equation}
    \begin{aligned}
    a_{\rm XUV+IR}^{(2)}=&-\frac{i}{\sqrt{2\pi}}\int_{-\infty}^\infty d\omega_{\rm R}\tilde{C}_{2s}(\omega_{\rm R};\tau)\int_{0}^{\infty}d\Omega e^{i(\omega_f-\Omega-\omega_{2s}-\omega_{\rm R})\tau}\times\\
    &\tilde{E}_{\rm IR}(\omega_f-\Omega-\omega_{2s}-\omega_{\rm R})\tilde{E}_{\Omega_{2q+1}}(\Omega)\sum_{\nu_1}\kern-1.5em \int\frac{\langle\psi_f\vert\hat{O}_\omega\vert\psi_{\nu_1}\rangle\langle\psi_{\nu_1}\vert \hat{O}_\Omega\vert\psi_{2s}\rangle}{\omega_{2s}+\omega_{\rm R}+\Omega-\omega_{\nu_1}}\\
    &-\frac{i}{\sqrt{2\pi}}\int_{-\infty}^\infty d\omega_{\rm R}\tilde{C}_{2p}(\omega_{\rm R};\tau)\int_{0}^{\infty}d\Omega e^{i(\omega_f-\Omega-\omega_{2p}-\omega_{\rm R})\tau}\times\\
    &\tilde{E}_{\rm IR}(\omega_f-\Omega-\omega_{2p}-\omega_{\rm R})\tilde{E}_{\Omega_{2q+1}}(\Omega)\sum_{\nu_1}\kern-1.5em \int\frac{\langle\psi_f\vert\hat{O}_\omega\vert\psi_{\nu_1}\rangle\langle\psi_{\nu_1}\vert \hat{O}_\Omega\vert\psi_{2p}\rangle}{\omega_{2p}+\omega_{\rm R}+\Omega-\omega_{\nu_1}},
    \end{aligned}
    \label{2nd_order}
  \end{equation}
  \begin{equation}
    \begin{aligned}
    a_{\rm XUV+IR+IR}^{(3)}=&-\frac{i}{2\pi}\int_{-\infty}^\infty d\omega_{\rm R}\tilde{C}_{2s}(\omega_{\rm R};\tau)\int_{0}^{\infty}d\omega\int_{0}^{\infty}d\Omega e^{i(\omega_f-\Omega-\omega_{2s}-\omega_{\rm R})\tau}\times\\
    &\tilde{E}_{\rm IR}(\omega_f-\omega-\Omega-\omega_{2s}-\omega_{\rm R})\tilde{E}_{\rm IR}(\omega)\tilde{E}_{\Omega_{2q+1}}(\Omega)\sum_{\nu_1,\nu_2}\kern-1.5em \int\frac{\langle\psi_f\vert\hat{O}_\omega\vert\psi_{\nu_2}\rangle\langle\psi_{\nu_2}\vert\hat{O}_\omega\vert\psi_{\nu_1}\rangle\langle\psi_{\nu_1}\vert \hat{O}_\Omega\vert\psi_{2s}\rangle}{(\omega_{2s}+\omega_{\rm R}+\Omega+\omega-\omega_{\nu_2})(\omega_{2s}+\omega_{\rm R}+\Omega-\omega_{\nu_1})}\\
    &-\frac{i}{2\pi}\int_{-\infty}^\infty d\omega_{\rm R}\tilde{C}_{2p}(\omega_{\rm R};\tau)\int_{0}^{\infty}d\omega\int_{0}^{\infty}d\Omega e^{i(\omega_f-\Omega-\omega_{2p}-\omega_{\rm R})\tau}\times\\
    &\tilde{E}_{\rm IR}(\omega_f-\omega-\Omega-\omega_{2p}-\omega_{\rm R})\tilde{E}_{\rm IR}(\omega)\tilde{E}_{\Omega_{2q+1}}(\Omega)\sum_{\nu_1,\nu_2}\kern-1.5em \int\frac{\langle\psi_f\vert\hat{O}_\omega\vert\psi_{\nu_2}\rangle\langle\psi_{\nu_2}\vert\hat{O}_\omega\vert\psi_{\nu_1}\rangle\langle\psi_{\nu_1}\vert \hat{O}_\Omega\vert\psi_{2p}\rangle}{(\omega_{2p}+\omega_{\rm R}+\Omega+\omega-\omega_{\nu_2})(\omega_{2p}+\omega_{\rm R}+\Omega-\omega_{\nu_1})},
    \end{aligned}
    \label{3rd_order}
  \end{equation}
  \label{3ionAmp}
\end{subequations}
\end{widetext}
where $\nu_i$ identifies the intermediate unperturbed states $\vert\psi_{\nu_i}\rangle$ of the energy $\omega_{\nu_i}$ as $H_0\vert\psi_{\nu_i}\rangle=\omega_{\nu_i}\vert\psi_{\nu_i}\rangle$; $\omega$ and $\Omega$ are the frequencies of the IR photon and the $(2q+1)$th-order XUV harmonics, respectively. Here, we use the relation of Delta-function $\delta(\omega)=\int_{-\infty}^\infty e^{i\omega t}dt /(2\pi)$ and the Fourier transform (FT): $\tilde{f}(\omega)=\frac{1}{\sqrt{2\pi}}\int^\infty_{-\infty}f(t)e^{i\omega t}dt$, $f(t)=\frac{1}{\sqrt{2\pi}}\int^\infty_{-\infty}\tilde{f}(\omega)e^{-i\omega t}d\omega$, and $\tilde{f}(\omega;\tau)\coloneqq  \frac{1}{\sqrt{2\pi}}\int^\infty_{-\infty}f(t;\tau)e^{i\omega t}dt=\tilde{f}(\omega)e^{i\omega\tau}$. The FT of the Rabi amplitudes [Eq.\,(\ref{coef})] are $\tilde{C}_{2s,2p}(\omega_{\rm R};\tau)\coloneqq  \frac{1}{\sqrt{2\pi}}\int_{-\infty}^\infty C_{2s,2p}(t;\tau)e^{-i\omega_{\rm R}t}d\omega_{\rm R}=\tilde{C}_{2s,2p}(\omega_{\rm R})e^{i\omega_{\rm R}\tau}$ (see Appendix \ref{appB:RabiWF}), where $\omega_{\rm R}$ is the frequency component of the Rabi amplitudes. For the long IR field of flat-top envelope [Eq.\,(\ref{IR_Aenv})], the Rabi amplitudes mainly distribute at $\omega_{\rm R}\approx \pm\frac{\Omega_{\rm R}^0}{2}$ in the frequency domain. Eq.\,(\ref{1st_order}) describes the one-photon transition from the zeroth-order Rabi wavefunction $\vert\Psi_{\mathcal{R}}\rangle$ to the continuum state $\vert \phi_f\rangle$ by absorbing an XUV photon of the frequency $\Omega=\omega_f-\omega_{2s,2p}-\omega_{\rm R}$; Eq.\,(\ref{2nd_order}) describes the two-photon transition through the absorption of an XUV photon followed by the absorption of an IR photon, where the energy-preserving condition $\omega_f=\omega_{2s,2p}+\omega_{\rm R}+\Omega+\omega$ is satisfied; and Eq.\,(\ref{3rd_order}) describes the three-photon transition through the absorption of an XUV photon followed by the absorption of two IR photons, where the energy-preserving condition $\omega_f=\omega_{2s,2p}+\omega_{\rm R}+\Omega+2\omega$ is satisfied.

As indicated by the term $\omega_{\rm R}$ in the denominators in Eqs.\,(\ref{3ionAmp}), the ionization from the Rabi state $\vert\Psi_{\mathcal{R}}\rangle$ can be understood to start from the dressed states of $\rm \vert \psi_{2s}\rangle$ and $\rm \vert \psi_{2p}\rangle$, which spilt into two quasienergies considering the sign of $\omega_{\rm R}$. More specifically, the first (second) term in Eqs.\,(\ref{3ionAmp}) corresponds to the ionization from the dressed 2s (2p) states with the quasienergies of $\omega_{2s}+\omega_{\rm R}$ ($\omega_{2p}+\omega_{\rm R}$), where $\omega_{\rm R}\approx -\frac{\Omega_{\rm R}^0}{2}$ for $\rm\vert \phi_{2s}^l\rangle$ ($\rm\vert \phi_{2p}^l\rangle$) and $\omega_{\rm R}\approx \frac{\Omega_{\rm R}^0}{2}$ for $\rm\vert \phi_{2s}^h\rangle$ ($\rm\vert \phi_{2p}^h\rangle$), respectively. In the photoelectron spectra, the peak of a lower (higher) photoelectron energy in the AT doublet, namely $\rm P^l$ ($\rm P^h$), is formed by the interference among all the ionization pathways initiated from $\rm\vert \phi_{2s}^l\rangle$ and $\rm\vert \phi_{2p}^l\rangle$ ($\rm\vert \phi_{2s}^h\rangle$ and $\rm\vert \phi_{2p}^h\rangle$).

\subsection{\label{subsec:path}Ionization pathways}
As an example, we display the ionization pathways contributing to $\rm P^h$ with the photoelectron energy of $2q\omega-I^{2s}_p+\omega_{\rm R}$ ($\omega_{\rm R}\approx \frac{\Omega_{\rm R}^0}{2}$) in Fig.\,\ref{fig:schem}. Pathways 1 to 4 are initiated from $\rm\vert \phi_{2s}^h\rangle$, among which pathways 1 and 3 are reminiscent of the well-known two-photon transitions in the traditional RABBIT cases \cite{PhysRevLett.106.143002}. Pathways 1 and 2 denote the absorption of one $(2q-1)$th-order XUV harmonic $\rm\Omega_{2q-1}$ and one IR photon $\omega$ in different time orders. Note that, as indicated by the term $H_{\rm int}^{\perp \mathcal{R}}$ in Eq.\,(\ref{psin}), $\psi_{2p}$ needs to be excluded from the intermediate summation in calculating the transition amplitude of pathway 2 since this transition has been handled in the non-perturbative treatment of the Rabi oscillations. Pathways 3 and 4 denote the absorption of one $(2q+1)$th-order XUV harmonic $\rm\Omega_{2q+1}$ and the emission of one IR photon $\omega$ in different time orders. In our case, because the 2p state is also populated during Rabi oscillations, there are also pathways initiated from $\rm\vert \phi_{2p}^h\rangle$, as denoted by pathways 5 to 11 in Fig.\,\ref{fig:schem}. Pathway 5 indicates the absorption of one $(2q-1)$th-order XUV harmonic $\rm\Omega_{2q-1}$. Pathways 6 to 8 refer to the absorption of one $(2q-3)$th-order XUV harmonic $\rm\Omega_{2q-3}$ and two IR photons in different time orders. Pathways 9 to 11 denote the absorption of one XUV photon $\rm\Omega_{2q+1}$ and the emission of two IR photons in different time orders. Similarly as pathway 2, in calculating the transition amplitude of pathways 10 and 11, $\psi_{2s}$ needs to be excluded from the first intermediate summation. Pathways of essential contribution are pathways 1, 3, 5, 6, and 9, whose numbers are highlighted in red in Fig.\,(\ref{fig:schem}).

In our calculations, the XUV and flat-top IR fields are both spectrally narrow enough and display no spectral overlap. Therefore, the convolution can be reasonably dismissed and both laser fields are approximated as monochromatic: let $E_{\rm IR}(t;\tau)=E_{\omega}\cos[\omega(t-\tau)]$ and $E_{\Omega_{\rm 2q+1}}(t)=E_{\rm 2q+1}\cos(\Omega_{\rm 2q+1} t-\phi_{\rm 2q+1})$, with the amplitude $E_{\rm 2q+1}$ and the phase $\phi_{\rm 2q+1}$ associated to the $(2q+1)$th-order XUV harmonic; and let $\tilde{C}_{2s}(\omega_{\rm R};\tau)=\sqrt{\frac{\pi}{2}}e^{i\omega_{\rm R}\tau}[\delta(\omega_{\rm R}+\frac{\vert\Omega_{\rm R}^0\vert}{2})+\delta(\omega_{\rm R}-\frac{\vert\Omega_{\rm R}^0\vert}{2})]$ and $\tilde{C}_{2p}(\omega_{\rm R};\tau)=-\sqrt{\frac{\pi}{2}}\frac{\Omega_{\rm R}^0}{\vert \Omega_{\rm R}^0\vert}e^{i(\omega+\omega_{\rm R})\tau}[\delta(\omega_{\rm R}+\frac{\vert\Omega_{\rm R}^0\vert}{2})-\delta(\omega_{\rm R}-\frac{\vert\Omega_{\rm R}^0\vert}{2})]$. Then each term in Eqs.\,(\ref{3ionAmp}) can be approximately calculated by multiplying the corresponding IR ($E_{\omega}$) and XUV ($E_{\rm 2q+1}$) electric field amplitudes to the residual integration of wavefunctions. For example, the ionization amplitudes $\mathcal{A}_{\mathcal{P}}$ for SB$_{2q}$, where $\mathcal{P}$ characterizes the specific pathway, i.e., $\mathcal{P}\in \{\textrm{pathway}\,\,(i)\vert i=1,\,\cdots,11\}$, are given as (see Appendix \ref{appC:OtherIonAmp} for the other pathways)
\begin{widetext}
\begin{subequations}
  \begin{equation}
  \mathcal{A}_{(1)}(\vec{k}_\pm,\tau)=- \frac{i\pi }{4}e^{i\left[\left(\omega\pm \frac{\vert\Omega_{\textrm{R}}^0\vert}{2}\right)\tau+\phi_{2q-1}\right]}E_{\omega}E_{\rm 2q-1}\sum_{\nu_1}\kern-1.5em \int\frac{\langle \psi_{f\pm} \vert \hat{O}_\omega \vert \psi_{\nu_1} \rangle \langle \psi_{\nu_1} \vert \hat{O}_\Omega \vert \psi_{2s} \rangle}{\omega_{2s}\pm \frac{\vert\Omega_{\textrm{R}}^0\vert}{2}+\Omega_{2q-1}-\omega_{\nu_1}},
  \end{equation}
  \begin{equation}
  \mathcal{A}_{(3)}(\vec{k}_\pm,\tau) =- \frac{i\pi}{4} e^{-i\left[\left(\omega\mp \frac{\vert\Omega_{\textrm{R}}^0\vert}{2}\right)\tau-\phi_{2q+1}\right]}E_{\omega}E_{\rm 2q+1}\sum_{\nu_1}\kern-1.5em \int\frac{\langle \psi_{f\pm} \vert \hat{O}^{\dagger}_{\omega} \vert \psi_{\nu_1} \rangle \langle \psi_{\nu_1} \vert \hat{O}_\Omega \vert \psi_{2s} \rangle}{\omega_{2s}\pm \frac{\vert\Omega_{\textrm{R}}^0\vert}{2}+\Omega_{2q+1}-\omega_{\nu_1}},
  \end{equation}
  \begin{equation}
  \mathcal{A}_{(5)}(\vec{k}_\pm,\tau) = \mp \frac{i\pi\Omega_{\rm R}^0}{2\vert \Omega_{\rm R}^0\vert}e^{i\left[\left(\omega\pm \frac{\vert\Omega_{\textrm{R}}^0\vert}{2}\right)\tau+\phi_{2q-1}\right]}E_{\rm 2q-1} \langle \psi_{f\pm} \vert \hat{O}_{\Omega} \vert \psi_{2p} \rangle,
  \end{equation}
  \begin{equation}
    \begin{aligned}
  \mathcal{A}_{(6)}(\vec{k}_\pm,\tau) =&\mp \frac{i\pi\Omega_{\rm R}^0}{8\vert \Omega_{\rm R}^0\vert} e^{i\left[\left(3\omega\pm \frac{\vert\Omega_{\textrm{R}}^0\vert}{2}\right)\tau+\phi_{2q-3}\right]}E_{\omega}^2E_{\rm 2q-3}\times\\
  &\sum_{\nu_1,\nu_2}\kern-1.5em \int\frac{\langle \psi_{f\pm} \vert \hat{O}_{\omega} \vert \psi_{\nu_2}\rangle \langle \psi_{\nu_2} \vert \hat{O}_{\omega} \vert \psi_{\nu_1}\rangle \langle \psi_{\nu_1} \vert \hat{O}_{\Omega} \vert \psi_{2p}\rangle}{(\omega_{2p}\pm \frac{\vert\Omega_{\textrm{R}}^0\vert}{2}+\Omega_{2q-3}+\omega-\omega_{\nu_2})(\omega_{2p}\pm \frac{\vert\Omega_{\textrm{R}}^0\vert}{2}+\Omega_{2q-3}-\omega_{\nu_1})},
    \end{aligned}
  \end{equation}
  \begin{equation}
  \begin{aligned}
  \mathcal{A}_{(9)}(\vec{k}_\pm,\tau) =&\mp \frac{i\pi\Omega_{\rm R}^0}{8\vert \Omega_{\rm R}^0\vert} e^{-i\left[\left(\omega\mp \frac{\vert\Omega_{\textrm{R}}^0\vert}{2}\right)\tau-\phi_{2q+1}\right]}E_{\omega}^2E_{\rm 2q+1}\times\\
  &\sum_{\nu_1,\nu_2}\kern-1.5em \int\frac{\langle \psi_{f\pm} \vert \hat{O}^{\dagger}_{\omega} \vert \psi_{\nu_2}\rangle \langle \psi_{\nu_2} \vert \hat{O}^{\dagger}_{\omega} \vert \psi_{\nu_1}\rangle \langle \psi_{\nu_1} \vert \hat{O}_{\Omega} \vert \psi_{2p}\rangle}{(\omega_{2p}\pm \frac{\vert\Omega_{\textrm{R}}^0\vert}{2}+\Omega_{2q+1}-\omega-\omega_{\nu_2})(\omega_{2p}\pm \frac{\vert\Omega_{\textrm{R}}^0\vert}{2}+\Omega_{2q+1}-\omega_{\nu_1})},
  \end{aligned}
  \end{equation}
  \label{ionchan}
\end{subequations}
\end{widetext}
where $\psi_{f+}$ ($\psi_{f-}$) is the final continuum state related to the peak $\rm P^h$ ($\rm P^l$) of the AT doublet. The energy and the asymptotic momentum of $\psi_{f\pm}$ are $E_{\pm} = 2q\omega-I_p^{2s}\pm \frac{\vert\Omega_{\textrm{R}}^0\vert}{2}=k_\pm^2/2$ and $\vec{k}_\pm\coloneqq  k_\pm \hat{k}_\pm$, respectively. Here, as indicated by Eqs.\,(\ref{rabi_wfn}) and (\ref{coef}), the zeroth-order Rabi wavefunction $\Psi_{\mathcal{R}}$ evolving with the time-delayed IR field leads to the factors $e^{\pm i \frac{\vert\Omega_{\textrm{R}}^0\vert}{2}\tau}$ and $e^{i(\omega\pm \frac{\vert\Omega_{\textrm{R}}^0\vert}{2})\tau}$ in Eqs.\,(\ref{ionchan}), respectively for the ionization pathways from the 2s and 2p states. In addition, the initial phase of the electron wave packet generated from $\rm\vert \phi_{2p}^l\rangle$ has an additional term of $\pi$ compared to that for $\rm\vert \phi_{2p}^h\rangle$, as implied by the beginning sign $\mp$ of the ionization amplitudes for pathways 5 to 11. These different initial phases between the dressed 2p states stem from the sine-like Rabi amplitude of the 2p state, as indicated by Eq.\,(\ref{coef}).
\begin{table*}[htb]
  \centering
  \begin{tabular}{c c c c c c c c c c c c c c c c}
    \hline\hline
    \,\,Case\,\, & \,\,XUV field\,\, & \,\,IR field\,\,& \,\,P1\,\, &\,\, P2\,\, & \,\,P3\,\, & \,\,P4\,\, & \,\,P5\,\, & \,\,P6\,\, &\,\,P7\,\,&\,\,P8\,\, & \,\,P9\,\, & \,\,P10\,\, & \,\,P11\,\, &  \,\,$\varphi-$oscillations\,\, & \,\,Phase matched\,\,\\
    \hline
    1 & $\Delta m=+1$ & $\Delta m=+1$ & 2 & 2 & 0 & 0 & 2 & 4 & 4 & 4 & 0 & 0 & 0 & $2\varphi$,$4\varphi$ & Yes\\
    \hline
    2 & $\Delta m=+1$ & $\Delta m=-1$ & 0 & 0 & 2 & 2 & 0 & -2 & -2 & -2 & 2 & 2 & 2 & $2\varphi$,$4\varphi$ & Yes\\
    \hline
    3 & $\Delta m=-1$ & $\Delta m=+1$ & 0 & 0 & -2 & -2 & 0 & 2 & 2 & 2 & -2 & -2 & -2 & $2\varphi$,$4\varphi$ & Yes\\
    \hline    
    4 & $\Delta m=-1$ & $\Delta m=-1$ & -2 & -2 & 0 & 0 & -2 & -4 & -4 & -4 & 0 & 0 & 0 & $2\varphi$,$4\varphi$ & Yes\\
    \hline
    5 & $\Delta m=0$ & $\Delta m=+1$ & 1 & 1 & -1 & -1 & 1 & 3 & 3 & 3 & -1 & -1 & -1 & $2\varphi$,$4\varphi$ & No\\
    \hline
    6 & $\Delta m=0$ & $\Delta m=-1$ & -1 & -1 & 1 & 1 & -1 & -3 & -3 & -3 & 1 & 1 & 1 & $2\varphi$,$4\varphi$ & No\\
    \hline
    7 & $\Delta m=+1$ & $\Delta m=0$ & 1 & 1 & 1 & 1 & 1 & 1 & 1 & 1 & 1 & 1 & 1 & No & No\\
    \hline
    8 & $\Delta m=-1$ & $\Delta m=0$ & -1 & -1 & -1 & -1 & -1 & -1 & -1 & -1 & -1 & -1 & -1 &  No & No\\
    \hline
    9 & $\Delta m=0$ & $\Delta m=0$ & 0 & 0 & 0 & 0 & 0 & 0 & 0 & 0 & 0 & 0 & 0 & No & Yes\\
    \hline\hline    
  \end{tabular}
  \caption{The magnetic quantum number of the final continuum state ($M$) for pathways 1 to 11 (denoted as P1 to P11), the oscillation components in SBs as a function of azimuthal angle $\varphi$, and the phase matching situation in experiments, under nine combinations of the laser field polarizations (denoted as Cases 1 to 9). The polarization of the fields is characterized by the dipole selection rules for the magnetic quantum number ($\Delta m$): $\Delta m=0$, $\Delta m=+1$, and $\Delta m=-1$ denote the linearly, left-hand circularly, and right-hand circularly polarized fields, respectively.}
  \label{magQN}
\end{table*}

In calculation, the incoming final continuum state $\psi_{\vec{k}_\pm}^{-}(\vec{r})\coloneqq  \langle\vec{r} \vert\psi_{f\pm} \rangle$ \cite{PhysRev.93.888,Altshuler1956} can be further expanded on the partial wave series as \cite{staraceTheo}
\begin{equation}
  \begin{aligned}
  \psi_{\vec{k}_\pm}^{-}(\vec{r})&=\frac{1}{k_\pm^{1/2}}\sum_{L=0}^{\infty}\sum_{M=-L}^{L}i^Le^{-i(\sigma_{L\pm}+\delta_{L\pm})}\\
  &\times Y^{\ast}_{L,M}(\hat{k}_\pm)R_{E_\pm, L}(r)Y_{L,M}(\hat{r}),
  \end{aligned}
  \label{scawav}
\end{equation}
where $L$ and $M$ are the angular momentum quantum number and the magnetic quantum number of the partial wave, respectively. Here the phase shift due to the short-range potential is $\delta_{L\pm}$ in lithium atom \cite{SARSA2004163}. The Coulombic phase is $\sigma_{L\pm}\coloneqq  \arg[\Gamma(1+L-iZ/k_\pm)]$ with the effective nuclear charge $Z=1$. The energy-normalized radial wavefunction is $R_{E_\pm L}(r)$ with its asymptotic behavior of $\sqrt{2/(\pi k_\pm r)}\sin[k_\pm r-L\pi/2 -Z\ln(2k_\pm r)/k_\pm+\sigma_{L\pm}+\delta_{L\pm}]$ when $r\rightarrow \infty$.

\subsection{\label{subsec:sche}The Rabi-RABBIT scheme using circularly polarized laser fields}

In the traditional RABBIT scheme with linearly polarized XUV and IR fields \cite{PhysRevLett.106.143002}, absorbing (emitting) an IR photon from the retarded ($\tau>0$) IR field $\textbf{E}_{\rm IR}(t-\tau)$ with respect to the XUV field $\textbf{E}_{\rm XUV}(t)$ contributes a phase like $e^{+i\omega\tau}$ ($e^{-i\omega\tau}$) with $\omega>0$ (see footnote \footnotemark[1]) \footnotetext[1]{Alternatively, absorbing an XUV photon from the XUV field $\textbf{E}_{\rm XUV}(t+\tau)$ in advance of the IR field $\textbf{E}_{\rm IR}(t)$ corresponds to an interaction phase of $e^{-i\Omega_{\rm 2q+1}\tau}$ with $\Omega_{\rm 2q+1}>0$. The equivalence of these two perspectives is established upon the energy-preserving condition during the ionization process.}. Therefore, the number of exchanged IR photons in each ionization pathway is imprinted in the $\tau$-related phase of the outgoing electron wave packet. Thus, the relative phase of the outgoing electron wave packets at the same energy via different pathways, named as \textit{RABBIT phase}, can be retrieved from the $2\omega\tau$-modulation of SBs \cite{E_S_Toma_2002}. As discussed in Ref.\,\cite{Sorngard_2020}, by using different combinations of the XUV and IR fields with either linear or circular polarizations (see footnote \footnotemark[2]) \footnotetext[2]{Note that in the coordinate system used here, the $xOy$ plane is defined as the polarization plane of the circularly polarized fields; and the $z$ axis is defined as the propagation direction of the circularly polarized fields and the polarization axis of the linearly polarized fields. This definition of the coordinate system for the circularly polarized fields differs from that used in our TDSE calculations [Eqs.\,(\ref{IR_A}) and (\ref{xuv})], however, the observed physical quantities are unchanged under the rotation [$\mathcal{C}:(\hat{e}_x,\hat{e}_y,\hat{e}_z)\rightarrow \mathcal{C}^\prime:(\hat{e}_z,\hat{e}_x,\hat{e}_y)$] of the coordinate system fixed with an observer.}, it is possible to extract the RABBIT phases from the modulation of SBs as a function of azimuthal angle $\varphi$. The mechanism behind this alternative way (see footnote \footnotemark[3]) \footnotetext[3]{According to the dipole selection rules, the involved ionization channels are different in the cases of linearly and circularly polarized fields and thus the retrieved RABBIT phases are inequivalent.} to extract the RABBIT phases is based on the dipole selection rules: exchanging one circularly polarized IR photon will increase or decrease the magnetic quantum number ($M$). Finally, the number of exchanged circularly polarized IR photons is encoded in the $\varphi$-related phase of the outgoing wave packet via $e^{iM\varphi}$ originating from $Y_{L,M}^\ast(\hat{k}_\pm)$ in Eq.\,(\ref{scawav}).

In principle, we have nine possible combinations of the polarizations of the XUV and IR fields, as listed by Cases 1 to 9 in Tab.\,\ref{magQN}. The corresponding transition operators $\hat{O}_\omega$ and $\hat{O}_\Omega$ used in Eqs.\,(\ref{ionchan}) are $r\cos\theta$, $r\sin\theta e^{i\varphi}/\sqrt{2}$, and $r\sin\theta e^{-i\varphi}/\sqrt{2}$ for linear, left-hand circular, and right-hand circular polarizations, respectively \cite{arfken1999mathematical}. Note that phase matching is only achievable in experiments for Cases 1, 2, 3, 4, and 9. Tab.\,\ref{magQN} gives that, \textit{if only} the circularly polarized IR field is used (Cases 1 to 6), the RABBIT phases can be successfully extracted from the $2\varphi-$oscillations of SBs without scan of the time delay between the XUV and IR fields. In addition, Cases 1 to 6 all give the same interference scheme as using both linearly polarized XUV and IR fields (Case 9). More specifically, $2\varphi$-signal ($4\varphi$-signal) results from the interference of pathways 1, 2, 5 versus 3, 4, 6, 7, 8, 9, 10, 11 (pathways 3, 4, 9, 10, 11 versus 6, 7, 8). As a demonstration, we show as follows how to extract the RABBIT phases from the $2\varphi-$oscillations in the Rabi-RABBIT scheme for \textit{Case 1} (Sec.\,\ref{circular}) and \textit{Case 2} (Sec.\,\ref{circular_coun}).

Substituting Eq.\,(\ref{scawav}) into Eqs.\,(\ref{ionchan}), then the ionization amplitude of pathway $i$ can be written as 
\begin{equation}
  \mathcal{A}_{(i)}(\vec{k}_\pm,\tau)=\sum_{\mathcal{Q}} \mathcal{A}_{(i), \mathcal{Q}}(\vec{k}_\pm,\tau).
  \label{sumIonChan}
\end{equation}
Here $\mathcal{A}_{(i),\mathcal{Q}} $ is the amplitude of a specified ionization channel for pathway $i\in \{1, 2,\,\cdots,11\}$, which is unambiguously characterized by the quantum numbers throughout all the states in transition. In $N-$photon transition, the ensemble of these quantum numbers is defined as $\mathcal{Q}= (l_i, m_i), (\lambda_1,\mu_1), \ldots, (\lambda_{N-1},\mu_{N-1}), (L,M)$, where $l_i$, $\lambda_{N-1}$, and $L$ ($m_i$, $\mu_{N-1}$, and $M$) respectively label the angular (magnetic) quantum numbers of the initial, the $(N-1)$th intermediate, and the final states. The specific quantum numbers in $\mathcal{Q}$ are determined by the selection rules, taking \textit{Case 1} as an example, we have $\Delta m=+1$ ($\Delta m=-1$) for absorption (emission) with $l\geq m$. For a single time-delay ($\tau=0$) RABBIT measurement in \textit{Case 1}, $\mathcal{A}_{(i),\mathcal{Q}}$ can be written as
\begin{equation}
  \mathcal{A}_{(i), \mathcal{Q}}(\vec{k}_\pm,\tau=0)=\frac{\pi}{k_\pm^{1/2}}Y_{L,M}(\hat{k}_\pm)\mathcal{M}_{(i), \mathcal{Q}}(E_\pm),
\end{equation}
where $\hat{k}_\pm=(\theta,\varphi)$ indicates the emission direction of photoelectrons. The amplitudes $\mathcal{M}_{(i), \mathcal{Q}}(E_\pm)$ can be further separated into its angular and radial parts in coordinate representation as (see Appendix \ref{appC:OtherIonAmp} for the other pathways)
\begin{widetext}
\begin{subequations}
  \begin{equation}
    \begin{aligned}
  \mathcal{M}_{(1), \mathcal{Q}}(E_{\pm}) &= \frac{\pi}{3}E_{\omega}E_{\rm 2q-1}i^{-(L+1)}e^{i(\sigma_{L\pm}+\delta_{L\pm}+\phi_{2q-1})}\langle Y_{L,M}\vert Y_{1,1}\vert Y_{\lambda_1,\mu_1}\rangle\langle Y_{\lambda_1,\mu_1}\vert Y_{1,1}\vert Y_{0,0}\rangle\\
  &\times\sum_{\nu_1}\kern-1.5em \int\frac{\langle R_{E_\pm,L} \vert r \vert R_{\nu_1,\lambda_1} \rangle \langle R_{\nu_1,\lambda_1} \vert r \vert R_{2,0} \rangle}{\omega_{2s}\pm \frac{\vert\Omega_{\textrm{R}}^0\vert}{2}+\Omega_{2q-1}-\omega_{\nu_1}},
    \end{aligned}
  \end{equation}
  \begin{equation}
    \begin{aligned}
  \mathcal{M}_{(3), \mathcal{Q}}(E_{\pm}) &= -\frac{\pi}{3}E_{\omega}E_{\rm 2q+1}i^{-(L+1)}e^{i(\sigma_{L\pm}+\delta_{L\pm}+\phi_{2q+1})}\langle Y_{L,M}\vert Y_{1,-1}\vert Y_{\lambda_1,\mu_1}\rangle\langle Y_{\lambda_1,\mu_1}\vert Y_{1,1}\vert Y_{0,0}\rangle\\
  &\times\sum_{\nu_1}\kern-1.5em \int\frac{\langle R_{E_\pm,L}  \vert r \vert R_{\nu_1,\lambda_1} \rangle \langle R_{\nu_1,\lambda_1} \vert r \vert R_{2,0} \rangle}{\omega_{2s}\pm \frac{\vert\Omega_{\textrm{R}}^0\vert}{2}+\Omega_{2q+1}-\omega_{\nu_1}},
    \end{aligned}
  \end{equation}
  \begin{equation}
  \mathcal{M}_{(5), \mathcal{Q}}(E_{\pm}) = -\left(\frac{\pi}{3}\right)^{\frac{1}{2}}\frac{\Omega_{\rm R}^0}{\vert \Omega_{\rm R}^0\vert} E_{\rm 2q-1}i^{-(L\pm 1)}e^{i(\sigma_{L\pm}+\delta_{L\pm}+\phi_{2q-1})}\langle Y_{L,M}\vert Y_{1,1}\vert Y_{1,1}\rangle\langle R_{E_\pm,L}  \vert r \vert R_{2,1} \rangle,
  \end{equation}
  \begin{equation}
  \begin{aligned}
  \mathcal{M}_{(6), \mathcal{Q}}(E_{\pm}) &=-\left(\frac{\pi}{3}\right)^{\frac{3}{2}}\frac{\Omega_{\rm R}^0}{\vert \Omega_{\rm R}^0\vert}E_{\omega}^2E_{\rm 2q-3}i^{-(L\pm 1)}e^{i(\sigma_{L\pm}+\delta_{L\pm}+\phi_{2q-3})}\langle Y_{L,M}\vert Y_{1,1}\vert Y_{\lambda_2,\mu_2}\rangle\langle Y_{\lambda_2,\mu_2}\vert Y_{1,1}\vert Y_{\lambda_1,\mu_1}\rangle\langle Y_{\lambda_1,\mu_1}\vert Y_{1,1}\vert Y_{1,1}\rangle \\
  &\times\sum_{\nu_1,\nu_2}\kern-1.5em \int\frac{\langle R_{E_\pm,L}  \vert r \vert R_{\nu_2,\lambda_2}\rangle \langle R_{\nu_2,\lambda_2} \vert r \vert R_{\nu_1,\lambda_1}\rangle \langle R_{\nu_1,\lambda_1} \vert r \vert R_{2,1}\rangle}{(\omega_{2p}\pm \frac{\vert\Omega_{\textrm{R}}^0\vert}{2}+\Omega_{2q-3}+\omega-\omega_{\nu_2})(\omega_{2p}\pm \frac{\vert\Omega_{\textrm{R}}^0\vert}{2}+\Omega_{2q-3}-\omega_{\nu_1})},
  \end{aligned}
  \end{equation}
  \begin{equation}
  \begin{aligned}
  \mathcal{M}_{(9), \mathcal{Q}}(E_{\pm}) &=  -\left(\frac{\pi}{3}\right)^{\frac{3}{2}}\frac{\Omega_{\rm R}^0}{\vert \Omega_{\rm R}^0\vert}E_{\omega}^2E_{\rm 2q+1}i^{-(L\pm 1)}e^{i(\sigma_{L\pm}+\delta_{L\pm}+\phi_{2q+1})}\langle Y_{L,M}\vert Y_{1,-1}\vert Y_{\lambda_2,\mu_2}\rangle\langle Y_{\lambda_2,\mu_2}\vert Y_{1,-1}\vert Y_{\lambda_1,\mu_1}\rangle\langle Y_{\lambda_1,\mu_1}\vert Y_{1,1}\vert Y_{1,1}\rangle \\
  &\times\sum_{\nu_1,\nu_2}\kern-1.5em \int\frac{\langle R_{E_\pm,L}  \vert r \vert R_{\nu_2,\lambda_2}\rangle \langle R_{\nu_2,\lambda_2} \vert r \vert R_{\nu_1,\lambda_1}\rangle \langle R_{\nu_1,\lambda_1} \vert r \vert R_{2,1}\rangle}{(\omega_{2p}\pm \frac{\vert\Omega_{\textrm{R}}^0\vert}{2}+\Omega_{2q+1}-\omega-\omega_{\nu_2})(\omega_{2p}\pm \frac{\vert\Omega_{\textrm{R}}^0\vert}{2}+\Omega_{2q+1}-\omega_{\nu_1})},
  \end{aligned}
  \end{equation}
  \label{ionchanAng}
\end{subequations}
\end{widetext}
where $\hat{O}_\Omega$ and  $\hat{O}_\omega$ [$\hat{O}^\dagger_\Omega$ and $\hat{O}^\dagger_\omega$] are replaced by $-\sqrt{\frac{4\pi}{3}}rY_{1,1}(\hat{r})$ [$\sqrt{\frac{4\pi}{3}}rY_{1,-1}(\hat{r})$] for absorbing [emitting] each photon. Here the wavefunctions in Eqs.\,(\ref{ionchan}) are separated into their radial part and spherical harmonics in coordinate presentation: $\langle \vec{r}\vert \psi_{2s}\rangle=Y_{0,0}(\hat{r})R_{2,0}(r)$, $\langle \vec{r}\vert \psi_{2p}\rangle=Y_{1,1}(\hat{r})R_{2,1}(r)$, and $\langle \vec{r}\vert \psi_{\nu_i}\rangle=Y_{\lambda_i,\mu_i}(\hat{r})R_{\nu_i,\lambda_i}(r)$ with $\hat{r}=(\theta_r,\varphi_r)$.

In our calculation, for the two- and three-photon transition amplitudes in Eqs.\,(\ref{ionchanAng}), the infinite summation in the radial part is evaluated with the Dalgarno-Lewis method \cite{doi:10.1098/rspa.1955.0246}. Then the radial part can be calculated using perturbed wavefunctions \cite{PhysRevA.90.023403}, which satisfy the inhomogeneous equation with its boundary conditions described in Ref.\,\cite{1981}. Particularly, the integration of two continuum wavefunctions appearing in Eqs.\,(\ref{ionchanAng}) is calculated by using the complex coordinate rotation method \cite{doi:10.1063/1.4903436}. Additionally, verification of the matrix elements has been performed by the extrapolation method \cite{E_Cormier_1995}. We select some numerical results of \textit{dipole transition matrix elements} in Appendix\,\ref{appD:DME}, where their radial and angular parts are separately shown.
\begin{figure*}[htb]
  \centering
  \includegraphics[width=0.8\textwidth]{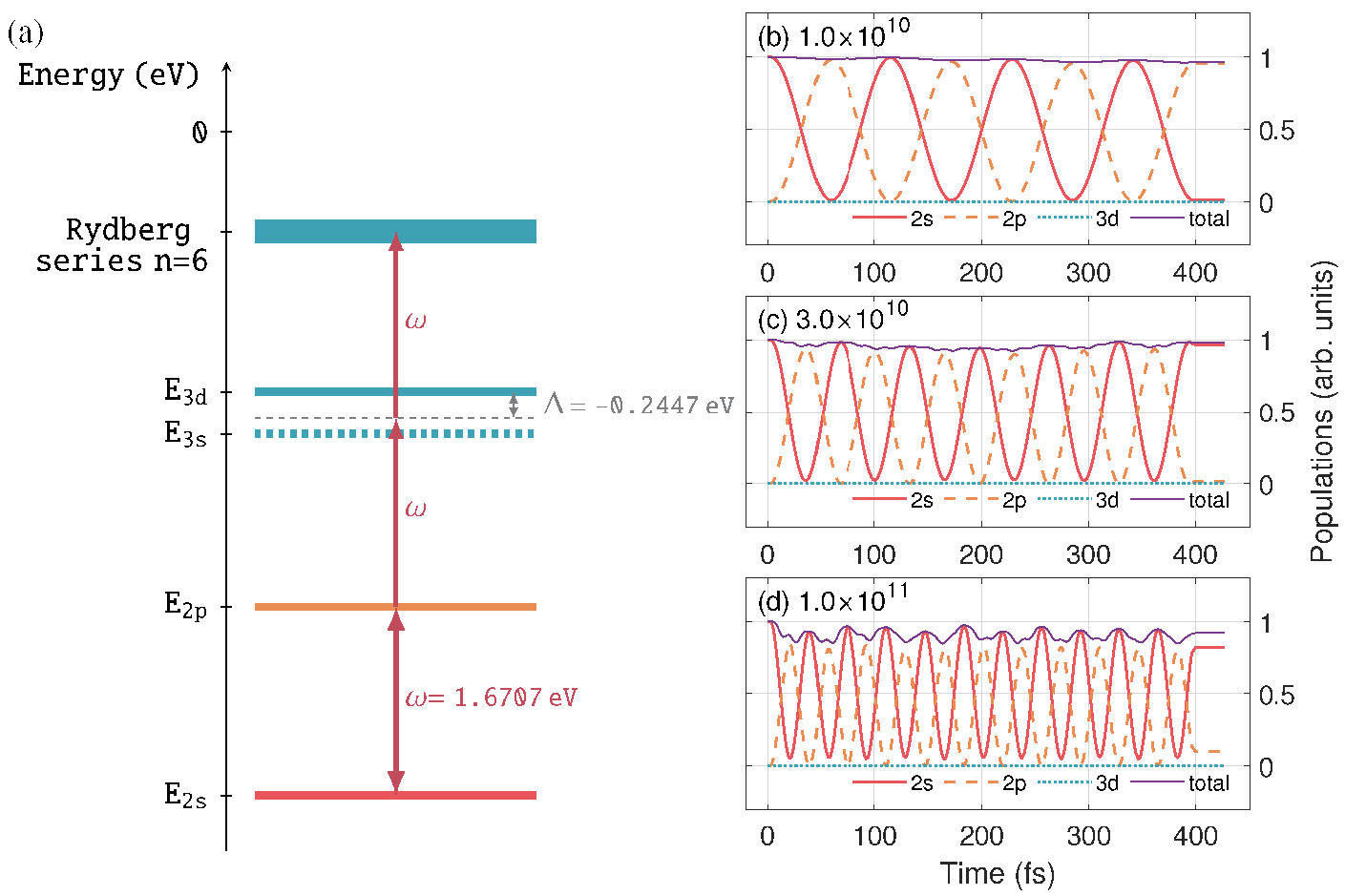}
  \caption{(a) The schematic of the essential states of the lithium atom coupled to the IR field. The double-side (single-side) red arrows denote the strong (weak) coupling. The thick horizontal solid (dashed) lines indicate the states that are allowed (forbidden) to be populated according to the dipole selection rules. The detuning of the 2p-3d transition is $\Delta\coloneqq \omega-E_{3d}+E_{2p}$, with $E_{2p}$ and $E_{3d}$ the energies of the 2p and 3d states of the lithium atom, respectively. (b, c, d) The populations of the 2s (red-solid lines), 2p (orange-dashed lines), 3d states of lithium (bluish-grey-dotted lines) and their summation (thin-purple-solid lines) as a function of the atom-field interaction time (the beginning time is set as zero). The intensities of the left-hand circularly polarized IR field are (b) $1\times 10^{10}$ (corresponding to the energy spacing of 0.0367\,eV), (c) $3\times 10^{10}$ (corresponding to the energy spacing of 0.0642\,eV), and (d) $1\times 10^{11}\rm W/cm^2$ (corresponding to the energy spacing of 0.1154\,eV).}
  \label{fig:Cpop}
\end{figure*}

\section{\label{res}Results}
\subsection{\label{circular}\textit{Case 1}: Rabi-RABBIT using the left-hand circularly polarized XUV and IR fields}
\begin{figure*}[htb]
  \centering
  \includegraphics[width=0.65\textwidth]{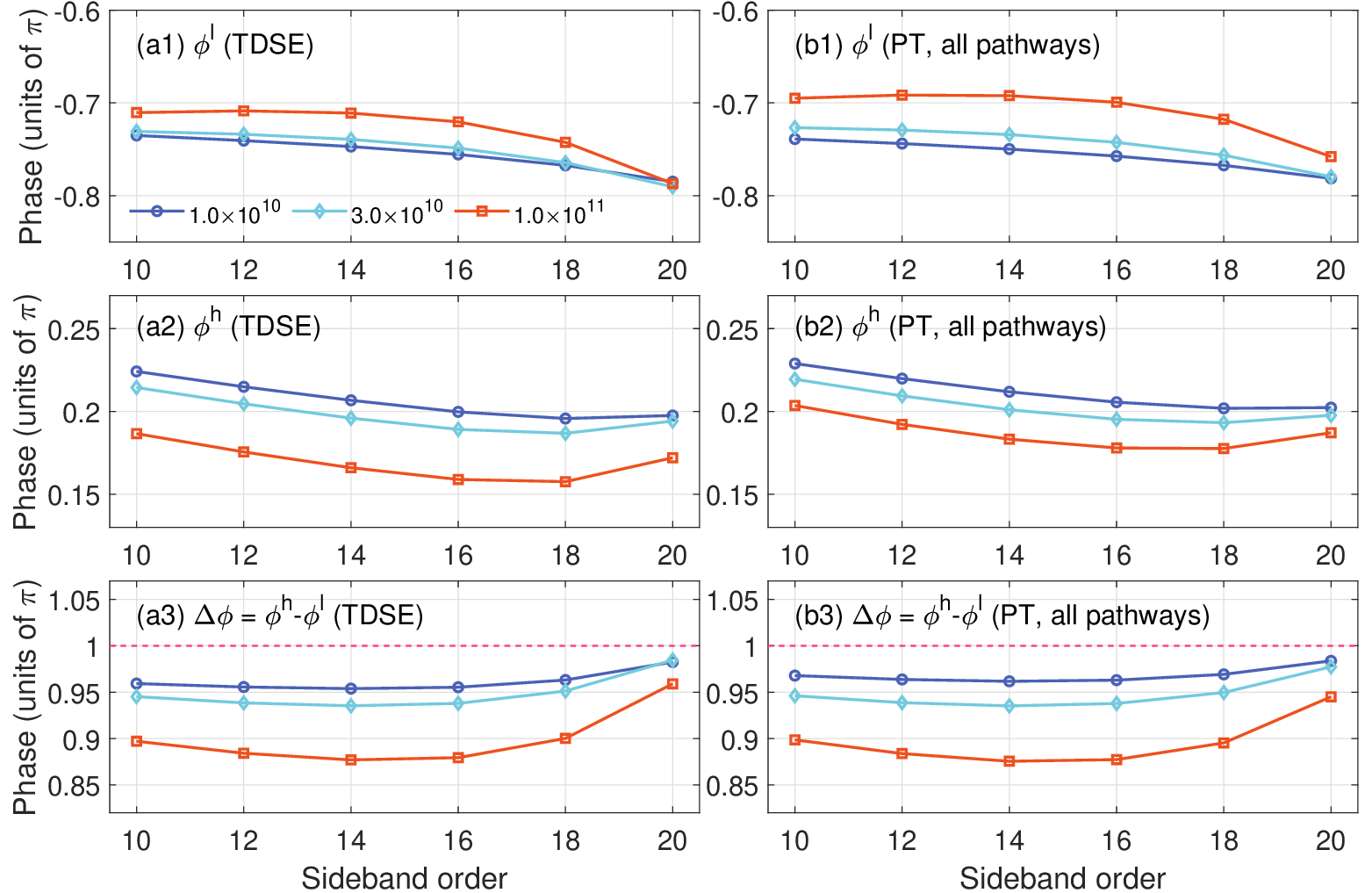}
  \caption{From top to bottom: (a1, b1) The RABBIT phases extracted from the photoelectron spectra within the polarization plane of the left-hand circularly polarized XUV and IR fields for $\rm P^l$ as a function of the photoelectron energy. (a2, b2) The same as (a1, b1), but for $\rm P^h$. (a3, b3) The relative RABBIT phases between $\rm P^h$ and $\rm P^l$, as a function of the photoelectron energy. The different columns correspond to the results obtained (a1, a2, a3) by solving the TDSE and (b1, b2, b3) by perturbation theory including all pathways, respectively. The purple circles, blue rhombuses and orange squares correspond to the IR intensities of $1\times 10^{10}$, $3\times 10^{10}$ and $1\times 10^{11}\rm W/cm^2$, respectively.}
  \label{fig:CpolALL}
\end{figure*}
Figure \ref{fig:Cpop}(b) shows the populations of the 2s ($m=0$), 2p ($m=1$), and 3d ($m=2$) states of the lithium atom as a function of the atom-field interaction time, which are approximately obtained by projecting the time-dependent TDSE solution to the eigenstates of the field-free Hamiltonian $H_0$ near the zero values of the vector potential of the external field. Within the dipole approximation, the time-dependent coefficients obtained by projecting the wavefunction onto the unperturbed eigenstates can be interpreted as probability amplitudes when using the full Hamiltonian written in the length gauge \cite{10.1119/1.11264}. Moreover, when using the velocity gauge, the calculated projection coefficients coincide with the ones employing the length gauge \textit{near the zeros of the vector potential} \cite{PhysRevA.106.053115}. Here the intensities of the left-hand circularly polarized XUV and IR fields are $\rm 1\times10^{13}$ and $\rm 1\times10^{10} W/cm^2$, respectively. As shown in Fig.\,\ref{fig:Cpop}(b), the populations of the 2s and 2p states oscillate out of phase with the interaction time, as a demonstration of Rabi oscillations \cite{PhysRev.51.652}. As the IR intensity increases, the Rabi oscillations between the 2s and 2p states become faster, as shown in Figs.\,\ref{fig:Cpop}(c) and \ref{fig:Cpop}(d) for the IR intensities of $\rm 3\times10^{10}$ and $\rm 1\times10^{11} W/cm^2$, respectively. As displayed in Fig.\,\ref{fig:Cpop}(a), the 3d ($m=2$) state [the 3s state ($m=0$)] is allowed [forbidden] to be populated through absorbing two IR photons from the 2s state ($m=0$) according to the dipole selection rule $\Delta m=+1$ for the left-hand circular polarization. Due to the large detuning in the 2p-3d transition, adopting the circular polarized IR field prevents the population leakage from the Rabi subspace $\mathcal{R}$, as shown in Figs.\,\ref{fig:Cpop}(b)-\ref{fig:Cpop}(d).

During the entire interaction time, the Rabi oscillations shown in Figs.\,\ref{fig:Cpop}(b)-\ref{fig:Cpop}(d) are smooth and uniform for the three IR intensities, respectively validating the rotating wave approximation in solving the Rabi amplitudes [Eq.\,(\ref{Hr_int})] and the monochromatic approximation for the IR field in Eqs.\,(\ref{ionchan}). Moreover, there is nearly no dampening in the Rabi oscillations for the three IR intensities due to the relatively negligible transition (including ionization) from $\mathcal{R}$ to $\mathcal{S}$, as indicated by the total population of the 2s, 2p, and 3d states after the laser pulses end in Figs.\,\ref{fig:Cpop}(b)-\ref{fig:Cpop}(d). Therefore, it is a good approximation to treat the ionization \textit{perturbatively} on the top of the Rabi wavefunction $\Psi_{\mathcal{R}}$ [Eq.\,(\ref{rabi_wfn})] which describes \textit{perfectly} ongoing Rabi oscillations within $\mathcal{R}$ subspace. In addition, the incomplete Rabi oscillations are owing to the energy shifts of the 2s and 2p states in the presence of the external fields, which are neglected in our analytical treatment owing to their small effects.

\subsubsection{\label{circular:pol}Comparison of TDSE and perturbation theory (PT): RABBIT phases extracted from the photoelectron spectra within the polarization plane}
Figure \ref{fig:CpolALL} shows the RABBIT phases as a function of photoelectron energy, which are extracted from the $2\varphi$-oscillation of SBs in the photoelectron spectra within the polarization plane of the XUV and IR fields. Note that the TDSE simulations and the PT calculations are performed in different coordinates without changing the physical observables (see footnote \footnotemark[2]). Here and hereafter the polarization plane refers to the $xOy$ plane with $\theta=\pi/2$ as defined in the PT calculations. Figures\,\ref{fig:CpolALL}(a1) and \ref{fig:CpolALL}(a2) respectively show the RABBIT phases $\rm \phi^l(\pi/2)$ and $\rm \phi^h(\pi/2)$ of the peaks with the lower ($\rm P^l$) and higher ($\rm P^h$) energies in the AT doublet for the three IR intensities, which are calculated by solving the TDSE. Despite the tiny energy spacing between the AT doublet $\rm P^l$ and $\rm P^h$, $\rm \phi^l(\pi/2)$ and $\rm \phi^h(\pi/2)$ behave quite differently as a function of the photoelectron energy, which reveals the influence of the Rabi dynamics on the phase of the ionized electron wave packets. For both $\rm \phi^l(\pi/2)$ and $\rm \phi^h(\pi/2)$, their specific variations with the photoelectron energy rely on the IR intensity. In addition, $\rm \phi^l(\pi/2)$ and $\rm \phi^h(\pi/2)$ obviously show a change in its slope before and after SB 16. Figure \ref{fig:CpolALL}(a3) shows the relative RABBIT phase $\rm \Delta\phi(\pi/2)=\phi^h(\pi/2)-\phi^l(\pi/2)$ as a function of the photoelectron energy for the three IR intensities, which are obtained by solving the TDSE. For all the three IR intensities here, $\Delta\phi(\pi/2)$ varies in a similar way with the photoelectron energy and it is close to $\pi$. In addition, $\Delta\phi(\pi/2)$ deviates more from $\pi$ for the higher IR intensity. Likewise, there is an obvious change in the slope of $\Delta\phi(\pi/2)$ around SB 16. Figures \ref{fig:CpolALL}(b1), \ref{fig:CpolALL}(b2), and \ref{fig:CpolALL}(b3) respectively show the RABBIT phases $\rm\phi^l(\pi/2)$, $\rm\phi^h(\pi/2)$, and $\Delta\phi(\pi/2)$ calculated by PT including \textit{all} pathways shown in Fig.\,\ref{fig:schem} and using the same laser intensities and polarizations as the TDSE simulations. By comparing Figs.\,\ref{fig:CpolALL}(b1), \ref{fig:CpolALL}(b2), and \ref{fig:CpolALL}(b3) separately with Figs.\,\ref{fig:CpolALL}(a1), \ref{fig:CpolALL}(a2), and \ref{fig:CpolALL}(a3), their agreement validates the PT in Sec.\,\ref{subsec:pert} on describing the physics process in the Rabi-RABBIT scheme for all the three IR intensities. In addition, the agreement between the PT and TDSE results are better at lower IR intensities when the Rabi oscillations within $\mathcal{R}$ are more accurately described by the zeroth-order wavefunction $\Psi_{\mathcal{R}}$ [Eq.\,(\ref{rabi_wfn})] due to smaller energy shifts of the 2s and 2p states, less population leakage to $\mathcal{S}$, and less dampening in the Rabi oscillations [Fig.\,\ref{fig:Cpop}].

\begin{figure*}[htb]
  \centering
  \includegraphics[width=\textwidth]{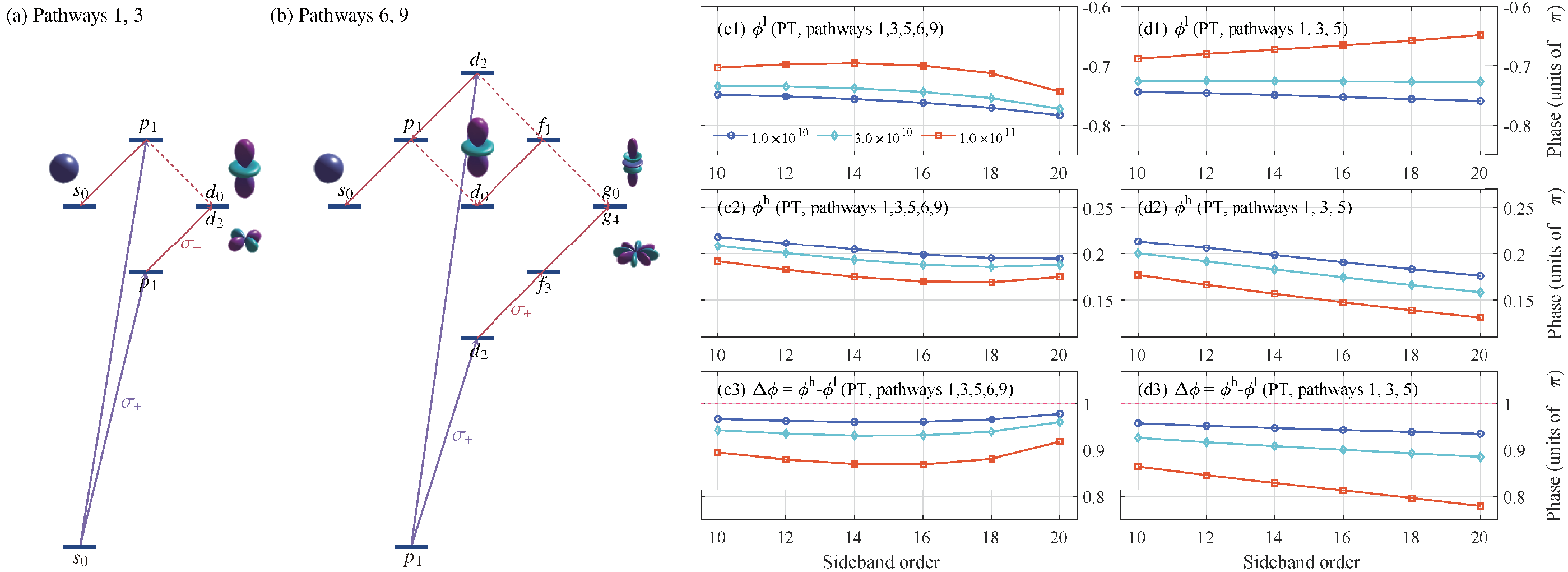}
  \caption{(a) The schematic of the ionization channels for pathways 1 and 3 when using the left-hand circularly polarized (denoted as $\sigma_{+}$) XUV and IR fields. (b) The same as (a), but for pathways 6 and 9. The purple and red arrows indicate the transition via exchanging the XUV and IR photons, respectively. The solid (dashed) arrows denote the relatively more (less) probable transition according to the propensity rule in laser-assisted photoionization when comparing each step of absorption and emission from the same state. The partial waves of each ionization channel are illustrated as the real spherical harmonics represented on polar plots. Pathway 5, having the same partial waves as pathway 1, is not shown. From top to bottom of the two rightmost columns: (c1, d1) The RABBIT phases extracted from the photoelectron spectra within the polarization plane of the circularly polarized fields for $\rm P^l$ as a function of the photoelectron energy. (c2, d2) The same as (c1, d1), but for $\rm P^h$. (c3, d3) The relative RABBIT phases between $\rm P^h$ and $\rm P^l$, as a function of the photoelectron energy. The different columns correspond to the results obtained by perturbation theory only including (c1, c2, c3) pathways 1, 3, 5, 6, 9 and (d1, d2, d3) pathways 1, 3, 5, respectively. The purple circles, blue rhombuses and orange squares correspond to the IR intensities of $1\times 10^{10}$, $3\times 10^{10}$ and $1\times 10^{11}\rm W/cm^2$, respectively.}
  \label{fig:CpolMCTP9_P135}
\end{figure*}

To specifically reveal the underlying physics in the Rabi-RABBIT scheme, we analyze the PT results based on the two reduced subsets containing important ionization pathways in Fig.\,\ref{fig:schem}: \textit{Subset five} consists of the \textit{essential} pathways 1, 3, 5, 6, and 9; and \textit{Subset three} is composed of pathways 1, 3, and 5. As a demonstration, Figs.\,\ref{fig:CpolMCTP9_P135}(a) and \ref{fig:CpolMCTP9_P135}(b) show the possible ionization channels of several essential pathways in \textit{Case 1} and their relative strengths, which are determined by the dipole selection rules and the propensity rule in laser-assisted photoionization \cite{PhysRevLett.123.133201,Bertolino_2020}. Pathway 1 has only one partial wave $\varepsilon d_2$ (here and hereafter the continuum partial wave is denoted as $\varepsilon l_m$ with $l$ the angular quantum number and $m$ the magnetic quantum number) via one channel and pathway 3 has two partial waves $\varepsilon s_0$ and $\varepsilon d_0$ via two possible channels, as shown in Fig.\,\ref{fig:CpolMCTP9_P135}(a). Pathway 5 (not shown) has the same partial wave as pathway 1. Pathway 6 has only one partial wave $\varepsilon g_4$ via one channel and pathway 9 has three partial waves $\varepsilon s_0$, $\varepsilon d_0$, and $\varepsilon g_0$ via four possible channels, as shown in Fig.\,\ref{fig:CpolMCTP9_P135}(b). 

The two rightmost columns in Fig.\,\ref{fig:CpolMCTP9_P135} show the RABBIT phases extracted from the $2\varphi$-oscillations of SBs in the photoelectron spectra within the polarization plane of the XUV and IR fields ($xOy$ plane with $\theta=\pi/2$) as a function of the photoelectron energy, which are calculated by PT including part of pathways in Fig.\,\ref{fig:schem} and using the same laser intensities and polarizations as the TDSE simulations. Figures \ref{fig:CpolMCTP9_P135}(c1), \ref{fig:CpolMCTP9_P135}(c2), and \ref{fig:CpolMCTP9_P135}(c3) respectively show the RABBIT phases $\rm\phi^l(\pi/2)$, $\rm\phi^h(\pi/2)$, and $\Delta\phi(\pi/2)$ calculated only including pathways of \textit{Subset five}, which exhibits an excellent agreement with the PT results including \textit{all} pathways plotted in Figs.\,\ref{fig:CpolALL}(b1), \ref{fig:CpolALL}(b2), and \ref{fig:CpolALL}(b3). Their further agreement with the TDSE results plotted in Figs.\,\ref{fig:CpolALL}(a1), \ref{fig:CpolALL}(a2), and \ref{fig:CpolALL}(a3) indicates that the TDSE results can be accurately interpreted by only analyzing the interference among pathways of \textit{Subset five}. Our calculations show that, in the formation of the 2$\varphi$-oscillation signal, the interference of pathways 3 and 5 has the most dominant contribution throughout all the photoelectron energy for the three IR intensities, followed by the interference of the usual RABBIT pathways 3 and 1 which has the second most important contribution. Furthermore, when comparing the phase of the 2$\varphi$-oscillation signal between the AT doublet $\rm P^l$ and $\rm P^h$, the interference of pathways 3 and 5 (3 and 1) will give a $\pi$ (zero) difference between the AT doublet $\rm P^l$ and $\rm P^h$, recalling that there is a $\pi$ difference between the initial phases of the electrons ionized from the dressed 2p states $\rm\vert \phi_{2p}^l\rangle$ and $\rm\vert \phi_{2p}^h\rangle$ due to the sine-like Rabi amplitude $\sin(\Omega_{\rm R}^0t/2)\sim e^{i\Omega_{\rm R}^0t/2}-e^{-i\Omega_{\rm R}^0t/2}$ of the 2p state [Eq.\,(\ref{coef})]. Therefore, the dominance of the interference of pathways 3 and 5 explains why $\Delta\phi(\pi/2)$ is very close to $\pi$ throughout all photoelectron energies for all the three IR intensities as shown in Figs.\,\ref{fig:CpolALL}(a3), \ref{fig:CpolALL}(b3), and \ref{fig:CpolMCTP9_P135}(c3), which is in qualitative accordance with the explanations only based on \textit{Subset three} in Ref.\,\cite{PhysRevA.105.063110}. In contrast to the Rabi-RABBIT scheme using the linearly polarized fields in Ref.\,\cite{PhysRevA.105.063110}, adopting the circularly polarized IR field prevents the 3s ($m=0$) from being populated and thus we have a more perfect two-level Rabi system within $\mathcal{R}$, as mentioned in the previous section. So here we aim to make a quantitative interpretation of the energy- and polar-angle-dependent RABBIT phases (time-delays) including more processes.

For comparison, Figs.\,\ref{fig:CpolMCTP9_P135}(d1), \ref{fig:CpolMCTP9_P135}(d2), and \ref{fig:CpolMCTP9_P135}(d3) respectively show the RABBIT phases $\rm\phi^l(\pi/2)$, $\rm\phi^h(\pi/2)$, and $\Delta\phi(\pi/2)$ calculated with only pathways of \textit{Subset three} included. The RABBIT phases $\rm\phi^l(\pi/2)$, $\rm\phi^h(\pi/2)$, and $\Delta\phi(\pi/2)$ all change monotonously with the photoelectron energy for the three IR intensities. In particular, the relative phase $\Delta\phi(\pi/2)$ deviates more from $\pi$ at higher photoelectron energies and for higher IR intensities because the relative contribution of the interference of pathways 3 and 1 with respect to that between pathways 3 and 5 increases with the photoelectron energy and with the IR intensity. This relatively increasing contribution of the interference between pathways 3 and 1 with \textit{the photoelectron energy} results from a slower decline in the radial integral of the dipole moment matrix element associated with pathway 1 starting from the 2s state compared to that of pathway 5 initiated from the 2p state, as implied by the one-photon ionization cross sections from the 2s and 2p states for the photoelectron energies from 10\,eV to 30\,eV shown in Fig.\,\ref{fig:CrossSec}. The relatively increasing contribution of the interference between pathways 3 and 1 with \textit{the IR intensity} is attributed to a higher order of exchanging IR photons in the continuum for pathway 1 than pathway 5. Note that the photoelectrons with higher kinetic energies also interact more strongly with the laser field.

Compared to the monotonous tendencies of $\rm\phi^l(\pi/2)$, $\rm\phi^h(\pi/2)$, and $\Delta\phi(\pi/2)$ in the PT results for \textit{Subset three}, the PT results for \textit{Subset five} show some extra “bending” structures from SB 14 to SB 20 in the cases of $\rm\phi^l(\pi/2)$, $\rm\phi^h(\pi/2)$, and $\Delta\phi(\pi/2)$, which implies the non-negligible influence of the interference of pathways 5 with pathways 6 and 9. Indeed, our calculations indicate that these "bending" structures are attributed to the competition between the interference of pathways 5 and 6 versus the interference of pathways 1 and 3. Similar to the interference of pathways 1 and 3 both starting from the 2s state, the interference of pathways 5 and 6 both initiated from the 2p state also gives a zero difference in the phase of the $2\varphi$-oscillations between the AT doublet $\rm P^l$ and $\rm P^h$. However, the interference between the zeroth-IR-order pathway 5 and the third-IR-order pathway 6 has an additional $\sim\pi$ phase compared to the interference between the two first-IR-order usual RABBIT pathways 1 and 3 due to the interaction phase shift accumulated during the absorption or emission of IR photons in multiphoton above threshold ionization \cite{PhysRevResearch.3.013270}. In this sense, the interference of pathways 5 and 6 will cancel in strength with the interference of pathways 1 and 3. As the photoelectron energy increases, the strength of the interference between pathways 5 and 6 becomes more comparable with (but still \textit{weaker} than) the interference of pathways 1 and 3 from SB 16 to SB 20 for the three IR intensities, which is mainly determined by the relative amplitudes of the absorbed XUV harmonics in pathways 3 and 6 [e.g., $E_{2q-3}$, $E_{2q-1}$, and $E_{2q+1}$ for SB$_{\rm 2q}$ in Eqs.\,(\ref{ionchan})]. Hence, the summed strength of the interferences of pathways 1, 3 and of pathways 5, 6 increases from SB 10 to SB 14 and then decreases from SB 16 to SB 20. Recall that the most dominant interference of pathways 3 and 5 contributes a \textit{Rabi $\pi$ phase} in the formation of the $2\varphi$-oscillations between the AT doublet $\rm P^l$ and $\rm P^h$. As a superposition effect of the three pairs of interferences among pathways 1, 3, 5, and 6, $\Delta\phi(\pi/2)$ is near $\pi$ while it is driven further from $\pi$ from SB 10 to SB 14 and then becomes closer to $\pi$ from SB 16 to SB 20, as shown in Figs.\,\ref{fig:CpolALL}(a3), \ref{fig:CpolALL}(b3), and \ref{fig:CpolMCTP9_P135}(c3).
\begin{figure*}[htb]
  \centering
  \includegraphics[width=0.8\textwidth]{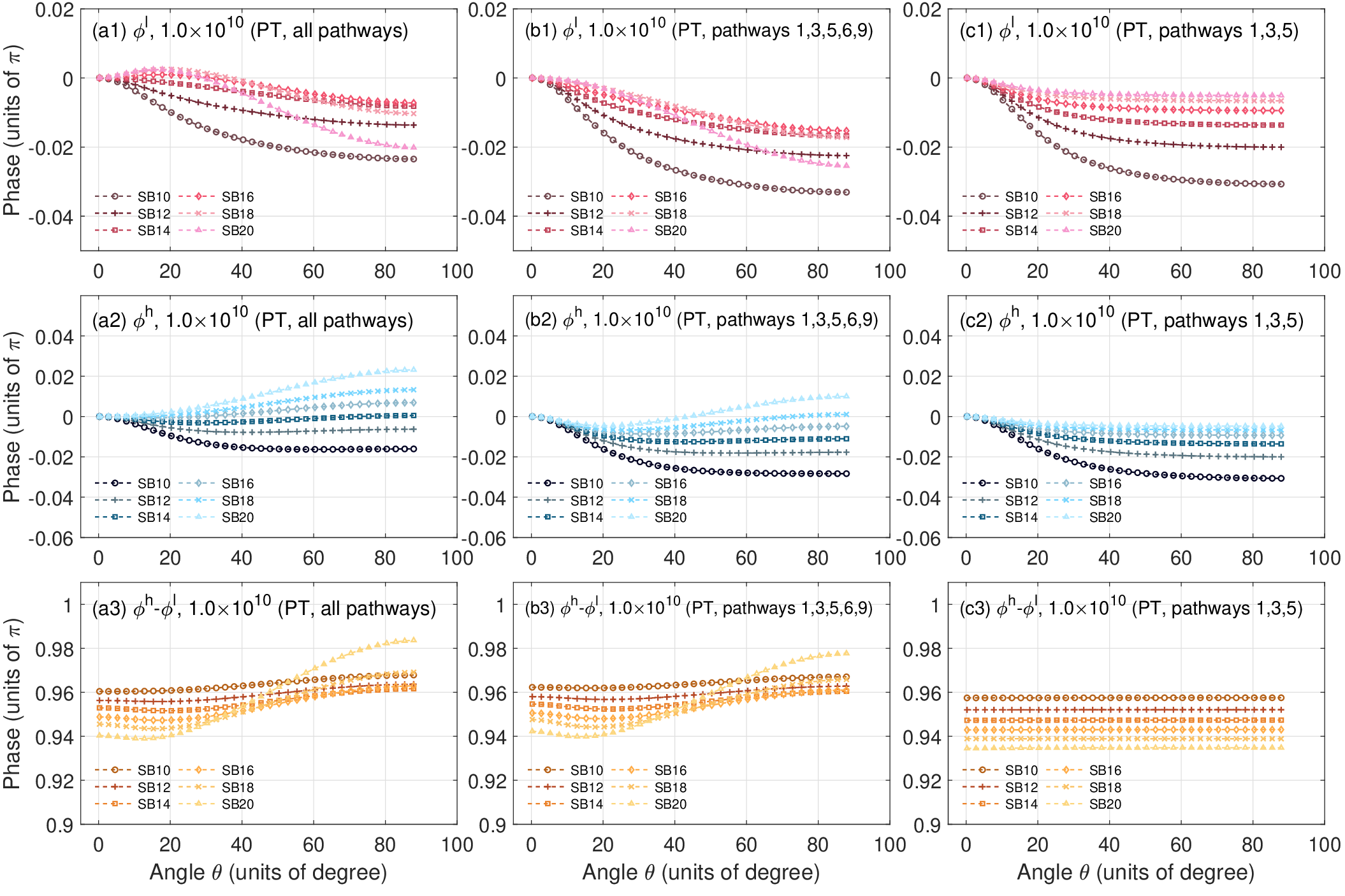}
  \caption{From top to bottom: (a1, b1, c1) The relative polar-angle-resolved RABBIT phases for $\rm P^l$ as a function of polar emission angle of photoelectrons, with the intensities of the left-hand circularly polarized XUV and IR fields $\rm 1\times10^{13}$ and $\rm 1\times10^{10} W/cm^2$, respectively. (a2, b2, c2) The same as (a1, b1, c1), but for $\rm P^h$. (a3, b3, c3) The relative phases between $\rm P^h$ and $\rm P^l$, as a function of the polar emission angle of photoelectrons. The different columns respectively correspond to the results obtained by perturbation theory including (a1, a2, a3) all pathways; (b1, b2, b3) only pathways 1, 3, 5, 6, and 9; and (c1, c2, c3) only pathways 1, 3, and 5. The circles, pluses, squares, rhombuses, crosses and triangles correspond to SBs 10, 12, 14, 16, 18 and 20, respectively.}
  \label{fig:Cang1e10MCTP9_P135}
\end{figure*}

\subsubsection{\label{circular:thetaRes}Detailed analysis of PT results: polar-angle-resolved RABBIT phases}
As shown in Sec.\,\ref{circular:pol}, our PT in Sec.\,\ref{subsec:pert} is reliable to quantitatively reproduce the TDSE results, especially for low IR intensities in \textit{Case 1}. Compared to the TDSE simulations, PT has the advantage of specifying the amplitude and the phase of each ionization channel so that the essential processes can be accurately identified in order to uncover the underlying mechanism. In the following, we apply PT for more detailed examinations on the polar-angle-integrated and polar-angle-resolved RABBIT phases in \textit{Case 1} (Sec.\,\ref{circular:thetaRes}) and \textit{Case 2} (Sec.\,\ref{circular_coun}). Our PT calculations for \textit{Case 1} show that, the phases extracted from the $2\varphi$-oscillations in SBs of the polar-angle-integrated photoelectron spectra (polar-angle-integrated RABBIT phases) have similar behaviors as the RABBIT phases $\rm\phi^l(\pi/2)$, $\rm\phi^h(\pi/2)$, and $\Delta\phi(\pi/2)$ due to the similar competition among the essential pathways of \textit{Subset five}. In this sense, only the polar-angle-resolved RABBIT phases for the lowest IR intensity will be discussed in this section.
\begin{figure*}[htb]
  \centering
  \includegraphics[width=\textwidth]{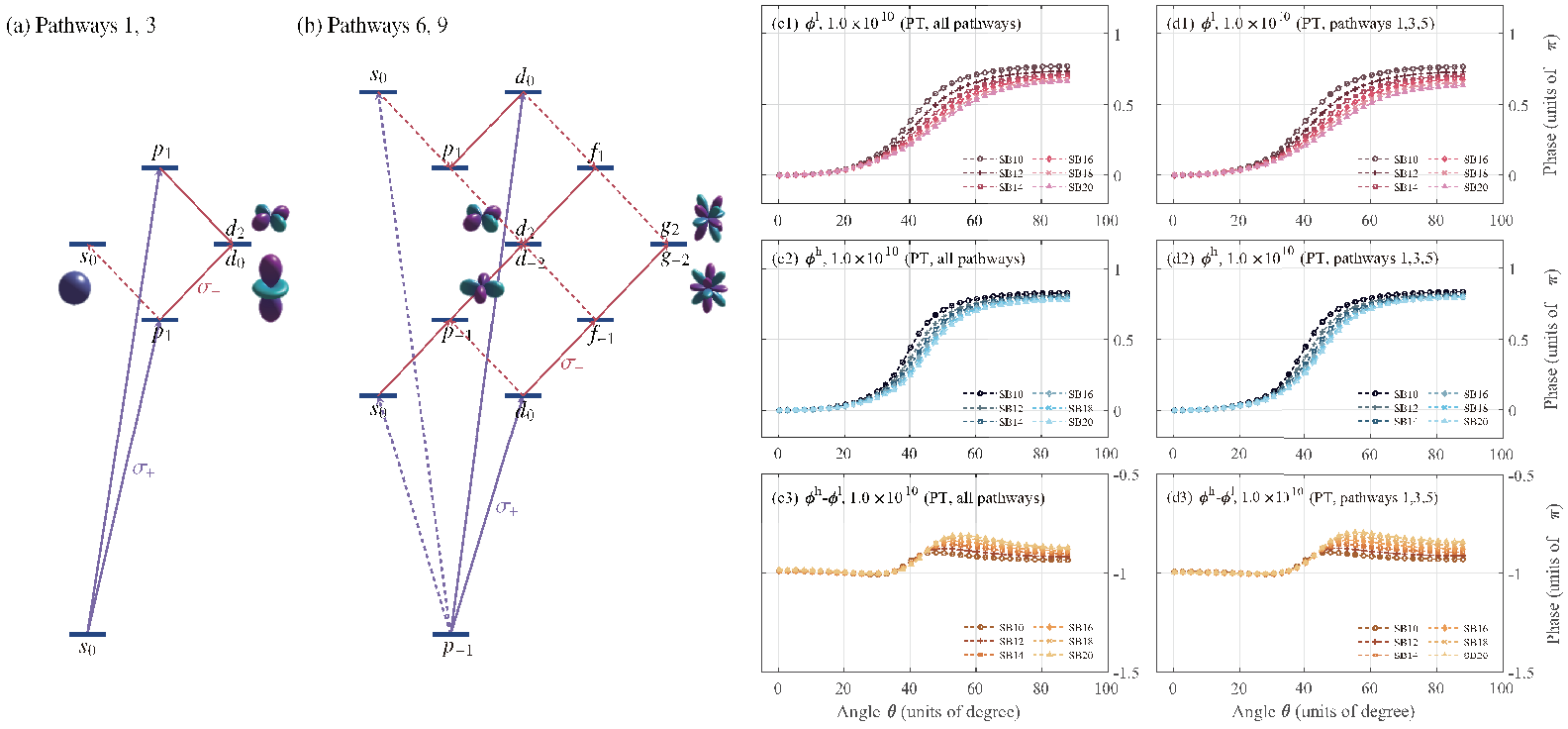}
  \caption{(a) The schematic of the ionization channels for pathways 1 and 3 when using the left-hand circularly polarized (denoted as $\sigma_{+}$) XUV field and the right-hand circularly polarized (denoted as $\sigma_{-}$) IR field. (b) The same as (a), but for pathways 6 and 9. The purple and red arrows indicate the transition via exchanging the XUV and IR photons, respectively. The solid (dashed) arrows denote the relatively more (less) probable transition according to Fano's propensity rule and the propensity rule in laser-assisted photoionization when comparing each step of absorption and emission from the same state. The partial waves of each ionization channel are illustrated as the real spherical harmonics represented on polar plots. Pathway 5, having the same partial waves as pathway 1, is not shown. From top to bottom of the two rightmost columns: (c1, d1) The polar-angle-resolved RABBIT phases for $\rm P^l$ as a function of polar emission angle of photoelectrons, with the intensities of the XUV and the IR fields $\rm 1\times10^{13}$ and $\rm 1\times10^{10} W/cm^2$, respectively. (c2, d2) The same as (c1, d1), but for $\rm P^h$. (c3, d3) The relative RABBIT phases between $\rm P^h$ and $\rm P^l$, as a function of the polar emission angle of photoelectrons. The different columns correspond to the results obtained by perturbation theory including (c1, c2, c3) all pathways and (d1, d2, d3) only pathways 1, 3, 5, respectively. The circles, pluses, squares, rhombuses, crosses and triangles correspond to SBs 10, 12, 14, 16, 18 and 20, respectively.}
  \label{fig:Coun_Cang1e10MCTP9_P135}
\end{figure*}

Figure \ref{fig:Cang1e10MCTP9_P135} shows the RABBIT phases as a function of polar emission angle $\theta$ of photoelectrons, which are extracted from the $2\varphi$-oscillation of SBs in the three-dimensional photoelectron spectra. Here the intensities of the left-hand circularly polarized XUV and IR fields are $\rm 1\times10^{13}$ and $\rm 1\times10^{10} W/cm^2$, respectively. Figure \ref{fig:Cang1e10MCTP9_P135}(a1) [Fig.\,\ref{fig:Cang1e10MCTP9_P135}(a2)] shows the polar-angle-resolved RABBIT phases $\rm\phi^l(\theta)$ [$\rm\phi^h(\theta)$] of the peak $\rm P^l$ [$\rm P^h$] in the AT doublet for SBs 10 to 20, which are calculated with \textit{all} pathways in Fig.\,\ref{fig:schem} included. The phases $\rm\phi^l(\theta)$ and $\rm\phi^h(\theta)$ are given relative to a small polar angle $\theta_0=0.25^\circ$. Note that only the partial waves associated with the zero magnetic quantum number ($M=0$) contribute to the signal along the propagation direction ($\theta=0^\circ$) in the photoelectron spectra and thus there is no $2\varphi$-oscillation for $\theta=0^\circ$. The RABBIT phases $\rm\phi^l(\theta)$ and $\rm\phi^h(\theta)$ both gently vary with polar angle $\theta$ for all SBs. For each SB, the obvious and counterintuitive discrepancy between the behaviors of $\rm\phi^l(\theta)$ and $\rm\phi^h(\theta)$ as a function of polar angle $\theta$ uncovers the influence of the Rabi dynamics on the phase of the ionized electron wave packets. Besides, the specific behaviors of both $\rm\phi^l(\theta)$ and $\rm\phi^h(\theta)$ depend on the photoelectron energy. For SBs 10 to 14, $\rm\phi^l(\theta)$ and $\rm\phi^h(\theta)$ both change smoothly and monotonously with polar angle $\theta$. For SBs 16 to 20, however, $\rm\phi^l(\theta)$ and $\rm\phi^h(\theta)$ both exhibit some fluctuations and bendings in the vicinities of $\theta=20^\circ$ and $\theta=80^\circ$ as a function of polar angle $\theta$. In addition, the curvatures of their bendings depend on the photoelectron energy. Figure \ref{fig:Cang1e10MCTP9_P135}(a3) shows the relative RABBIT phases $\rm \Delta\phi(\theta)=\rm\phi^h(\theta)-\rm\phi^l(\theta)$ as a function of polar angle $\theta$ for SBs 10 to 20. The relative phase $\Delta\phi(\theta)$ is close to $\pi$ at all polar emission angles $\theta$ for all SBs and it increases with polar angle $\theta$ for all SBs. Similarly, $\Delta\phi(\theta)$ varies smoothly with polar angle $\theta$ for SBs 10 to 14 while $\Delta\phi(\theta)$ shows a downward and an upward bendings separately near $\theta=20^\circ$ and $\theta=80^\circ$ for SBs 14 to 20. In addition, the slope of the increase of $\Delta\phi(\theta)$ becomes sharper and the curvatures of the twice bendings become bigger as the photoelectron energy increases. Figures\,\ref{fig:Cang1e10MCTP9_P135}(b1), \ref{fig:Cang1e10MCTP9_P135}(b2), and \ref{fig:Cang1e10MCTP9_P135}(b3) respectively show the RABBIT phases $\rm\phi^l(\theta)$, $\rm\phi^h(\theta)$, and $\Delta\phi(\theta)$ as a function of polar angle $\theta$ for SBs 10 to 20, which are calculated with only pathways of \textit{Subset five} included. By comparing the first two columns in Fig.\,\ref{fig:Cang1e10MCTP9_P135}, it is seen that \textit{Subset five} can already reproduce the results in the first column (including \textit{all} pathways in Fig.\,\ref{fig:schem}) well, for both the location of the “bending” structures and their dependence on the photoelectron energy.

For comparison, we calculate the RABBIT phases $\rm\phi^l(\theta)$, $\rm\phi^h(\theta)$, and $\Delta\phi(\theta)$ for SBs 10 to 20 by only including pathways in \textit{Subset three}, which are respectively shown in Figs.\,\ref{fig:Cang1e10MCTP9_P135}(c1), \ref{fig:Cang1e10MCTP9_P135}(c2), and \ref{fig:Cang1e10MCTP9_P135}(c3). In Figs.\,\ref{fig:Cang1e10MCTP9_P135}(c1) and \ref{fig:Cang1e10MCTP9_P135}(c2), $\rm\phi^l(\theta)$ and $\rm\phi^h(\theta)$ both decrease monotonously with polar angle $\theta$. In Fig.\,\ref{fig:Cang1e10MCTP9_P135}(c3), $\Delta\phi(\theta)$ is flat as a function of polar angle $\theta$ for all SBs. Our calculations show that the interference between pathways 3 and 5 (3 and 1) has the most (second most) important contribution in the formation of the $2\varphi$-oscillation signal throughout all emission polar angles $\theta$ for all SBs. This dominance of the interference between pathways 3 and 5 explains why $\Delta\phi(\theta)$ is close to a \textit{Rabi $\pi$ phase} throughout all emission polar angles $\theta$ for all SBs in Figs.\,\ref{fig:Cang1e10MCTP9_P135}(a3), \ref{fig:Cang1e10MCTP9_P135}(b3), and \ref{fig:Cang1e10MCTP9_P135}(c3), rather than the usual near zero RABBIT phase at high photoelectron energies in consideration of the interaction phase shift \cite{PhysRevResearch.3.013270}. Furthermore, the relative strength of the interference between pathways 3 and 1 compared to that between pathways 3 and 5 increases with the photoelectron energy and thus $\Delta\phi(\theta)$ deviates more from $\pi$ at higher photoelectron energies in Fig.\,\ref{fig:Cang1e10MCTP9_P135}(c3) [and at higher IR intensities (not shown)]. For each SB, because pathways 1 and 5 have the same partial wave $\varepsilon d_2$, the relative strength of the interference between pathways 3 and 1 with respect to that between pathways 3 and 5 keeps constant as a function of polar angle $\theta$ and thus $\Delta\phi(\theta)$ is flat with polar angle $\theta$ in Fig.\,\ref{fig:Cang1e10MCTP9_P135}(c3).

The discrepancy between the two rightmost columns suggests that the interferences between pathways 5, 6 and between pathways 5, 9 are responsible for the observed “bending” structures in Figs.\,\ref{fig:Cang1e10MCTP9_P135}(a3) and \ref{fig:Cang1e10MCTP9_P135}(b3). Similarly to the interference of pathways 5 and 6, the interference of pathways 5 and 9 also gives a zero difference in the phase of the $2\varphi$-oscillations between the AT doublet $\rm P^l$ and $\rm P^h$ and it also cancels in strength with the interference of pathways 1 and 3 due to the $\pi$ interaction phase existing in the third-IR-order pathway 9. At small polar angles $\theta$, the amplitude of the partial wave $\varepsilon g_4$ of pathway 6 is nearly zero [Fig.\,\ref{fig:CpolMCTP9_P135}(b)] and so as the strength of the interference of pathways 5 and 6. Therefore, the behavior of $\Delta\phi(\theta)$ near $\theta=20^\circ$ in Figs.\,\ref{fig:Cang1e10MCTP9_P135}(a3) and \ref{fig:Cang1e10MCTP9_P135}(b3) is mainly determined by the cancellation on the interference of pathways 1 and 3 by the interference of pathways 5 and 9, which depends on the relative strength of these two pairs of interferences. For SBs 10 and 12, the relative strength of the interference between pathways 5 and 9 monotonously increases with polar angle $\theta$ and thus there is no bending for SBs 10 and 12. For SBs 14 to 20, their relative strength shows a dip in the vicinity of $\theta=20^\circ$ and this dip becomes deeper with the increasing photoelectron energy. Correspondingly, the curvatures of the downward bending around $\theta=20^\circ$ are bigger at higher energies for SBs 14 to 20. At big polar angles $\theta$, the interference of pathways 5 and 6 dominates over that of pathways 5 and 9 for SBs 16 to 20 (which is mainly determined by the relative amplitudes of the absorbed XUV harmonics). Hence, the upward shape of the bending structure near $\theta=80^\circ$ for SBs 16 to 20 in Figs.\,\ref{fig:Cang1e10MCTP9_P135}(a3) and \ref{fig:Cang1e10MCTP9_P135}(b3) is mainly attributed to the cancellation on the interference of pathways 1 and 3 by the interference of pathways 5 and 6. For SBs 16 to 20, the relative strength of the interference between pathways 5 and 6 monotonously increases with polar angle $\theta$ and its increase becomes more gentle as $\theta$ approaches $80^\circ$. Moreover, the slope of its increase in their relative strength is sharper for the SBs at higher energies and thus the curvature of the upward bending near $\theta=80^\circ$ is bigger with increasing photoelectron energy, as shown in Figs.\,\ref{fig:Cang1e10MCTP9_P135}(a3) and \ref{fig:Cang1e10MCTP9_P135}(b3). Note that for the emission pathways 3 and 9, the partial wave $\varepsilon d_0$ ($\varepsilon g_0$) has a node at $\theta\approx 54.58^\circ$ (two nodes at $\theta\approx 34.54^\circ$ and $\theta\approx 69.85^\circ$). Nevertheless, no abrupt phase jumps occur in the interference of pathway 3 with pathway 1 nor 5 because the $\varepsilon s_0$ wave dominates over the $\varepsilon d_0$ wave in pathway 3 according to the propensity rule in laser-assisted photoionization \cite{PhysRevLett.123.133201}. Likewise, in pathway 9 where two IR photons are emitted in the continuum, 
each step of emission favors decreasing the electron angular momentum \cite{Bertolino_2020} and thus there is also no phase jump in the interference between pathways 5 and 9.

\subsection{\label{circular_coun}\textit{Case 2}: Rabi-RABBIT using the left-hand circularly polarized XUV field and right-hand circularly polarized IR field}
As shown in Tab.\,\ref{magQN}, the RABBIT phases can be retrieved from the $2\varphi-$oscillations of SBs when the circularly polarized IR field is used. According to the dipole selection rules, the ionization channels differ when changing the polarizations of the laser fields in the Rabi-RABBIT scheme and so do the behaviors of the RABBIT phases. As demonstrated in Sec.\,\ref{circular}, when the circularly polarized XUV and IR fields are co-rotating, the behaviors of the RABBIT phases are mainly determined by the interferences among pathways of \textit{Subset five}. As a comparison, in this section, we apply PT to examine the polar-angle-resolved (Sec.\,\ref{circular_coun:thetaRes}) and polar-angle-integrated (Sec.\,\ref{circular_coun:thetaInt}) RABBIT phases in \textit{Case 2} of Tab.\,\ref{magQN} where the XUV and IR fields have opposite circular polarizations (counter-rotating). 
\subsubsection{\label{circular_coun:thetaRes}Polar-angle-resolved RABBIT phases}
\begin{figure*}[htb]
  \centering
  \includegraphics[width=0.65\textwidth]{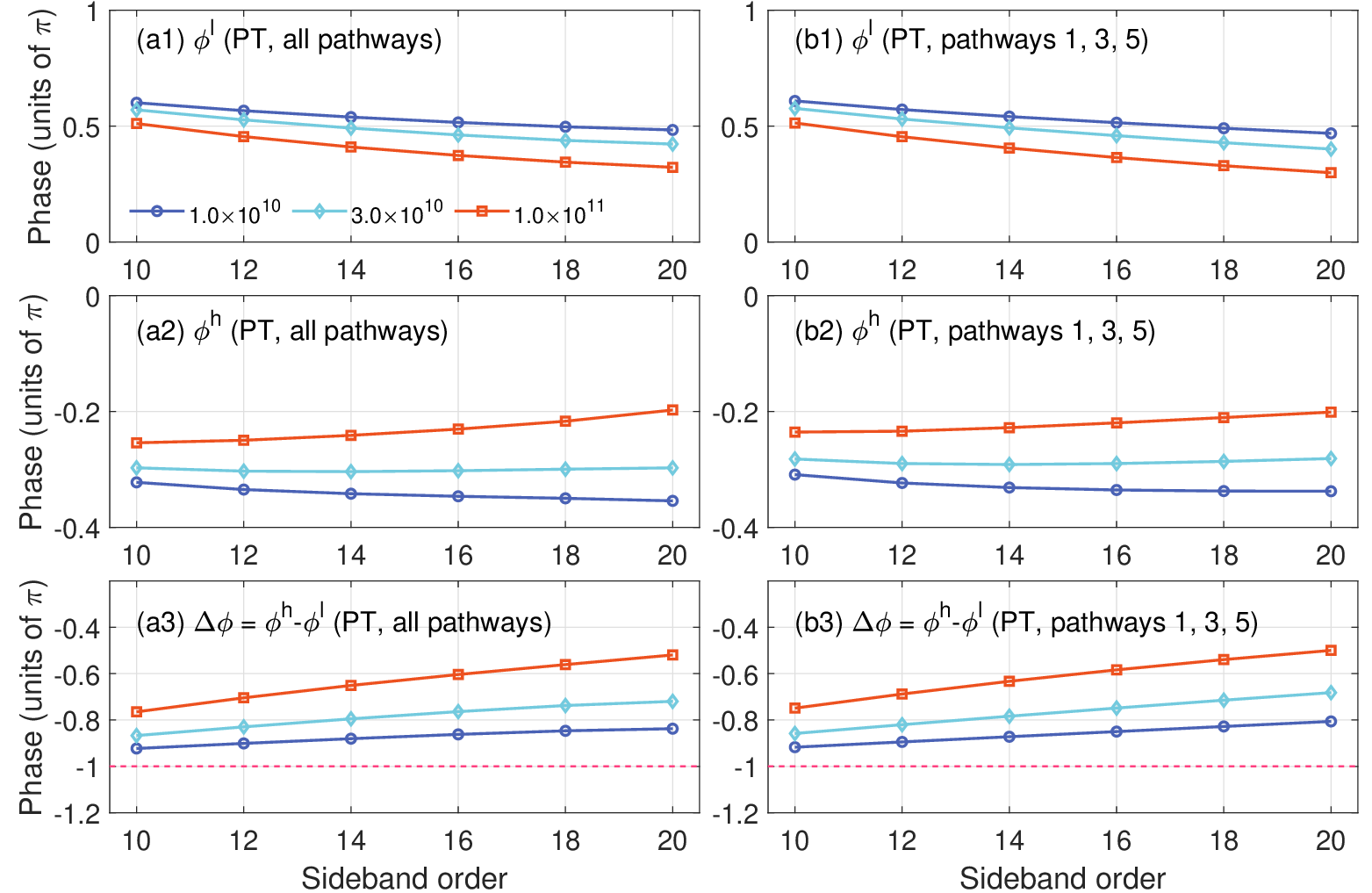}
  \caption{From top to bottom: (a1, b1) The polar-angle-integrated RABBIT phases for $\rm P^l$ as a function of the photoelectron energy. (a2, b2) The same as (a1, b1), but for $\rm P^h$. (a3, b3) The relative RABBIT phases between $\rm P^h$ and $\rm P^l$, as a function of the photoelectron energy. The different columns correspond to the results obtained by perturbation theory including (a1, a2, a3) all pathways and (b1, b2, b3) only pathways 1, 3, and 5, respectively. The purple circles, blue rhombuses and orange squares correspond to the IR intensities of $1\times 10^{10}$, $3\times 10^{10}$ and $1\times 10^{11}\rm W/cm^2$, respectively.}
  \label{fig:Coun_CthetaIntMCTP9_P135}
\end{figure*}
As a demonstration, Figs.\,\ref{fig:Coun_Cang1e10MCTP9_P135} (a) and \ref{fig:Coun_Cang1e10MCTP9_P135}(b) show the possible ionization channels of several pathways in \textit{Case 2} and their relative strengths, which are determined by the dipole selection rules, Fano's propensity rule \cite{PhysRevA.32.617}, and the propensity rule in laser-assisted photoionization \cite{PhysRevLett.123.133201,Bertolino_2020}. Pathway 1 has two partial waves $\varepsilon s_0$ and $\varepsilon d_0$ via two possible channels and pathway 3 only has one partial wave $\varepsilon d_2$ via one channel, as shown in Fig.\,\ref{fig:Coun_Cang1e10MCTP9_P135} (a). Pathway 5 (not shown) has the same two partial waves as pathway 1. Pathway 6 has two partial waves $\varepsilon d_{-2}$ and $\varepsilon g_{-2}$ via four possible channels and pathway 9 has two partial waves $\varepsilon d_2$ and $\varepsilon g_2$ via four possible channels, as shown in Fig.\,\ref{fig:Coun_Cang1e10MCTP9_P135} (b). 

The two rightmost columns in Fig.\,\ref{fig:Coun_Cang1e10MCTP9_P135} show the RABBIT phases as a function of polar emission angle $\theta$ of photoelectrons, which are extracted from the $2\varphi$-oscillations of SBs in the three-dimensional photoelectron spectra. Here the intensities of the left-hand circularly polarized XUV field and the right-hand circularly polarized IR field are $1\times 10^{13}$ and $1\times 10^{10}\rm W/cm^2$, respectively. Figure \ref{fig:Coun_Cang1e10MCTP9_P135}(c1) [Fig.\,\ref{fig:Coun_Cang1e10MCTP9_P135}(c2)] shows the polar-angle-resolved RABBIT phases $\rm\phi^l(\theta)$ [$\rm\phi^h(\theta)$] of the peak $\rm P^l$ [$\rm P^h$] in the AT doublet for SBs 10 to 20, which are calculated by including \textit{all} pathways in Fig.\,\ref{fig:schem}. Here $\rm\phi^l(\theta)$ and $\rm\phi^h(\theta)$ are given relative to a small polar angle $\theta_0=0.25^\circ$. In contrast to co-rotating \textit{Case 1}, when adopting the counter-rotating XUV and IR fields in \textit{Case 2}, $\rm\phi^l(\theta)$ and $\rm\phi^h(\theta)$ both exhibit an obvious phase jump around $\theta\approx 40^\circ$ for all SBs, as shown in Figs.\,\ref{fig:Coun_Cang1e10MCTP9_P135}(c1) and \ref{fig:Coun_Cang1e10MCTP9_P135}(c2). In addition, the steepness of the phase jump in $\rm\phi^h(\theta)$ is slightly sharper than that of $\rm\phi^l(\theta)$ for all SBs. For both $\rm\phi^l(\theta)$ and $\rm\phi^h(\theta)$, the steepness of the phase jump becomes more gentle and the amplitude of the phase jump becomes smaller with the increasing photoelectron energy. Figure \ref{fig:Coun_Cang1e10MCTP9_P135}(c3) shows the relative RABBIT phases $\rm \Delta\phi(\theta)=\rm\phi^h(\theta)-\rm\phi^l(\theta)$ as a function of polar angle $\theta$ for SBs 10 to 20. For all SBs, the relative phase $\Delta\phi(\theta)$ is close to $\pi$ and exhibits a phase jump of around 0.25$\pi$ near $\theta\approx 40^\circ$. As the photoelectron energy increases, the amplitude of the phase jump in $\Delta\phi(\theta)$ becomes bigger. 

Figures\,\ref{fig:Coun_Cang1e10MCTP9_P135}(d1), \ref{fig:Coun_Cang1e10MCTP9_P135}(d2), and \ref{fig:Coun_Cang1e10MCTP9_P135}(d3) respectively show the RABBIT phases $\rm\phi^l(\theta)$, $\rm\phi^h(\theta)$, and $\Delta\phi(\theta)$ calculated by only involving pathways in \textit{Subset three}. The results agree excellently with the PT results including \textit{all} pathways. This agreement implies that, in \textit{Case 2}, only the interferences of pathway 3 with pathways 1 and 5 have the determinant influence on the behaviors of the RABBIT phases. Indeed, different from \textit{Case 1} where pathways 5 and 6 only have a single partial wave (Fig.\,\ref{fig:CpolMCTP9_P135}), pathways 5, 6, and 9 all have two partial waves in \textit{Case 2}. More importantly, for each of pathways 5, 6, and 9 in \textit{Case 2}, their two partial waves will interfere destructively due to the opposite signs of their dipole moment matrix elements originating from the angular integrals. Thus, the interferences of pathway 5 with pathways 6 and 9, as coherent summations of all the ionization channels, are negligible compared to that of the interferences of pathway 3 with pathways 1 and 5. Focusing on the interferences among pathways in \textit{Subset three}, the interference of pathways 3 and 5 dominates over that of pathways 3 and 1 throughout all polar angles $\theta$ for all SBs except in the vicinity of $\theta\approx 40^\circ$ where the interference of pathways 3 and 5 has a minimum in strength. Correspondingly, the relative phase $\Delta\phi(\theta)$ is close to a \textit{Rabi $\pi$ phase} in Figs.\,\ref{fig:Coun_Cang1e10MCTP9_P135}(c3) and \ref{fig:Coun_Cang1e10MCTP9_P135}(d3). In addition, the abrupt jumps of $\rm\phi^l(\theta)$ and $\rm\phi^h(\theta)$ respectively in Figs.\,\ref{fig:Coun_Cang1e10MCTP9_P135}(d1) and \ref{fig:Coun_Cang1e10MCTP9_P135}(d2) for all SBs are attributed to the dominance of the $\varepsilon d_0$ wave over the $\varepsilon s_0$ wave in pathway 5 according to Fano's propensity rule favoring to increase angular quantum number in absorption of one photon from the dressed 2p states \cite{PhysRevA.32.617}. As the $\varepsilon d_0$ wave changes its sign as a function of polar angle around its node $\theta\approx 54.5^\circ$, the $\varepsilon d_0$ wave destructively (constructively) interferes with the nodeless $\varepsilon s_0$ wave in pathway 5 at the angles smaller (bigger) than $54.5^\circ$, in reminiscent of the \textit{opposite} signs of the dipole moment matrix elements related to the $\varepsilon s_0$ and $\varepsilon d_0$ waves. Therefore, the phase jumps occur in the vicinity of $\theta=40^\circ$ smaller than the $\varepsilon d_0$ node at $54.5^\circ$ where the interference between pathways 3 and 5 has a minimum in strength. Note that this differs from the case in Ref.\,\cite{PhysRevLett.123.133201} where the phase jumps locate around $\theta=75^\circ$ larger than the $\varepsilon d_0$ node at $54.5^\circ$ because their $\varepsilon s_0$ and $\varepsilon d_0$ waves have the \textit{same} signs when using the linearly polarized XUV and IR fields. In Figs.\,\ref{fig:Coun_Cang1e10MCTP9_P135}(d1) and \ref{fig:Coun_Cang1e10MCTP9_P135}(d2), no phase jump is observed for neither $\rm\phi^l(\theta)$ nor $\rm\phi^h(\theta)$ in the vicinity of $\theta=15^\circ$ where the interference of pathways 3 and 1 has a minimum in strength due to its negligible contribution compared to the interference of pathways 3 and 5. Furthermore, the exact behaviors of the phase jumps are mainly determined by the relative strength and the phase of the interferences of pathway 3 with pathways 1 and 5. In particular, the slightly sharper steepness of the phase jump in $\rm\phi^h(\theta)$ compared to that of $\rm\phi^l(\theta)$ emphasizes the fact that there is a $\pi$ difference between the initial phases of the electrons ionized from the dressed 2p states $\rm\vert \phi_{2p}^l\rangle$ and $\rm\vert \phi_{2p}^h\rangle$ [Eq.\,(\ref{coef})].

\subsubsection{\label{circular_coun:thetaInt}Polar-angle-integrated RABBIT phases}
Figure \ref{fig:Coun_CthetaIntMCTP9_P135} shows the RABBIT phases as a function of the photoelectron energy, which are extracted from the 2$\varphi$-oscillations in the photoelectron spectra integrated along polar emission angle $\theta$ of photoelectrons. Figures\,\ref{fig:Coun_CthetaIntMCTP9_P135}(a1) and \ref{fig:Coun_CthetaIntMCTP9_P135}(a2) respectively show the polar-angle-integrated RABBIT phases $\rm \bar{\phi}^l$ and $\rm \bar{\phi}^h$ of the peaks the AT doublet $\rm P^l$ and $\rm P^h$ in the AT doublet, which are calculated by including all pathways in Fig.\,\ref{fig:schem}. Here the intensity of the left-hand circularly polarized XUV field is $1\times 10^{13}\rm W/cm^2$. The intensities of the right-hand circularly polarized IR field are $1\times 10^{10}$, $3\times 10^{10}$, and $1\times 10^{11}\rm W/cm^2$, respectively. In Fig.\,\ref{fig:Coun_CthetaIntMCTP9_P135}(a1), $\rm\bar{\phi}^l$ decreases with the photoelectron energy for the three IR intensities and it is less positive for the higher IR intensity. In Fig.\,\ref{fig:Coun_CthetaIntMCTP9_P135}(a2), $\rm \bar{\phi}^h$ is less negative for the higher IR intensity and it shows different dependence on the photoelectron energy for different IR intensities. At the highest (lowest) IR intensity, $\rm \bar{\phi}^h$ increases (decreases) with the photoelectron energy. At the moderate IR intensity, $\rm \bar{\phi}^h$ keeps nearly constant with the photoelectron energy. Figure\,\ref{fig:Coun_CthetaIntMCTP9_P135}(a3) shows the relative RABBIT phase $\rm \Delta\bar{\phi}=\bar{\phi}^h-\bar{\phi}^l$ as a function of the photoelectron energy for the three IR intensities. For all the three IR intensities here, $\rm \Delta\bar{\phi}$ is close to a \textit{Rabi $\pi$ phase} for all the photoelectron energies, which is explained by the dominance of the interference between pathways 3 and 5. 

As a comparison, Figs.\,\ref{fig:Coun_CthetaIntMCTP9_P135}(b1), \ref{fig:Coun_CthetaIntMCTP9_P135}(b2), and \ref{fig:Coun_CthetaIntMCTP9_P135}(b3) respectively show the RABBIT phases $\rm\bar{\phi}^l$, $\rm\bar{\phi}^h$, and $\Delta\bar{\phi}$ calculated by only including pathways of \textit{Subset three}, which shows an excellent agreement with the PT results including \textit{all} pathways in Figs.\,\ref{fig:Coun_CthetaIntMCTP9_P135}(a1), \ref{fig:Coun_CthetaIntMCTP9_P135}(a2), and \ref{fig:Coun_CthetaIntMCTP9_P135}(a3). This agreement stresses the determinant roles of the interferences of pathway 3 with pathways 1 and 5 in \textit{Case 2}. Furthermore, the relative contribution of the interference of pathways 1 and 3 with respect to that of pathways 3 and 5 increases with the photoelectron energy and with the IR intensity. Correspondingly, the relative phase $\Delta\bar{\phi}$ deviates more from a \textit{Rabi $\pi$ phase} at higher photoelectron energies and for higher IR intensities in Figs.\,\ref{fig:Coun_CthetaIntMCTP9_P135}(a3) and \ref{fig:Coun_CthetaIntMCTP9_P135}(b3). Moreover, the opposite dependence of the $\rm\bar{\phi}^l$ and $\rm\bar{\phi}^h$ on the photoelectron energy shown in Figs.\,\ref{fig:Coun_CthetaIntMCTP9_P135}(a2) and \ref{fig:Coun_CthetaIntMCTP9_P135}(b2) is attributed to the $\pi$ difference between the initial phases of the electrons ionized from the dressed 2p states $\rm\vert \phi_{2p}^l\rangle$ and $\rm\vert \phi_{2p}^h\rangle$.

\section{\label{sec:con}CONCLUSION}
We have carefully examined the RABBIT technique applied to a Rabi-cycling atom by utilizing circularly polarized XUV and IR fields. In the RABBIT measurements, the circularly polarized IR field induced Rabi oscillations between the 2s and 2p states of lithium, which ensured \textit{minimal} population leakage to the other bound states. Besides, the use of circularly polarized laser fields circumvented repetitive RABBIT measurements by scanning the time delay between the XUV and IR fields, which may make experimental verification more feasible for the proposed phase-matched \textit{Cases 1 and 2}. In a single time-delay RABBIT measurement, the interference phase of the ionized electron wave packets was retrieved from the photoemission anisotropy along the azimuthal direction. As expected, both polar-angle-integrated and polar-angle-resolved photoelectron spectra exhibited a near $\pi$ phase difference between each AT doublet, as a manifestation of the modulating populations of the two Rabi states \cite{PhysRevA.105.063110}. In addition, adopting the circular polarizations exhibits a potential to steer the phase of the emitted electron wave packets by controlling the ionization channels. By using co- or counter-rotating XUV and IR fields, the polar-angle-integrated and polar-angle-resolved RABBIT phases exhibited different behaviors, which was traced back to different competition among partial waves determined by propensity rules \cite{PhysRevA.32.617,PhysRevLett.123.133201,Bertolino_2020}. In the co-rotating case, the polar-angle-integrated RABBIT phases showed a bending structure as a function of the photoelectron energy while the polar-angle-resolved RABBIT phases showed no phase jumps as a function of polar angle. These complex phenomena were captured by the competition among the essential ionization channels of \textit{Subset five}. In the counter-rotating case, the polar-angle-integrated RABBIT phases varied monotonously with the photoelectron energy while the polar-angle-resolved RABBIT phases showed phase jumps near $\theta=40^\circ$, which were explained by the competition among the essential ionization channels of \textit{Subset three}. 

We discussed how the interference pattern in photoelectron spectra encoded the information of the modulating populations of the two Rabi states. We believe the underlying physics can be generalized to the Rabi process studied by other similar interferometric schemes \cite{PhysRevA.108.053117} and in more complicated molecular \cite{Pan2023,PhysRevLett.91.023002} and solid systems \cite{PhysRevA.58.R2648,PhysRevA.106.053103}. Besides, the pump-probe interferometry used here can also be adopted to investigate the modulating population in a multi-level system \cite{PhysRevA.54.1586,PhysRevA.13.1962,PhysRevA.93.052324}. In addition, the analysis of the interference among multiple pathways is also applicable in other interferometric schemes \cite{PhysRevLett.109.083001,PhysRevA.103.022834,Maroju2023}.

Furthermore, the periodicity of the RABBIT process is determined by both the physical process of interest and the measurement process. In the RABBIT measurement on the autoionization process \cite{PhysRevA.71.060702}, the broken periodicity leads to a broadening \cite{Gruson734} and a frequency modulation \cite{PhysRevLett.113.263001,PhysRevA.93.023429} of the spectral SB peaks. In the RABBIT measurement of Rabi oscillations, the periodic Rabi process manifests itself as a splitting of the observed SB peaks in the energy domain, and the periodicity of the RABBIT process is characterized commonly by the Rabi frequency and the laser probe frequency. When these two frequencies are incommensurable, there is no repetitive interference pattern as a function of the time delay between the two fields in the photoelectron spectra due to the non-periodic RABBIT process. This breakdown of the periodicity becomes visible when the two SB peaks overlap in the AT doublet by using the laser fields with broader spectral widths (e.g., the Rainbow RABBIT method \cite{Gruson734}). In this case, assuming the IR field is monochromatic, the modulations are like $\vert\Omega_{\textrm{R}}^0\vert\tau$ within the overlap of the two SB peaks \cite{PhysRevResearch.3.L032052} while the modulations are like $2\omega\tau$ outside of their overlap according to Eqs.\,(\ref{ionchan}). Furthermore, the modulation frequency of SB peaks should include the finite-pulse effect \cite{PhysRevLett.113.263001,PhysRevA.93.023429} when both laser fields are spectrally broad.

\begin{acknowledgments}
Y. L. acknowledges stimulating discussions with Zdeněk Mašín, Serguei Patchkovskii, Zhen Zhao, Mattias Bertolino, David Busto, Dominik Hoff, and Anne L'Huillier. This work was supported by National Key Research and Development Program of China (Grant No.\,2019YFA0308300), National Natural Science Foundation of China (Grants No.\,12374264, No.\,12021004 and No.\,92250303), and Shaanxi Science Foundation (Grant No.\,2022JC-DW-05). J. M. D. acknowledges support from the Knut and Alice Wallenberg Foundation (Grant No.\,2019.0154) and the Olle Engkvist's Foundation (Grant No.\,194-0734). The computing work in this paper is supported by the Public Service Platform of High Performance Computing provided by Network and Computing Center of HUST. 
\end{acknowledgments}

\appendix
\section{\label{appB:RabiWF}Solving the Schr\"odinger equation within the Rabi subspace}
In the two-level Rabi subspace $\mathcal{R}$, the Schr\"odinger equation is written as
\begin{equation}
  i \frac{d}{dt} \vert\Psi_{\mathcal{R}}(t)\rangle = H_{\mathcal{R}} \vert\Psi_{\mathcal{R}}(t)\rangle,
  \label{r_equ_ap}
\end{equation}
where the Hamiltonian $H_{\mathcal{R}}$ is defined in Eq.\,(\ref{RabiH}). The ansatz of this equation is a superposition of the 2s and 2p states
\begin{equation}
  \vert\Psi_{\mathcal{R}}(t)\rangle=C_{2s}(t;\tau)e^{-i\omega_{2s}t}\vert \psi_{2s}\rangle+C_{2p}(t;\tau)e^{-i\omega_{2p}t}\vert \psi_{2p}\rangle,
  \label{rabi_wfn_ap}
\end{equation}
which encodes the unitary transform $U_r=\exp(i\omega_{2s}t)\vert \psi_{2s} \rangle\langle\psi_{2s}\vert+\exp(i\omega_{2p}t)\vert \psi_{2p} \rangle\langle\psi_{2p}\vert$. In the condition at resonance $\omega=\omega_{2p}-\omega_{2s}$, the full transform can be decomposed into two successive unitary transforms as $U_r=S_0 S_r$, where $S_0=\exp[i(\omega_{2s}+\omega_{2p})t/2]\mathbb{I}$ subtracts the global energy offset $(\omega_{2s}+\omega_{2p})/2$ with $\mathbb{I}=\vert \psi_{2s} \rangle\langle\psi_{2s}\vert+\vert \psi_{2p} \rangle\langle\psi_{2p}\vert$ and where $S_r=\exp(i\omega\hat{\sigma}_z/2)$ encodes the change into a rotating frame \cite{RevModPhys.26.167,1986qmv1.book.....C} with $\hat{\sigma}_z=\vert \psi_{2s} \rangle\langle\psi_{2s}\vert-\vert \psi_{2p} \rangle\langle\psi_{2p}\vert$ the pauli matrix for $\mathcal{R}$ subspace.

Inserting Eq.\,(\ref{rabi_wfn_ap}) into Eq.\,(\ref{r_equ_ap}), the coefficient equations are obtained as 
\begin{subequations}
  \begin{equation}
    \frac{d}{dt}\bar{Y}(t)=\bar{A}(t)\bar{Y}(t),
    \label{eq:coe}
  \end{equation}
where the coefficient matrix is
\begin{equation}
  \bar{Y}(t)\coloneqq
  \begin{pmatrix}
    C_{2s}(t;\tau)\\
    C_{2p}(t;\tau)
  \end{pmatrix},
\end{equation}
and the interaction matrix is
\begin{equation}
  \bar{A}(t)\coloneqq -\frac{i}{2}\Omega_{\rm R}(t;\tau)\bar{M}
\end{equation}
with the skew hermitian matrix
\begin{equation}
  \bar{M}\coloneqq
  \begin{pmatrix}
    0 & e^{-i\omega\tau} \\
    e^{i\omega\tau} & 0
  \end{pmatrix}.
\end{equation}
\end{subequations}
Note here that $\Omega_{\rm R}(t;\tau)=\Omega^\ast_{\rm R}(t;\tau)$ is used considering that $\Omega_{\rm R}(t;\tau)$ is real number. According to Magnus expansion \cite{BLANES2009151}, the solution of Eq.\,(\ref{eq:coe}) is 
\begin{subequations}
\begin{equation}
  \bar{Y}(t)=U^\prime_{\mathcal{R}}(t,t_0)\bar{Y}(t_0),
\end{equation}
with the propagator
\begin{equation}
  \begin{aligned}
  &U^\prime_{\mathcal{R}}(t,t_0)= \exp\\
  &\left(\int_{t_0}^t\bar{A}(t_1)dt_1-\frac{1}{2}\int_{t_0}^tdt_1\int_{t_0}^{t_1}dt_2\left[\bar{A}(t_2),\bar{A}(t_1)\right]+\cdots\right).
  \end{aligned}
\end{equation}
\end{subequations}
Here $U^\prime_{\mathcal{R}}$ is the propagator corresponding to a unitary-transformed Hamiltonian $H^\prime_{\mathcal{R}}=U_rH_{\mathcal{R}}U_r^\dagger -iU_r\partial_tU_r^\dagger$. Due to the commutation relation $\left[\bar{A}(t),\bar{A}(t^\prime)\right]=0$, only the first-order term is kept in the exponential, i.e.,
\begin{equation}
  \begin{aligned}
  &U^\prime_{\mathcal{R}}(t,t_0)= \exp\left(\int_{t_0}^t\bar{A}(t^\prime)dt^\prime\right)\\
  &=\sum_{n=0}^\infty \frac{1}{n!}\left(\int_{t_0}^t\bar{A}(t^\prime)dt^\prime\right)^n\\
  &=\sum_{n=0}^\infty \frac{1}{n!}\left(-\frac{i}{2}\int_{t_0}^t\Omega_{\rm R}(t^\prime;\tau)dt^\prime\right)^n\bar{M}^n.
  \end{aligned}
  \label{U_r_ap}
\end{equation}
Note that $\bar{M}$ satisfies 
\begin{equation}
  \bar{M}^n=
  \begin{cases}
    \bar{M},&\rm for\,\,odd\,\,n\\
    \mathbb{I},&\rm for\,\,even\,\,n,
  \end{cases}
\end{equation}
and thus
\begin{equation}
  \begin{aligned}
  U^\prime_{\mathcal{R}}(t,t_0)=&\cos\left[\frac{1}{2}\int_{t_0}^t\Omega_{\rm R}(t^\prime;\tau)\ dt^\prime\right]\mathbb{I}\\
  &-i\sin\left[\frac{1}{2}\int_{t_0}^t\Omega_{\rm R}(t^\prime;\tau)\ dt^\prime\right]\bar{M}.
  \label{pro_r}
  \end{aligned}
\end{equation}
Note that Eq.\,(\ref{pro_r}) can also be derived from Eq.\,(\ref{U_r_ap}) by directly using the relations 
\begin{subequations}
  \begin{equation}
    \bar{M}=\vec{\sigma}\cdot\hat{n},
  \end{equation}
  \begin{equation}
    \begin{aligned}
    &\exp\left[-i\frac{1}{2}\int_{t_0}^t\Omega_{\rm R}(t^\prime;\tau)\ dt^\prime\vec{\sigma}\cdot\hat{n}\right]=\\
  &\cos\left[\frac{1}{2}\int_{t_0}^t\Omega_{\rm R}(t^\prime;\tau)\ dt^\prime\right]\mathbb{I}\\
  &-i\sin\left[\frac{1}{2}\int_{t_0}^t\Omega_{\rm R}(t^\prime;\tau)\ dt^\prime\right]\vec{\sigma}\cdot\hat{n},
    \end{aligned}
  \end{equation}
\end{subequations}
where $\vec{\sigma}=[\hat{\sigma}_x,\hat{\sigma}_y,\hat{\sigma}_z]$ is related to pauli matrices for $\mathcal{R}$ subspace and where $\hat{n}=[\cos(\omega\tau),\sin(\omega\tau),0]$ is the unity vector within the $xOy$ plane in the representation of Bloch sphere \cite{RevModPhys.26.167,PhysRevLett.131.200801,PhysRevA.99.013402}.

Then the Rabi amplitudes in accordance with the area theorem \cite{Eberly:98} are given as
\begin{subequations}
  \begin{equation}
    C_{2s}(t;\tau)=\cos\left[\frac{1}{2}\int_{t_0}^t\Omega_{\rm R}(t^\prime;\tau)dt^\prime\right],
  \end{equation}
  \begin{equation}
    C_{2p}(t;\tau)=- i e^{i\omega\tau}\sin\left[\frac{1}{2}\int_{t_0}^t\Omega_{\rm R}(t^\prime;\tau) dt^\prime\right],
  \end{equation}
\end{subequations}
which is obtained by using $\bar{Y}(t)=U^\prime_{\mathcal{R}}(t,t_0)\bar{Y}(t_0)$ with $\bar{Y}(t_0)$ corresponding the boundary condition of $C_{2s}(t_0)=1$ and $C_{2p}(t_0)=0$. Here we take the initial time to infinity $t_0\rightarrow -\infty$. Furthermore, the Fourier transform of the Rabi coefficients is defined as 
\begin{subequations}
  \begin{equation}
    \begin{aligned}
    &\tilde{C}_{2s}(\omega_{\rm R};\tau)\coloneqq  \frac{1}{\sqrt{2\pi}}\int_{-\infty}^\infty C_{2s}(t;\tau)e^{-i\omega_{\rm R}t}d\omega_{\rm R}\\
    &=\frac{1}{\sqrt{2\pi}}\int_{-\infty}^\infty \cos\left[\frac{1}{2}\int_{-\infty}^t\Omega_{\rm R}(t^\prime-\tau) dt^\prime\right]e^{-i\omega_{\rm R}t}d\omega_{\rm R}\\
    &=\frac{1}{\sqrt{2\pi}}\int_{-\infty}^\infty \cos\left[\frac{1}{2}\int_{-\infty}^{t-\tau}\Omega_{\rm R}(t^\prime)dt^\prime\right]e^{-i\omega_{\rm R}(t-\tau)}d\omega_{\rm R}e^{i\omega_{\rm R}\tau}\\
    &=\tilde{C}_{2s}(\omega_{\rm R})e^{i\omega_{\rm R}\tau},
    \end{aligned}
  \end{equation}
  \begin{equation}
    \begin{aligned}
    &\tilde{C}_{2p}(\omega_{\rm R};\tau)\coloneqq  \frac{1}{\sqrt{2\pi}}\int_{-\infty}^\infty C_{2p}(t;\tau)e^{-i\omega_{\rm R}t}d\omega_{\rm R}\\
    &=-\frac{ie^{i\omega\tau}}{\sqrt{2\pi}}\int_{-\infty}^\infty  \sin\left[\frac{1}{2}\int_{-\infty}^t\Omega_{\rm R}(t^\prime-\tau)dt^\prime\right]e^{-i\omega_{\rm R}t}d\omega_{\rm R}\\
    &=\frac{-ie^{i(\omega+\omega_{\rm R})\tau}}{\sqrt{2\pi}}\int_{-\infty}^\infty \sin\left[\frac{1}{2}\int_{-\infty}^{t-\tau}\Omega_{\rm R}(t^\prime) dt^\prime\right]e^{-i\omega_{\rm R}(t-\tau)}d\omega_{\rm R}\\
    &=\tilde{C}_{2p}(\omega_{\rm R})e^{i\omega_{\rm R}\tau}.
    \end{aligned}
  \end{equation}
\end{subequations}

\section{\label{appA:DysonSer}Derivations of the Dyson series in the Rabi-RABBIT scheme}
In verifying that the wavefunction $\vert\Psi(t)\rangle$ in Eq.\,(\ref{dyson1}) satisfies the Schr\"odinger equation 
\begin{equation}
	i\frac{d }{d t}\vert \Psi (t)\rangle = H(t) \vert \Psi(t)\rangle,
\end{equation}
we directly substitute into the Schr\"odinger equation by the ansatz \cite{10.1063/1.4797476,PhysRevA.107.043105}
\begin{subequations}
	\begin{equation}
	\vert \Psi (t)\rangle = U(t,t_0)\vert \Psi(t_0)\rangle.
\end{equation}
\begin{equation}
	\begin{aligned}
	U(t,t_0) =U_{\mathcal{R}}(t,t_0) -i\int_{t_0}^t dt^\prime U(t,t^\prime) H_{\rm int}^{\perp \mathcal{R}}(t^\prime)U_{\mathcal{R}}(t^\prime,t_0).
	\end{aligned}
\end{equation}
\end{subequations}

Considering the initial condition $\vert\Psi_{\mathcal{R}}(t_0)\rangle=\vert\Psi(t_0)\rangle=e^{-i\omega_{2s}t_0}\vert \psi_{2s}\rangle$, then we obtain 
\begin{widetext}
\begin{equation}
	\begin{aligned}
	i\frac{d }{d t}\vert \Psi (t)\rangle &= i\frac{d }{d t}U(t,t_0)\vert \Psi(t_0)\rangle\\
	&=i\frac{d }{d t}U_{\mathcal{R}}(t,t_0)\vert \Psi_{\mathcal{R}}(t_0)\rangle+U(t,t) H_{\rm int}^{\perp \mathcal{R}}(t)U_{\mathcal{R}}(t,t_0)\vert \Psi_{\mathcal{R}}(t_0)\rangle\\
	&-i\int_{t_0}^t dt^\prime i\frac{d }{d t}U(t,t^\prime) H_{\rm int}^{\perp \mathcal{R}}(t^\prime)U_{\mathcal{R}}(t^\prime,t_0)\vert \Psi_{\mathcal{R}}(t_0)\rangle-i\int_{t_0}^t dt^\prime H(t)U(t,t^\prime) H_{\rm int}^{\perp \mathcal{R}}(t^\prime)U_{\mathcal{R}}(t^\prime,t_0)\vert \Psi_{\mathcal{R}}(t_0)\rangle\\
	&=[H_{\mathcal{R}}(t)+ H_{\rm int}^{\perp \mathcal{R}}(t)] U_{\mathcal{R}}(t,t_0)\vert \Psi_{\mathcal{R}}(t_0)\rangle-iH(t)\int_{t_0}^t dt^\prime U(t,t^\prime) H_{\rm int}^{\perp \mathcal{R}}(t^\prime)U_{\mathcal{R}}(t^\prime,t_0)\vert \Psi_{\mathcal{R}}(t_0)\rangle\\
	&=H(t)[U_{\mathcal{R}}(t,t_0)-i\int_{t_0}^t dt^\prime U(t,t^\prime) H_{\rm int}^{\perp \mathcal{R}}(t^\prime)U_{\mathcal{R}}(t^\prime,t_0)]\vert \Psi_{\mathcal{R}}(t_0)\rangle\\
	&=H(t)U(t,t_0)\vert \Psi(t_0)\rangle\\
	&=H(t)\vert \Psi(t)\rangle.
	\end{aligned}
\end{equation}
\end{widetext}
Note that in our derivation, we use 
\begin{subequations}
	\begin{equation}
		i\frac{d}{dt} \vert\Psi_{\mathcal{R}}(t)\rangle = H_{\mathcal{R}}(t) \vert\Psi_{\mathcal{R}}(t)\rangle,
	\end{equation}
	\begin{equation}
		\vert\Psi_{\mathcal{R}}(t)\rangle= U_{\mathcal{R}}(t,t_0)\vert\Psi_{\mathcal{R}}(t_0)\rangle,
	\end{equation}
	\begin{equation}
		i\frac{d}{dt} U_{\mathcal{R}}(t,t_0)\vert\Psi_{\mathcal{R}}(t_0)\rangle = H_{\mathcal{R}}(t) U_{\mathcal{R}}(t,t_0)\vert\Psi_{\mathcal{R}}(t_0)\rangle,
	\end{equation}
	\begin{equation}
		i\frac{d}{dt} U(t,t^\prime) = H(t) U(t,t^\prime),
	\end{equation}
\end{subequations}
where the propagator of the full Hamiltonian is $U(t,t^\prime)\coloneqq  \hat{\mathcal{T}} \exp[-i\int^t_{t^\prime}H(\tau)d\tau]$ with $\hat{\mathcal{T}}$ the time-ordering operator and $U(t,t)=1$.

\begin{widetext}
\section{\label{appC:OtherIonAmp}The ionization amplitudes of the other pathways}
The ionization amplitudes $\mathcal{A}_{(i)}$ and their reduced amplitudes $\mathcal{M}_{(i), \mathcal{Q}}$ of pathways 2, 4, 7, 8, 10, and 11 in Fig.\,\ref{fig:schem} are 
\begin{subequations}
	\begin{equation}
		\mathcal{A}_{(2)}(\vec{k}_\pm,\tau) =- \frac{i\pi}{4}e^{i\left[\left(\omega\pm \frac{\vert\Omega_{\textrm{R}}^0\vert}{2}\right)\tau+\phi_{2q-1}\right]}E_{\rm 2q-1}E_{\omega}\sum_{\nu_1\neq 2p}\kern-1.5em \int\frac{\langle \psi_{f\pm} \vert \hat{O}_\Omega \vert \psi_{\nu_1} \rangle \langle \psi_{\nu_1} \vert \hat{O}_\omega \vert \psi_{2s} \rangle}{\omega_{2s}\pm \frac{\vert\Omega_{\textrm{R}}^0\vert}{2}+\omega-\omega_{\nu_1}},
		\end{equation}
	\begin{equation}
		\mathcal{A}_{(4)}(\vec{k}_\pm,\tau) =-\frac{i\pi}{4} e^{-i\left[\left(\omega\mp \frac{\vert\Omega_{\textrm{R}}^0\vert}{2}\right)\tau-\phi_{2q+1}\right]}E_{\rm 2q+1}E_{\omega}\sum_{\nu_1}\kern-1.5em \int\frac{\langle \psi_{f\pm} \vert \hat{O}_\Omega \vert \psi_{\nu_1} \rangle \langle \psi_{\nu_1} \vert \hat{O}^{\dagger}_{\omega} \vert \psi_{2s} \rangle}{\omega_{2s}\pm \frac{\vert\Omega_{\textrm{R}}^0\vert}{2}-\omega-\omega_{\nu_1}},
		\end{equation}
	\begin{equation}
		\begin{aligned}
		\mathcal{A}_{(7)}(\vec{k}_\pm,\tau) =&\mp \frac{i\pi}{8}\frac{\Omega_{\rm R}^0}{\vert \Omega_{\rm R}^0\vert} e^{i\left[\left(3\omega\pm \frac{\vert\Omega_{\textrm{R}}^0\vert}{2}\right)\tau+\phi_{2q-3}\right]}E_{\omega}^2E_{\rm 2q-3}\times\\
		&\sum_{\nu_1,\nu_2}\kern-1.5em \int\frac{\langle \psi_{f\pm} \vert \hat{O}_{\omega} \vert \psi_{\nu_2}\rangle \langle \psi_{\nu_2} \vert \hat{O}_{\Omega} \vert \psi_{\nu_1}\rangle \langle \psi_{\nu_1} \vert \hat{O}_{\omega} \vert \psi_{2p}\rangle}{(\omega_{2p}\pm \frac{\vert\Omega_{\textrm{R}}^0\vert}{2}+\omega+\Omega_{2q-3}-\omega_{\nu_2})(\omega_{2p}\pm \frac{\vert\Omega_{\textrm{R}}^0\vert}{2}+\omega-\omega_{\nu_1})},
		\end{aligned}
		\end{equation}
		\begin{equation}
			\begin{aligned}
			\mathcal{A}_{(8)}(\vec{k}_\pm,\tau) =&\mp \frac{i\pi}{8}\frac{\Omega_{\rm R}^0}{\vert \Omega_{\rm R}^0\vert} e^{i\left[\left(3\omega\pm \frac{\vert\Omega_{\textrm{R}}^0\vert}{2}\right)\tau+\phi_{2q-3}\right]}E_{\omega}^2E_{\rm 2q-3}\times\\
			&\sum_{\nu_1,\nu_2}\kern-1.5em \int\frac{\langle \psi_{f\pm} \vert \hat{O}_{\Omega} \vert \psi_{\nu_2}\rangle \langle \psi_{\nu_2} \vert \hat{O}_{\omega} \vert \psi_{\nu_1}\rangle \langle \psi_{\nu_1} \vert \hat{O}_{\omega} \vert \psi_{2p}\rangle}{(\omega_{2p}\pm \frac{\vert\Omega_{\textrm{R}}^0\vert}{2}+\omega+\omega-\omega_{\nu_2})(\omega_{2p}\pm \frac{\vert\Omega_{\textrm{R}}^0\vert}{2}+\omega-\omega_{\nu_1})},
			\end{aligned}
		\end{equation}
	\begin{equation}
		\begin{aligned}
		\mathcal{A}_{(10)}(\vec{k}_\pm,\tau) =&\mp \frac{i\pi}{8}\frac{\Omega_{\rm R}^0}{\vert \Omega_{\rm R}^0\vert} e^{-i\left[\left(\omega\mp \frac{\vert\Omega_{\textrm{R}}^0\vert}{2}\right)\tau-\phi_{2q+1}\right]}E_{\omega}^2E_{\rm 2q+1}\times\\
		&\sum_{\nu_1\neq 2s,\nu_2}\kern-2.2em \int\,\,\,\,\frac{\langle \psi_{f\pm} \vert \hat{O}^{\dagger}_{\omega} \vert \psi_{\nu_2}\rangle \langle \psi_{\nu_2} \vert \hat{O}_{\Omega} \vert \psi_{\nu_1}\rangle \langle \psi_{\nu_1} \vert \hat{O}^{\dagger}_{\omega} \vert \psi_{2p}\rangle}{(\omega_{2p}\pm \frac{\vert\Omega_{\textrm{R}}^0\vert}{2}-\omega+\Omega_{2q+1}-\omega_{\nu_2})(\omega_{2p}\pm \frac{\vert\Omega_{\textrm{R}}^0\vert}{2}-\omega-\omega_{\nu_1})},
		\end{aligned}
		\end{equation}
		\begin{equation}
			\begin{aligned}
			\mathcal{A}_{(11)}(\vec{k}_\pm,\tau) =&\mp \frac{i\pi}{8}\frac{\Omega_{\rm R}^0}{\vert \Omega_{\rm R}^0\vert} e^{-i\left[\left(\omega\mp \frac{\vert\Omega_{\textrm{R}}^0\vert}{2}\right)\tau-\phi_{2q+1}\right]}E_{\omega}^2E_{\rm 2q+1}\times\\
			&\sum_{\nu_1\neq 2s,\nu_2}\kern-2.2em \int\,\,\,\,\frac{\langle \psi_{f\pm} \vert \hat{O}_{\Omega} \vert \psi_{\nu_2}\rangle \langle \psi_{\nu_2} \vert \hat{O}^{\dagger}_{\omega} \vert \psi_{\nu_1}\rangle \langle \psi_{\nu_1} \vert \hat{O}^{\dagger}_{\omega} \vert \psi_{2p}\rangle}{(\omega_{2p}\pm \frac{\vert\Omega_{\textrm{R}}^0\vert}{2}-\omega-\omega-\omega_{\nu_2})(\omega_{2p}\pm \frac{\vert\Omega_{\textrm{R}}^0\vert}{2}-\omega-\omega_{\nu_1})},
			\end{aligned}
			\end{equation}
	\label{Otherionchan}
\end{subequations}
\begin{subequations}
	\begin{equation}
		\begin{aligned}
		\mathcal{M}_{(2), \mathcal{Q}}(E_{\pm}) &=\frac{\pi}{3}E_{\rm 2q-1}E_{\omega}i^{-(L+1)}e^{i(\sigma_{L\pm}+\delta_{L\pm}+\phi_{2q-1})} \langle Y_{L,M}\vert Y_{1,1}\vert Y_{\lambda_1,\mu_1}\rangle\langle Y_{\lambda_1,\mu_1}\vert Y_{1,1}\vert Y_{0,0}\rangle\\
		& \times\sum_{(\nu_1,\lambda_1)\neq 2p}\kern-2.6em \int\frac{\langle R_{E_\pm,L}  \vert r \vert R_{\nu_1,\lambda_1} \rangle \langle R_{\nu_1,\lambda_1} \vert r \vert R_{2,0} \rangle}{\omega_{2s}\pm \frac{\vert\Omega_{\textrm{R}}^0\vert}{2}+\omega-\omega_{\nu_1}},
		\end{aligned}
		\end{equation}
	\begin{equation}
		\begin{aligned}
		\mathcal{M}_{(4), \mathcal{Q}}(E_{\pm}) &=-\frac{\pi}{3}E_{\rm 2q+1}E_{\omega}i^{-(L+1)}e^{i(\sigma_{L\pm}+\delta_{L\pm}+\phi_{2q+1})}\langle Y_{L,M}\vert Y_{1,1}\vert Y_{\lambda_1,\mu_1}\rangle\langle Y_{\lambda_1,\mu_1}\vert Y_{1,-1}\vert Y_{0,0}\rangle\\
		&\times \sum_{\nu_1}\kern-1.3em \int\frac{\langle R_{E_\pm,L}  \vert r \vert R_{\nu_1,\lambda_1} \rangle \langle R_{\nu_1,\lambda_1} \vert r \vert R_{2,0} \rangle}{\omega_{2s}\pm \frac{\vert\Omega_{\textrm{R}}^0\vert}{2}-\omega-\omega_{\nu_1}},
		\end{aligned}
		\end{equation}
	\begin{equation}
		\begin{aligned}
		\mathcal{M}_{(7), \mathcal{Q}}(E_{\pm}) &=-\left(\frac{\pi}{3}\right)^{\frac{3}{2}}\frac{\Omega_{\rm R}^0}{\vert \Omega_{\rm R}^0\vert}E_{\omega}^2E_{\rm 2q-3}i^{-(L\pm 1)}e^{i(\sigma_{L\pm}+\delta_{L\pm}+\phi_{2q-3})}\langle Y_{L,M}\vert Y_{1,1}\vert Y_{\lambda_2,\mu_2}\rangle\langle Y_{\lambda_2,\mu_2}\vert Y_{1,1}\vert Y_{\lambda_1,\mu_1}\rangle\langle Y_{\lambda_1,\mu_1}\vert Y_{1,1}\vert Y_{1,1}\rangle \\
		&\times\sum_{\nu_1,\nu_2}\kern-1.5em \int\frac{\langle R_{E_\pm,L}  \vert r \vert R_{\nu_2,\lambda_2}\rangle \langle R_{\nu_2,\lambda_2} \vert r \vert R_{\nu_1,\lambda_1}\rangle \langle R_{\nu_1,\lambda_1} \vert r \vert R_{2,1}\rangle}{(\omega_{2p}\pm \frac{\vert\Omega_{\textrm{R}}^0\vert}{2}+\omega+\Omega_{2q-3}-\omega_{\nu_2})(\omega_{2p}\pm \frac{\vert\Omega_{\textrm{R}}^0\vert}{2}+\omega-\omega_{\nu_1})},
		\end{aligned}
		\end{equation}
		\begin{equation}
			\begin{aligned}
			\mathcal{M}_{(8), \mathcal{Q}}(E_{\pm}) &=-\left(\frac{\pi}{3}\right)^{\frac{3}{2}}\frac{\Omega_{\rm R}^0}{\vert \Omega_{\rm R}^0\vert}E_{\omega}^2E_{\rm 2q-3}i^{-(L\pm 1)}e^{i(\sigma_{L\pm}+\delta_{L\pm}+\phi_{2q-3})}\langle Y_{L,M}\vert Y_{1,1}\vert Y_{\lambda_2,\mu_2}\rangle\langle Y_{\lambda_2,\mu_2}\vert Y_{1,1}\vert Y_{\lambda_1,\mu_1}\rangle\langle Y_{\lambda_1,\mu_1}\vert Y_{1,1}\vert Y_{1,1}\rangle \\
			&\times\sum_{\nu_1,\nu_2}\kern-1.5em \int\frac{\langle R_{E_\pm,L}  \vert r \vert R_{\nu_2,\lambda_2}\rangle \langle R_{\nu_2,\lambda_2} \vert r \vert R_{\nu_1,\lambda_1}\rangle \langle R_{\nu_1,\lambda_1} \vert r \vert R_{2,1}\rangle}{(\omega_{2p}\pm \frac{\vert\Omega_{\textrm{R}}^0\vert}{2}+\omega+\omega-\omega_{\nu_2})(\omega_{2p}\pm \frac{\vert\Omega_{\textrm{R}}^0\vert}{2}+\omega-\omega_{\nu_1})},
			\end{aligned}
			\end{equation}
	\begin{equation}
		\begin{aligned}
		\mathcal{M}_{(10), \mathcal{Q}}(E_{\pm}) &=	-\left(\frac{\pi}{3}\right)^{\frac{3}{2}}\frac{\Omega_{\rm R}^0}{\vert \Omega_{\rm R}^0\vert}E_{\omega}^2E_{\rm 2q+1}i^{-(L\pm 1)}e^{i(\sigma_{L\pm}+\delta_{L\pm}+\phi_{2q+1})}\langle Y_{L,M}\vert Y_{1,-1}\vert Y_{\lambda_2,\mu_2}\rangle\langle Y_{\lambda_2,\mu_2}\vert Y_{1,1}\vert Y_{\lambda_1,\mu_1}\rangle\langle Y_{\lambda_1,\mu_1}\vert Y_{1,-1}\vert Y_{1,1}\rangle \\
		&\times\sum_{(\nu_1,\lambda_1)\neq 2s,\nu_2}\kern-3.1em \int\,\,\,\,\,\,\,\,\,\,\frac{\langle R_{E_\pm,L}   \vert r \vert R_{\nu_2,\lambda_2}\rangle \langle R_{\nu_2,\lambda_2} \vert r \vert R_{\nu_1,\lambda_1}\rangle \langle R_{\nu_1,\lambda_1} \vert r \vert R_{2,1}\rangle}{(\omega_{2p}\pm \frac{\vert\Omega_{\textrm{R}}^0\vert}{2}-\omega+\Omega_{2q+1}-\omega_{\nu_2})(\omega_{2p}\pm \frac{\vert\Omega_{\textrm{R}}^0\vert}{2}-\omega-\omega_{\nu_1})},
		\end{aligned}
		\end{equation}
		\begin{equation}
		\begin{aligned}
		\mathcal{M}_{(11), \mathcal{Q}}(E_{\pm}) &=	-\left(\frac{\pi}{3}\right)^{\frac{3}{2}}\frac{\Omega_{\rm R}^0}{\vert \Omega_{\rm R}^0\vert}E_{\omega}^2E_{\rm 2q+1}i^{-(L\pm 1)}e^{i(\sigma_{L\pm}+\delta_{L\pm}+\phi_{2q+1})}\langle Y_{L,M}\vert Y_{1,1}\vert Y_{\lambda_2,\mu_2}\rangle\langle Y_{\lambda_2,\mu_2}\vert Y_{1,-1}\vert Y_{\lambda_1,\mu_1}\rangle\langle Y_{\lambda_1,\mu_1}\vert Y_{1,-1}\vert Y_{1,1}\rangle \\
		&\times\sum_{(\nu_1,\lambda_1)\neq 2s,\nu_2}\kern-3.1em \int\,\,\,\,\,\,\,\,\,\,\frac{\langle R_{E_\pm,L}  \vert r \vert R_{\nu_2,\lambda_2}\rangle \langle R_{\nu_2,\lambda_2} \vert r \vert R_{\nu_1,\lambda_1}\rangle \langle R_{\nu_1,\lambda_1} \vert r \vert R_{2,1}\rangle}{(\omega_{2p}\pm \frac{\vert\Omega_{\textrm{R}}^0\vert}{2}-\omega-\omega-\omega_{\nu_2})(\omega_{2p}\pm \frac{\vert\Omega_{\textrm{R}}^0\vert}{2}-\omega-\omega_{\nu_1})}.
		\end{aligned}
		\end{equation}
	\label{OtherionchanAng}
\end{subequations}

\section{\label{appD:DME}The selected numerical results of the dipole transition matrix elements}
Tab.\,\ref{DME} gives the radial and angular integrals of the dipole transition matrix elements of pathways 1 to 11 in Fig.\,\ref{fig:schem} for the photoelectron energy of $2q\omega-I^{2s}_p+\frac{\Omega_{\rm R}^0}{2}$ with $q=5$ and $\Omega_{\rm R}^0=0.0367$\,eV (the higher-energy peak of SB 10) in \textit{Case 1}.
\begin{table*}[htb]
	\centering
	\begin{tabular}{|c|c|c|c|c|c|c|c|c|c|c|c|c|c|c|c|}
		\hline\hline
		\,\,Pathway\,\, & \,\,Ionization channel\,\, & \,\,Radial integral\,\, & \,\,Angular integral\,\,\\
		\hline
		1 & $(l_i,\lambda_1,L)=(0,1,2)$ & 41.031722524570355 + 53.198466268618446i & 0.365148371670111 \\
		\hline
		2 & $(l_i,\lambda_1,L)=(0,1,2)$ & 3.823833748129400 & 0.365148371670111 \\
		\hline
		\multirow{2}*{3} & $(l_i,\lambda_1,L)=(0,1,0)$ & 13.544666666666670 + 44.780000000000015i & 0.333333333333333 \\
		\cline{2-4}
		~ & $(l_i,\lambda_1,L)=(0,1,2)$ & -12.984548602944278 - 11.427350863413425i & -0.149071198499986 \\
		\hline
		\multirow{2}*{4} & $(l_i,\lambda_1,L)=(0,1,0)$ & -1.398733333333334 & 0.333333333333333 \\
		\cline{2-4}
		~ & $(l_i,\lambda_1,L)=(0,1,2)$ & 0.852016435026670 & -0.149071198499986 \\
		\hline
		5 & $(l_i,L)=(1,2)$ & 0.213074268742145 & 0.632455532033676 \\
		\hline		
		6 & $(l_i,\lambda_1,\lambda_2,L)=(1,2,3,4)$ & 4656.010341994734 + 3491.179677824839i & 0.276026223736942 \\
		\hline
		7 & $(l_i,\lambda_1,\lambda_2,L)=(1,2,3,4)$ & 4082.151822845632 + 1288.738836005407i & 0.276026223736942 \\
		\hline
		8 & $(l_i,\lambda_1,\lambda_2,L)=(1,2,3,4)$ & -839.4509516287872 & 0.276026223736942 \\
		\hline
		\multirow{4}*{9} & $(l_i,\lambda_1,\lambda_2,L)=(1,2,1,0)$ & -2362.517301523949 - 1025.027667919262i & 0.230940107675850 \\
		\cline{2-4}
		~ & $(l_i,\lambda_1,\lambda_2,L)=(1,2,1,2)$ & 1004.238761782608 - 123.4397252103228i & -0.103279555898864 \\
		\cline{2-4}
		~ & $(l_i,\lambda_1,\lambda_2,L)=(1,2,3,2)$ & 277.7039715897753 - 23.54641086493669i & -0.044262666813799 \\
		\cline{2-4}
		~ & $(l_i,\lambda_1,\lambda_2,L)=(1,2,3,4)$ & 169.0151673747687 + 39.07506621875388i & 0.032991443953693 \\
		\hline
		\multirow{6}*{10} & $(l_i,\lambda_1,\lambda_2,L)=(1,0,1,0)$ & 276.6854940046418 + 869.2200752739551i & 0.192450089729875 \\
		\cline{2-4}
		~ & $(l_i,\lambda_1,\lambda_2,L)=(1,0,1,2)$ & -256.6496964086782 - 221.8186661817862i & -0.086066296582387 \\
		\cline{2-4}
		~ & $(l_i,\lambda_1,\lambda_2,L)=(1,2,1,0)$ & 20.244209838865040 + 63.312230519334360i & 0.038490017945975 \\
		\cline{2-4}
		~ & $(l_i,\lambda_1,\lambda_2,L)=(1,2,1,2)$ & -17.338916109487684 - 16.156881592225183i & -0.017213259316477 \\
		\cline{2-4}
		~ & $(l_i,\lambda_1,\lambda_2,L)=(1,2,3,2)$ & -219.8836499309096 + 75.62719251805706i & -0.044262666813799 \\
		\cline{2-4}
		~ & $(l_i,\lambda_1,\lambda_2,L)=(1,2,3,4)$ & -98.482759346168690 - 11.215441372057903i & 0.032991443953693 \\
		\hline
		\multirow{6}*{11} & $(l_i,\lambda_1,\lambda_2,L)=(1,0,1,0)$ & -21.007851794913194 & 0.192450089729875 \\
		\cline{2-4}
		~ & $(l_i,\lambda_1,\lambda_2,L)=(1,0,1,2)$ & 0.800098112918845 & -0.086066296582387 \\
		\cline{2-4}
		~ & $(l_i,\lambda_1,\lambda_2,L)=(1,2,1,0)$ & 0.049998533311822 & 0.038490017945975 \\
		\cline{2-4}
		~ & $(l_i,\lambda_1,\lambda_2,L)=(1,2,1,2)$ & -1.710240592647929 & -0.017213259316477 \\
		\cline{2-4}
		~ & $(l_i,\lambda_1,\lambda_2,L)=(1,2,3,2)$ & -4.063799702841705 & -0.044262666813799 \\
		\cline{2-4}
		~ & $(l_i,\lambda_1,\lambda_2,L)=(1,2,3,4)$ & -8.459336144166398 & 0.032991443953693 \\
		\hline\hline		
	\end{tabular}
	\caption{The radial and angular integrals of the dipole transition matrix elements of pathways 1 to 11 for the higher-energy peak of SB 10 in \textit{Case 1}.}
	\label{DME}
\end{table*}
\end{widetext}
\clearpage

\begin{thebibliography}{100}%
  \makeatletter
  \providecommand \@ifxundefined [1]{%
   \@ifx{#1\undefined}
  }%
  \providecommand \@ifnum [1]{%
   \ifnum #1\expandafter \@firstoftwo
   \else \expandafter \@secondoftwo
   \fi
  }%
  \providecommand \@ifx [1]{%
   \ifx #1\expandafter \@firstoftwo
   \else \expandafter \@secondoftwo
   \fi
  }%
  \providecommand \natexlab [1]{#1}%
  \providecommand \enquote  [1]{``#1''}%
  \providecommand \bibnamefont  [1]{#1}%
  \providecommand \bibfnamefont [1]{#1}%
  \providecommand \citenamefont [1]{#1}%
  \providecommand \href@noop [0]{\@secondoftwo}%
  \providecommand \href [0]{\begingroup \@sanitize@url \@href}%
  \providecommand \@href[1]{\@@startlink{#1}\@@href}%
  \providecommand \@@href[1]{\endgroup#1\@@endlink}%
  \providecommand \@sanitize@url [0]{\catcode `\\12\catcode `\$12\catcode
    `\&12\catcode `\#12\catcode `\^12\catcode `\_12\catcode `\%12\relax}%
  \providecommand \@@startlink[1]{}%
  \providecommand \@@endlink[0]{}%
  \providecommand \url  [0]{\begingroup\@sanitize@url \@url }%
  \providecommand \@url [1]{\endgroup\@href {#1}{\urlprefix }}%
  \providecommand \urlprefix  [0]{URL }%
  \providecommand \Eprint [0]{\href }%
  \providecommand \doibase [0]{http://dx.doi.org/}%
  \providecommand \selectlanguage [0]{\@gobble}%
  \providecommand \bibinfo  [0]{\@secondoftwo}%
  \providecommand \bibfield  [0]{\@secondoftwo}%
  \providecommand \translation [1]{[#1]}%
  \providecommand \BibitemOpen [0]{}%
  \providecommand \bibitemStop [0]{}%
  \providecommand \bibitemNoStop [0]{.\EOS\space}%
  \providecommand \EOS [0]{\spacefactor3000\relax}%
  \providecommand \BibitemShut  [1]{\csname bibitem#1\endcsname}%
  \let\auto@bib@innerbib\@empty
  \bibitem [{\citenamefont {Rabi}(1937)}]{PhysRev.51.652}%
    \BibitemOpen
    \bibfield  {author} {\bibinfo {author} {\bibfnamefont {I.~I.}\ \bibnamefont
    {Rabi}},\ }\href {\doibase 10.1103/PhysRev.51.652} {\bibfield  {journal}
    {\bibinfo  {journal} {Phys. Rev.}\ }\textbf {\bibinfo {volume} {51}},\
    \bibinfo {pages} {652} (\bibinfo {year} {1937})}\BibitemShut {NoStop}%
  \bibitem [{\citenamefont {Pileio}\ \emph {et~al.}(2009)\citenamefont {Pileio},
    \citenamefont {Carravetta},\ and\ \citenamefont
    {Levitt}}]{PhysRevLett.103.083002}%
    \BibitemOpen
    \bibfield  {author} {\bibinfo {author} {\bibfnamefont {G.}~\bibnamefont
    {Pileio}}, \bibinfo {author} {\bibfnamefont {M.}~\bibnamefont {Carravetta}},
    \ and\ \bibinfo {author} {\bibfnamefont {M.~H.}\ \bibnamefont {Levitt}},\
    }\href {\doibase 10.1103/PhysRevLett.103.083002} {\bibfield  {journal}
    {\bibinfo  {journal} {Phys. Rev. Lett.}\ }\textbf {\bibinfo {volume} {103}},\
    \bibinfo {pages} {083002} (\bibinfo {year} {2009})}\BibitemShut {NoStop}%
  \bibitem [{\citenamefont {Dudin}\ \emph {et~al.}(2012)\citenamefont {Dudin},
    \citenamefont {Li}, \citenamefont {Bariani},\ and\ \citenamefont
    {Kuzmich}}]{Dudin2012}%
    \BibitemOpen
    \bibfield  {author} {\bibinfo {author} {\bibfnamefont {Y.~O.}\ \bibnamefont
    {Dudin}}, \bibinfo {author} {\bibfnamefont {L.}~\bibnamefont {Li}}, \bibinfo
    {author} {\bibfnamefont {F.}~\bibnamefont {Bariani}}, \ and\ \bibinfo
    {author} {\bibfnamefont {A.}~\bibnamefont {Kuzmich}},\ }\href {\doibase
    10.1038/nphys2413} {\bibfield  {journal} {\bibinfo  {journal} {Nature
    Physics}\ }\textbf {\bibinfo {volume} {8}},\ \bibinfo {pages} {790} (\bibinfo
    {year} {2012})}\BibitemShut {NoStop}%
  \bibitem [{\citenamefont {Batalov}\ \emph {et~al.}(2008)\citenamefont
    {Batalov}, \citenamefont {Zierl}, \citenamefont {Gaebel}, \citenamefont
    {Neumann}, \citenamefont {Chan}, \citenamefont {Balasubramanian},
    \citenamefont {Hemmer}, \citenamefont {Jelezko},\ and\ \citenamefont
    {Wrachtrup}}]{PhysRevLett.100.077401}%
    \BibitemOpen
    \bibfield  {author} {\bibinfo {author} {\bibfnamefont {A.}~\bibnamefont
    {Batalov}}, \bibinfo {author} {\bibfnamefont {C.}~\bibnamefont {Zierl}},
    \bibinfo {author} {\bibfnamefont {T.}~\bibnamefont {Gaebel}}, \bibinfo
    {author} {\bibfnamefont {P.}~\bibnamefont {Neumann}}, \bibinfo {author}
    {\bibfnamefont {I.-Y.}\ \bibnamefont {Chan}}, \bibinfo {author}
    {\bibfnamefont {G.}~\bibnamefont {Balasubramanian}}, \bibinfo {author}
    {\bibfnamefont {P.~R.}\ \bibnamefont {Hemmer}}, \bibinfo {author}
    {\bibfnamefont {F.}~\bibnamefont {Jelezko}}, \ and\ \bibinfo {author}
    {\bibfnamefont {J.}~\bibnamefont {Wrachtrup}},\ }\href {\doibase
    10.1103/PhysRevLett.100.077401} {\bibfield  {journal} {\bibinfo  {journal}
    {Phys. Rev. Lett.}\ }\textbf {\bibinfo {volume} {100}},\ \bibinfo {pages}
    {077401} (\bibinfo {year} {2008})}\BibitemShut {NoStop}%
  \bibitem [{\citenamefont {Stievater}\ \emph {et~al.}(2001)\citenamefont
    {Stievater}, \citenamefont {Li}, \citenamefont {Steel}, \citenamefont
    {Gammon}, \citenamefont {Katzer}, \citenamefont {Park}, \citenamefont
    {Piermarocchi},\ and\ \citenamefont {Sham}}]{PhysRevLett.87.133603}%
    \BibitemOpen
    \bibfield  {author} {\bibinfo {author} {\bibfnamefont {T.~H.}\ \bibnamefont
    {Stievater}}, \bibinfo {author} {\bibfnamefont {X.}~\bibnamefont {Li}},
    \bibinfo {author} {\bibfnamefont {D.~G.}\ \bibnamefont {Steel}}, \bibinfo
    {author} {\bibfnamefont {D.}~\bibnamefont {Gammon}}, \bibinfo {author}
    {\bibfnamefont {D.~S.}\ \bibnamefont {Katzer}}, \bibinfo {author}
    {\bibfnamefont {D.}~\bibnamefont {Park}}, \bibinfo {author} {\bibfnamefont
    {C.}~\bibnamefont {Piermarocchi}}, \ and\ \bibinfo {author} {\bibfnamefont
    {L.~J.}\ \bibnamefont {Sham}},\ }\href {\doibase
    10.1103/PhysRevLett.87.133603} {\bibfield  {journal} {\bibinfo  {journal}
    {Phys. Rev. Lett.}\ }\textbf {\bibinfo {volume} {87}},\ \bibinfo {pages}
    {133603} (\bibinfo {year} {2001})}\BibitemShut {NoStop}%
  \bibitem [{\citenamefont {Merlin}(2021)}]{10.1119/10.0001897}%
    \BibitemOpen
    \bibfield  {author} {\bibinfo {author} {\bibfnamefont {R.}~\bibnamefont
    {Merlin}},\ }\href {\doibase 10.1119/10.0001897} {\bibfield  {journal}
    {\bibinfo  {journal} {American Journal of Physics}\ }\textbf {\bibinfo
    {volume} {89}},\ \bibinfo {pages} {26} (\bibinfo {year} {2021})}\BibitemShut
    {NoStop}%
  \bibitem [{\citenamefont {Knight}\ and\ \citenamefont
    {Milonni}(1980)}]{KNIGHT198021}%
    \BibitemOpen
    \bibfield  {author} {\bibinfo {author} {\bibfnamefont {P.}~\bibnamefont
    {Knight}}\ and\ \bibinfo {author} {\bibfnamefont {P.}~\bibnamefont
    {Milonni}},\ }\href {\doibase https://doi.org/10.1016/0370-1573(80)90119-2}
    {\bibfield  {journal} {\bibinfo  {journal} {Physics Reports}\ }\textbf
    {\bibinfo {volume} {66}},\ \bibinfo {pages} {21} (\bibinfo {year}
    {1980})}\BibitemShut {NoStop}%
  \bibitem [{\citenamefont {Autler}\ and\ \citenamefont
    {Townes}(1955)}]{PhysRev.100.703}%
    \BibitemOpen
    \bibfield  {author} {\bibinfo {author} {\bibfnamefont {S.~H.}\ \bibnamefont
    {Autler}}\ and\ \bibinfo {author} {\bibfnamefont {C.~H.}\ \bibnamefont
    {Townes}},\ }\href {\doibase 10.1103/PhysRev.100.703} {\bibfield  {journal}
    {\bibinfo  {journal} {Phys. Rev.}\ }\textbf {\bibinfo {volume} {100}},\
    \bibinfo {pages} {703} (\bibinfo {year} {1955})}\BibitemShut {NoStop}%
  \bibitem [{\citenamefont {Jiang}\ \emph {et~al.}(2021)\citenamefont {Jiang},
    \citenamefont {Liang}, \citenamefont {Wang}, \citenamefont {Peng},\ and\
    \citenamefont {Burgd\"orfer}}]{PhysRevResearch.3.L032052}%
    \BibitemOpen
    \bibfield  {author} {\bibinfo {author} {\bibfnamefont {W.-C.}\ \bibnamefont
    {Jiang}}, \bibinfo {author} {\bibfnamefont {H.}~\bibnamefont {Liang}},
    \bibinfo {author} {\bibfnamefont {S.}~\bibnamefont {Wang}}, \bibinfo {author}
    {\bibfnamefont {L.-Y.}\ \bibnamefont {Peng}}, \ and\ \bibinfo {author}
    {\bibfnamefont {J.}~\bibnamefont {Burgd\"orfer}},\ }\href {\doibase
    10.1103/PhysRevResearch.3.L032052} {\bibfield  {journal} {\bibinfo  {journal}
    {Phys. Rev. Research}\ }\textbf {\bibinfo {volume} {3}},\ \bibinfo {pages}
    {L032052} (\bibinfo {year} {2021})}\BibitemShut {NoStop}%
  \bibitem [{\citenamefont {Nandi}\ \emph {et~al.}(2022)\citenamefont {Nandi},
    \citenamefont {Olofsson}, \citenamefont {Bertolino}, \citenamefont
    {Carlstr{\"o}m}, \citenamefont {Zapata}, \citenamefont {Busto}, \citenamefont
    {Callegari}, \citenamefont {Di~Fraia}, \citenamefont {Eng-Johnsson},
    \citenamefont {Feifel}, \citenamefont {Gallician}, \citenamefont
    {Gisselbrecht}, \citenamefont {Maclot}, \citenamefont {Neori{\v{c}}i{\'{c}}},
    \citenamefont {Peschel}, \citenamefont {Plekan}, \citenamefont {Prince},
    \citenamefont {Squibb}, \citenamefont {Zhong}, \citenamefont {Demekhin},
    \citenamefont {Meyer}, \citenamefont {Miron}, \citenamefont {Badano},
    \citenamefont {Danailov}, \citenamefont {Giannessi}, \citenamefont
    {Manfredda}, \citenamefont {Sottocorona}, \citenamefont {Zangrando},\ and\
    \citenamefont {Dahlstr{\"o}m}}]{Nandi2022}%
    \BibitemOpen
    \bibfield  {author} {\bibinfo {author} {\bibfnamefont {S.}~\bibnamefont
    {Nandi}}, \bibinfo {author} {\bibfnamefont {E.}~\bibnamefont {Olofsson}},
    \bibinfo {author} {\bibfnamefont {M.}~\bibnamefont {Bertolino}}, \bibinfo
    {author} {\bibfnamefont {S.}~\bibnamefont {Carlstr{\"o}m}}, \bibinfo {author}
    {\bibfnamefont {F.}~\bibnamefont {Zapata}}, \bibinfo {author} {\bibfnamefont
    {D.}~\bibnamefont {Busto}}, \bibinfo {author} {\bibfnamefont
    {C.}~\bibnamefont {Callegari}}, \bibinfo {author} {\bibfnamefont
    {M.}~\bibnamefont {Di~Fraia}}, \bibinfo {author} {\bibfnamefont
    {P.}~\bibnamefont {Eng-Johnsson}}, \bibinfo {author} {\bibfnamefont
    {R.}~\bibnamefont {Feifel}}, \bibinfo {author} {\bibfnamefont
    {G.}~\bibnamefont {Gallician}}, \bibinfo {author} {\bibfnamefont
    {M.}~\bibnamefont {Gisselbrecht}}, \bibinfo {author} {\bibfnamefont
    {S.}~\bibnamefont {Maclot}}, \bibinfo {author} {\bibfnamefont
    {L.}~\bibnamefont {Neori{\v{c}}i{\'{c}}}}, \bibinfo {author} {\bibfnamefont
    {J.}~\bibnamefont {Peschel}}, \bibinfo {author} {\bibfnamefont
    {O.}~\bibnamefont {Plekan}}, \bibinfo {author} {\bibfnamefont {K.~C.}\
    \bibnamefont {Prince}}, \bibinfo {author} {\bibfnamefont {R.~J.}\
    \bibnamefont {Squibb}}, \bibinfo {author} {\bibfnamefont {S.}~\bibnamefont
    {Zhong}}, \bibinfo {author} {\bibfnamefont {P.~V.}\ \bibnamefont {Demekhin}},
    \bibinfo {author} {\bibfnamefont {M.}~\bibnamefont {Meyer}}, \bibinfo
    {author} {\bibfnamefont {C.}~\bibnamefont {Miron}}, \bibinfo {author}
    {\bibfnamefont {L.}~\bibnamefont {Badano}}, \bibinfo {author} {\bibfnamefont
    {M.~B.}\ \bibnamefont {Danailov}}, \bibinfo {author} {\bibfnamefont
    {L.}~\bibnamefont {Giannessi}}, \bibinfo {author} {\bibfnamefont
    {M.}~\bibnamefont {Manfredda}}, \bibinfo {author} {\bibfnamefont
    {F.}~\bibnamefont {Sottocorona}}, \bibinfo {author} {\bibfnamefont
    {M.}~\bibnamefont {Zangrando}}, \ and\ \bibinfo {author} {\bibfnamefont
    {J.~M.}\ \bibnamefont {Dahlstr{\"o}m}},\ }\href {\doibase
    10.1038/s41586-022-04948-y} {\bibfield  {journal} {\bibinfo  {journal}
    {Nature}\ }\textbf {\bibinfo {volume} {608}},\ \bibinfo {pages} {488}
    (\bibinfo {year} {2022})}\BibitemShut {NoStop}%
  \bibitem [{\citenamefont {Liao}\ \emph {et~al.}(2022)\citenamefont {Liao},
    \citenamefont {Zhou}, \citenamefont {Pi}, \citenamefont {Liang},
    \citenamefont {Ke}, \citenamefont {Zhao}, \citenamefont {Li},\ and\
    \citenamefont {Lu}}]{PhysRevA.105.063110}%
    \BibitemOpen
    \bibfield  {author} {\bibinfo {author} {\bibfnamefont {Y.}~\bibnamefont
    {Liao}}, \bibinfo {author} {\bibfnamefont {Y.}~\bibnamefont {Zhou}}, \bibinfo
    {author} {\bibfnamefont {L.-W.}\ \bibnamefont {Pi}}, \bibinfo {author}
    {\bibfnamefont {J.}~\bibnamefont {Liang}}, \bibinfo {author} {\bibfnamefont
    {Q.}~\bibnamefont {Ke}}, \bibinfo {author} {\bibfnamefont {Y.}~\bibnamefont
    {Zhao}}, \bibinfo {author} {\bibfnamefont {M.}~\bibnamefont {Li}}, \ and\
    \bibinfo {author} {\bibfnamefont {P.}~\bibnamefont {Lu}},\ }\href {\doibase
    10.1103/PhysRevA.105.063110} {\bibfield  {journal} {\bibinfo  {journal}
    {Phys. Rev. A}\ }\textbf {\bibinfo {volume} {105}},\ \bibinfo {pages}
    {063110} (\bibinfo {year} {2022})}\BibitemShut {NoStop}%
  \bibitem [{\citenamefont {Zhang}\ \emph {et~al.}(2022)\citenamefont {Zhang},
    \citenamefont {Zhou}, \citenamefont {Liao}, \citenamefont {Chen},
    \citenamefont {Liang}, \citenamefont {Ke}, \citenamefont {Li}, \citenamefont
    {Csehi},\ and\ \citenamefont {Lu}}]{PhysRevA.106.063114}%
    \BibitemOpen
    \bibfield  {author} {\bibinfo {author} {\bibfnamefont {X.}~\bibnamefont
    {Zhang}}, \bibinfo {author} {\bibfnamefont {Y.}~\bibnamefont {Zhou}},
    \bibinfo {author} {\bibfnamefont {Y.}~\bibnamefont {Liao}}, \bibinfo {author}
    {\bibfnamefont {Y.}~\bibnamefont {Chen}}, \bibinfo {author} {\bibfnamefont
    {J.}~\bibnamefont {Liang}}, \bibinfo {author} {\bibfnamefont
    {Q.}~\bibnamefont {Ke}}, \bibinfo {author} {\bibfnamefont {M.}~\bibnamefont
    {Li}}, \bibinfo {author} {\bibfnamefont {A.}~\bibnamefont {Csehi}}, \ and\
    \bibinfo {author} {\bibfnamefont {P.}~\bibnamefont {Lu}},\ }\href {\doibase
    10.1103/PhysRevA.106.063114} {\bibfield  {journal} {\bibinfo  {journal}
    {Phys. Rev. A}\ }\textbf {\bibinfo {volume} {106}},\ \bibinfo {pages}
    {063114} (\bibinfo {year} {2022})}\BibitemShut {NoStop}%
  \bibitem [{\citenamefont {Olofsson}\ and\ \citenamefont
    {Dahlstr\"om}(2023)}]{PhysRevResearch.5.043017}%
    \BibitemOpen
    \bibfield  {author} {\bibinfo {author} {\bibfnamefont {E.}~\bibnamefont
    {Olofsson}}\ and\ \bibinfo {author} {\bibfnamefont {J.~M.}\ \bibnamefont
    {Dahlstr\"om}},\ }\href {\doibase 10.1103/PhysRevResearch.5.043017}
    {\bibfield  {journal} {\bibinfo  {journal} {Phys. Rev. Res.}\ }\textbf
    {\bibinfo {volume} {5}},\ \bibinfo {pages} {043017} (\bibinfo {year}
    {2023})}\BibitemShut {NoStop}%
  \bibitem [{\citenamefont {Chu}\ and\ \citenamefont
    {Lin}(2013)}]{PhysRevA.87.013415}%
    \BibitemOpen
    \bibfield  {author} {\bibinfo {author} {\bibfnamefont {W.-C.}\ \bibnamefont
    {Chu}}\ and\ \bibinfo {author} {\bibfnamefont {C.~D.}\ \bibnamefont {Lin}},\
    }\href {\doibase 10.1103/PhysRevA.87.013415} {\bibfield  {journal} {\bibinfo
    {journal} {Phys. Rev. A}\ }\textbf {\bibinfo {volume} {87}},\ \bibinfo
    {pages} {013415} (\bibinfo {year} {2013})}\BibitemShut {NoStop}%
  \bibitem [{\citenamefont {Wu}\ \emph {et~al.}(2013)\citenamefont {Wu},
    \citenamefont {Chen}, \citenamefont {Gaarde},\ and\ \citenamefont
    {Schafer}}]{PhysRevA.88.043416}%
    \BibitemOpen
    \bibfield  {author} {\bibinfo {author} {\bibfnamefont {M.}~\bibnamefont
    {Wu}}, \bibinfo {author} {\bibfnamefont {S.}~\bibnamefont {Chen}}, \bibinfo
    {author} {\bibfnamefont {M.~B.}\ \bibnamefont {Gaarde}}, \ and\ \bibinfo
    {author} {\bibfnamefont {K.~J.}\ \bibnamefont {Schafer}},\ }\href {\doibase
    10.1103/PhysRevA.88.043416} {\bibfield  {journal} {\bibinfo  {journal} {Phys.
    Rev. A}\ }\textbf {\bibinfo {volume} {88}},\ \bibinfo {pages} {043416}
    (\bibinfo {year} {2013})}\BibitemShut {NoStop}%
  \bibitem [{\citenamefont {Ott}\ \emph {et~al.}(2014)\citenamefont {Ott},
    \citenamefont {Kaldun}, \citenamefont {Argenti}, \citenamefont {Raith},
    \citenamefont {Meyer}, \citenamefont {Laux}, \citenamefont {Zhang},
    \citenamefont {Bl{\"a}ttermann}, \citenamefont {Hagstotz}, \citenamefont
    {Ding}, \citenamefont {Heck}, \citenamefont {Madro{\~{n}}ero}, \citenamefont
    {Mart{\'i}n},\ and\ \citenamefont {Pfeifer}}]{Ott2014}%
    \BibitemOpen
    \bibfield  {author} {\bibinfo {author} {\bibfnamefont {C.}~\bibnamefont
    {Ott}}, \bibinfo {author} {\bibfnamefont {A.}~\bibnamefont {Kaldun}},
    \bibinfo {author} {\bibfnamefont {L.}~\bibnamefont {Argenti}}, \bibinfo
    {author} {\bibfnamefont {P.}~\bibnamefont {Raith}}, \bibinfo {author}
    {\bibfnamefont {K.}~\bibnamefont {Meyer}}, \bibinfo {author} {\bibfnamefont
    {M.}~\bibnamefont {Laux}}, \bibinfo {author} {\bibfnamefont {Y.}~\bibnamefont
    {Zhang}}, \bibinfo {author} {\bibfnamefont {A.}~\bibnamefont
    {Bl{\"a}ttermann}}, \bibinfo {author} {\bibfnamefont {S.}~\bibnamefont
    {Hagstotz}}, \bibinfo {author} {\bibfnamefont {T.}~\bibnamefont {Ding}},
    \bibinfo {author} {\bibfnamefont {R.}~\bibnamefont {Heck}}, \bibinfo {author}
    {\bibfnamefont {J.}~\bibnamefont {Madro{\~{n}}ero}}, \bibinfo {author}
    {\bibfnamefont {F.}~\bibnamefont {Mart{\'i}n}}, \ and\ \bibinfo {author}
    {\bibfnamefont {T.}~\bibnamefont {Pfeifer}},\ }\href {\doibase
    10.1038/nature14026} {\bibfield  {journal} {\bibinfo  {journal} {Nature}\
    }\textbf {\bibinfo {volume} {516}},\ \bibinfo {pages} {374} (\bibinfo {year}
    {2014})}\BibitemShut {NoStop}%
  \bibitem [{\citenamefont {Harkema}\ \emph {et~al.}(2021)\citenamefont
    {Harkema}, \citenamefont {Cariker}, \citenamefont {Lindroth}, \citenamefont
    {Argenti},\ and\ \citenamefont {Sandhu}}]{PhysRevLett.127.023202}%
    \BibitemOpen
    \bibfield  {author} {\bibinfo {author} {\bibfnamefont {N.}~\bibnamefont
    {Harkema}}, \bibinfo {author} {\bibfnamefont {C.}~\bibnamefont {Cariker}},
    \bibinfo {author} {\bibfnamefont {E.}~\bibnamefont {Lindroth}}, \bibinfo
    {author} {\bibfnamefont {L.}~\bibnamefont {Argenti}}, \ and\ \bibinfo
    {author} {\bibfnamefont {A.}~\bibnamefont {Sandhu}},\ }\href {\doibase
    10.1103/PhysRevLett.127.023202} {\bibfield  {journal} {\bibinfo  {journal}
    {Phys. Rev. Lett.}\ }\textbf {\bibinfo {volume} {127}},\ \bibinfo {pages}
    {023202} (\bibinfo {year} {2021})}\BibitemShut {NoStop}%
  \bibitem [{\citenamefont {{Ramsey}}\ and\ \citenamefont
    {{Hellwarth}}(1956)}]{1956PhT.....9k..37R}%
    \BibitemOpen
    \bibfield  {author} {\bibinfo {author} {\bibfnamefont {N.~F.}\ \bibnamefont
    {{Ramsey}}}\ and\ \bibinfo {author} {\bibfnamefont {R.~W.}\ \bibnamefont
    {{Hellwarth}}},\ }\href {\doibase 10.1063/1.3059818} {\bibfield  {journal}
    {\bibinfo  {journal} {Physics Today}\ }\textbf {\bibinfo {volume} {9}},\
    \bibinfo {pages} {37} (\bibinfo {year} {1956})}\BibitemShut {NoStop}%
  \bibitem [{\citenamefont {Xu}\ and\ \citenamefont
    {Heinzen}(1999)}]{PhysRevA.59.R922}%
    \BibitemOpen
    \bibfield  {author} {\bibinfo {author} {\bibfnamefont {G.}~\bibnamefont
    {Xu}}\ and\ \bibinfo {author} {\bibfnamefont {D.~J.}\ \bibnamefont
    {Heinzen}},\ }\href {\doibase 10.1103/PhysRevA.59.R922} {\bibfield  {journal}
    {\bibinfo  {journal} {Phys. Rev. A}\ }\textbf {\bibinfo {volume} {59}},\
    \bibinfo {pages} {R922} (\bibinfo {year} {1999})}\BibitemShut {NoStop}%
  \bibitem [{\citenamefont {Floquet}(1883)}]{ASENS_1883_2_12__47_0}%
    \BibitemOpen
    \bibfield  {author} {\bibinfo {author} {\bibfnamefont {G.}~\bibnamefont
    {Floquet}},\ }\href {\doibase 10.24033/asens.220} {\bibfield  {journal}
    {\bibinfo  {journal} {Annales scientifiques de l'\'Ecole Normale
    Sup\'erieure}\ }\textbf {\bibinfo {volume} {2e s{\'e}rie, 12}},\ \bibinfo
    {pages} {47} (\bibinfo {year} {1883})}\BibitemShut {NoStop}%
  \bibitem [{\citenamefont {Chu}\ and\ \citenamefont {Telnov}(2004)}]{CHU20041}%
    \BibitemOpen
    \bibfield  {author} {\bibinfo {author} {\bibfnamefont {S.-I.}\ \bibnamefont
    {Chu}}\ and\ \bibinfo {author} {\bibfnamefont {D.~A.}\ \bibnamefont
    {Telnov}},\ }\href {\doibase https://doi.org/10.1016/j.physrep.2003.10.001}
    {\bibfield  {journal} {\bibinfo  {journal} {Physics Reports}\ }\textbf
    {\bibinfo {volume} {390}},\ \bibinfo {pages} {1} (\bibinfo {year}
    {2004})}\BibitemShut {NoStop}%
  \bibitem [{\citenamefont {Jaynes}\ and\ \citenamefont
    {Cummings}(1963)}]{1443594}%
    \BibitemOpen
    \bibfield  {author} {\bibinfo {author} {\bibfnamefont {E.}~\bibnamefont
    {Jaynes}}\ and\ \bibinfo {author} {\bibfnamefont {F.}~\bibnamefont
    {Cummings}},\ }\href {\doibase 10.1109/PROC.1963.1664} {\bibfield  {journal}
    {\bibinfo  {journal} {Proceedings of the IEEE}\ }\textbf {\bibinfo {volume}
    {51}},\ \bibinfo {pages} {89} (\bibinfo {year} {1963})}\BibitemShut {NoStop}%
  \bibitem [{\citenamefont {Shore}\ and\ \citenamefont
    {Knight}(1993)}]{doi:10.1080/09500349314551321}%
    \BibitemOpen
    \bibfield  {author} {\bibinfo {author} {\bibfnamefont {B.~W.}\ \bibnamefont
    {Shore}}\ and\ \bibinfo {author} {\bibfnamefont {P.~L.}\ \bibnamefont
    {Knight}},\ }\href {\doibase 10.1080/09500349314551321} {\bibfield  {journal}
    {\bibinfo  {journal} {Journal of Modern Optics}\ }\textbf {\bibinfo {volume}
    {40}},\ \bibinfo {pages} {1195} (\bibinfo {year} {1993})}\BibitemShut
    {NoStop}%
  \bibitem [{\citenamefont {Hentschel}\ \emph {et~al.}(2001)\citenamefont
    {Hentschel}, \citenamefont {Kienberger}, \citenamefont {Spielmann},
    \citenamefont {Reider}, \citenamefont {Milosevic}, \citenamefont {Brabec},
    \citenamefont {Corkum}, \citenamefont {Heinzmann}, \citenamefont {Drescher},\
    and\ \citenamefont {Krausz}}]{Hentschel2001}%
    \BibitemOpen
    \bibfield  {author} {\bibinfo {author} {\bibfnamefont {M.}~\bibnamefont
    {Hentschel}}, \bibinfo {author} {\bibfnamefont {R.}~\bibnamefont
    {Kienberger}}, \bibinfo {author} {\bibfnamefont {C.}~\bibnamefont
    {Spielmann}}, \bibinfo {author} {\bibfnamefont {G.~A.}\ \bibnamefont
    {Reider}}, \bibinfo {author} {\bibfnamefont {N.}~\bibnamefont {Milosevic}},
    \bibinfo {author} {\bibfnamefont {T.}~\bibnamefont {Brabec}}, \bibinfo
    {author} {\bibfnamefont {P.}~\bibnamefont {Corkum}}, \bibinfo {author}
    {\bibfnamefont {U.}~\bibnamefont {Heinzmann}}, \bibinfo {author}
    {\bibfnamefont {M.}~\bibnamefont {Drescher}}, \ and\ \bibinfo {author}
    {\bibfnamefont {F.}~\bibnamefont {Krausz}},\ }\href {\doibase
    10.1038/35107000} {\bibfield  {journal} {\bibinfo  {journal} {Nature}\
    }\textbf {\bibinfo {volume} {414}},\ \bibinfo {pages} {509} (\bibinfo {year}
    {2001})}\BibitemShut {NoStop}%
  \bibitem [{\citenamefont {Paul}\ \emph {et~al.}(2001)\citenamefont {Paul},
    \citenamefont {Toma}, \citenamefont {Breger}, \citenamefont {Mullot},
    \citenamefont {Augé}, \citenamefont {Balcou}, \citenamefont {Muller},\ and\
    \citenamefont {Agostini}}]{doi:10.1126/science.1059413}%
    \BibitemOpen
    \bibfield  {author} {\bibinfo {author} {\bibfnamefont {P.~M.}\ \bibnamefont
    {Paul}}, \bibinfo {author} {\bibfnamefont {E.~S.}\ \bibnamefont {Toma}},
    \bibinfo {author} {\bibfnamefont {P.}~\bibnamefont {Breger}}, \bibinfo
    {author} {\bibfnamefont {G.}~\bibnamefont {Mullot}}, \bibinfo {author}
    {\bibfnamefont {F.}~\bibnamefont {Augé}}, \bibinfo {author} {\bibfnamefont
    {P.}~\bibnamefont {Balcou}}, \bibinfo {author} {\bibfnamefont {H.~G.}\
    \bibnamefont {Muller}}, \ and\ \bibinfo {author} {\bibfnamefont
    {P.}~\bibnamefont {Agostini}},\ }\href {\doibase 10.1126/science.1059413}
    {\bibfield  {journal} {\bibinfo  {journal} {Science}\ }\textbf {\bibinfo
    {volume} {292}},\ \bibinfo {pages} {1689} (\bibinfo {year}
    {2001})}\BibitemShut {NoStop}%
  \bibitem [{\citenamefont {Pazourek}\ \emph {et~al.}(2015)\citenamefont
    {Pazourek}, \citenamefont {Nagele},\ and\ \citenamefont
    {Burgd\"orfer}}]{RevModPhys.87.765}%
    \BibitemOpen
    \bibfield  {author} {\bibinfo {author} {\bibfnamefont {R.}~\bibnamefont
    {Pazourek}}, \bibinfo {author} {\bibfnamefont {S.}~\bibnamefont {Nagele}}, \
    and\ \bibinfo {author} {\bibfnamefont {J.}~\bibnamefont {Burgd\"orfer}},\
    }\href {\doibase 10.1103/RevModPhys.87.765} {\bibfield  {journal} {\bibinfo
    {journal} {Rev. Mod. Phys.}\ }\textbf {\bibinfo {volume} {87}},\ \bibinfo
    {pages} {765} (\bibinfo {year} {2015})}\BibitemShut {NoStop}%
  \bibitem [{\citenamefont {Isinger}\ \emph {et~al.}(2017)\citenamefont
    {Isinger}, \citenamefont {Squibb}, \citenamefont {Busto}, \citenamefont
    {Zhong}, \citenamefont {Harth}, \citenamefont {Kroon}, \citenamefont {Nandi},
    \citenamefont {Arnold}, \citenamefont {Miranda}, \citenamefont {Dahlström},
    \citenamefont {Lindroth}, \citenamefont {Feifel}, \citenamefont
    {Gisselbrecht},\ and\ \citenamefont
    {L’Huillier}}]{doi:10.1126/science.aao7043}%
    \BibitemOpen
    \bibfield  {author} {\bibinfo {author} {\bibfnamefont {M.}~\bibnamefont
    {Isinger}}, \bibinfo {author} {\bibfnamefont {R.~J.}\ \bibnamefont {Squibb}},
    \bibinfo {author} {\bibfnamefont {D.}~\bibnamefont {Busto}}, \bibinfo
    {author} {\bibfnamefont {S.}~\bibnamefont {Zhong}}, \bibinfo {author}
    {\bibfnamefont {A.}~\bibnamefont {Harth}}, \bibinfo {author} {\bibfnamefont
    {D.}~\bibnamefont {Kroon}}, \bibinfo {author} {\bibfnamefont
    {S.}~\bibnamefont {Nandi}}, \bibinfo {author} {\bibfnamefont {C.~L.}\
    \bibnamefont {Arnold}}, \bibinfo {author} {\bibfnamefont {M.}~\bibnamefont
    {Miranda}}, \bibinfo {author} {\bibfnamefont {J.~M.}\ \bibnamefont
    {Dahlström}}, \bibinfo {author} {\bibfnamefont {E.}~\bibnamefont
    {Lindroth}}, \bibinfo {author} {\bibfnamefont {R.}~\bibnamefont {Feifel}},
    \bibinfo {author} {\bibfnamefont {M.}~\bibnamefont {Gisselbrecht}}, \ and\
    \bibinfo {author} {\bibfnamefont {A.}~\bibnamefont {L’Huillier}},\ }\href
    {\doibase 10.1126/science.aao7043} {\bibfield  {journal} {\bibinfo  {journal}
    {Science}\ }\textbf {\bibinfo {volume} {358}},\ \bibinfo {pages} {893}
    (\bibinfo {year} {2017})}\BibitemShut {NoStop}%
  \bibitem [{\citenamefont {Beaulieu}\ \emph {et~al.}(2017)\citenamefont
    {Beaulieu}, \citenamefont {Comby}, \citenamefont {Clergerie}, \citenamefont
    {Caillat}, \citenamefont {Descamps}, \citenamefont {Dudovich}, \citenamefont
    {Fabre}, \citenamefont {Géneaux}, \citenamefont {Légaré}, \citenamefont
    {Petit}, \citenamefont {Pons}, \citenamefont {Porat}, \citenamefont {Ruchon},
    \citenamefont {Taïeb}, \citenamefont {Blanchet},\ and\ \citenamefont
    {Mairesse}}]{doi:10.1126/science.aao5624}%
    \BibitemOpen
    \bibfield  {author} {\bibinfo {author} {\bibfnamefont {S.}~\bibnamefont
    {Beaulieu}}, \bibinfo {author} {\bibfnamefont {A.}~\bibnamefont {Comby}},
    \bibinfo {author} {\bibfnamefont {A.}~\bibnamefont {Clergerie}}, \bibinfo
    {author} {\bibfnamefont {J.}~\bibnamefont {Caillat}}, \bibinfo {author}
    {\bibfnamefont {D.}~\bibnamefont {Descamps}}, \bibinfo {author}
    {\bibfnamefont {N.}~\bibnamefont {Dudovich}}, \bibinfo {author}
    {\bibfnamefont {B.}~\bibnamefont {Fabre}}, \bibinfo {author} {\bibfnamefont
    {R.}~\bibnamefont {Géneaux}}, \bibinfo {author} {\bibfnamefont
    {F.}~\bibnamefont {Légaré}}, \bibinfo {author} {\bibfnamefont
    {S.}~\bibnamefont {Petit}}, \bibinfo {author} {\bibfnamefont
    {B.}~\bibnamefont {Pons}}, \bibinfo {author} {\bibfnamefont {G.}~\bibnamefont
    {Porat}}, \bibinfo {author} {\bibfnamefont {T.}~\bibnamefont {Ruchon}},
    \bibinfo {author} {\bibfnamefont {R.}~\bibnamefont {Taïeb}}, \bibinfo
    {author} {\bibfnamefont {V.}~\bibnamefont {Blanchet}}, \ and\ \bibinfo
    {author} {\bibfnamefont {Y.}~\bibnamefont {Mairesse}},\ }\href {\doibase
    10.1126/science.aao5624} {\bibfield  {journal} {\bibinfo  {journal}
    {Science}\ }\textbf {\bibinfo {volume} {358}},\ \bibinfo {pages} {1288}
    (\bibinfo {year} {2017})}\BibitemShut {NoStop}%
  \bibitem [{\citenamefont {Lucchini}\ \emph {et~al.}(2015)\citenamefont
    {Lucchini}, \citenamefont {Castiglioni}, \citenamefont {Kasmi}, \citenamefont
    {Kliuiev}, \citenamefont {Ludwig}, \citenamefont {Greif}, \citenamefont
    {Osterwalder}, \citenamefont {Hengsberger}, \citenamefont {Gallmann},\ and\
    \citenamefont {Keller}}]{PhysRevLett.115.137401}%
    \BibitemOpen
    \bibfield  {author} {\bibinfo {author} {\bibfnamefont {M.}~\bibnamefont
    {Lucchini}}, \bibinfo {author} {\bibfnamefont {L.}~\bibnamefont
    {Castiglioni}}, \bibinfo {author} {\bibfnamefont {L.}~\bibnamefont {Kasmi}},
    \bibinfo {author} {\bibfnamefont {P.}~\bibnamefont {Kliuiev}}, \bibinfo
    {author} {\bibfnamefont {A.}~\bibnamefont {Ludwig}}, \bibinfo {author}
    {\bibfnamefont {M.}~\bibnamefont {Greif}}, \bibinfo {author} {\bibfnamefont
    {J.}~\bibnamefont {Osterwalder}}, \bibinfo {author} {\bibfnamefont
    {M.}~\bibnamefont {Hengsberger}}, \bibinfo {author} {\bibfnamefont
    {L.}~\bibnamefont {Gallmann}}, \ and\ \bibinfo {author} {\bibfnamefont
    {U.}~\bibnamefont {Keller}},\ }\href {\doibase
    10.1103/PhysRevLett.115.137401} {\bibfield  {journal} {\bibinfo  {journal}
    {Phys. Rev. Lett.}\ }\textbf {\bibinfo {volume} {115}},\ \bibinfo {pages}
    {137401} (\bibinfo {year} {2015})}\BibitemShut {NoStop}%
  \bibitem [{\citenamefont {Jordan}\ \emph {et~al.}(2020)\citenamefont {Jordan},
    \citenamefont {Huppert}, \citenamefont {Rattenbacher}, \citenamefont {Peper},
    \citenamefont {Jelovina}, \citenamefont {Perry}, \citenamefont {von Conta},
    \citenamefont {Schild},\ and\ \citenamefont
    {Wörner}}]{doi:10.1126/science.abb0979}%
    \BibitemOpen
    \bibfield  {author} {\bibinfo {author} {\bibfnamefont {I.}~\bibnamefont
    {Jordan}}, \bibinfo {author} {\bibfnamefont {M.}~\bibnamefont {Huppert}},
    \bibinfo {author} {\bibfnamefont {D.}~\bibnamefont {Rattenbacher}}, \bibinfo
    {author} {\bibfnamefont {M.}~\bibnamefont {Peper}}, \bibinfo {author}
    {\bibfnamefont {D.}~\bibnamefont {Jelovina}}, \bibinfo {author}
    {\bibfnamefont {C.}~\bibnamefont {Perry}}, \bibinfo {author} {\bibfnamefont
    {A.}~\bibnamefont {von Conta}}, \bibinfo {author} {\bibfnamefont
    {A.}~\bibnamefont {Schild}}, \ and\ \bibinfo {author} {\bibfnamefont {H.~J.}\
    \bibnamefont {Wörner}},\ }\href {\doibase 10.1126/science.abb0979}
    {\bibfield  {journal} {\bibinfo  {journal} {Science}\ }\textbf {\bibinfo
    {volume} {369}},\ \bibinfo {pages} {974} (\bibinfo {year}
    {2020})}\BibitemShut {NoStop}%
  \bibitem [{\citenamefont {Dahlström}\ \emph {et~al.}(2012)\citenamefont
    {Dahlström}, \citenamefont {L'Huillier},\ and\ \citenamefont
    {Maquet}}]{2012}%
    \BibitemOpen
    \bibfield  {author} {\bibinfo {author} {\bibfnamefont {J.~M.}\ \bibnamefont
    {Dahlström}}, \bibinfo {author} {\bibfnamefont {A.}~\bibnamefont
    {L'Huillier}}, \ and\ \bibinfo {author} {\bibfnamefont {A.}~\bibnamefont
    {Maquet}},\ }\href {\doibase 10.1088/0953-4075/45/18/183001} {\bibfield
    {journal} {\bibinfo  {journal} {J. Phys. B}\ }\textbf {\bibinfo {volume}
    {45}},\ \bibinfo {pages} {183001} (\bibinfo {year} {2012})}\BibitemShut
    {NoStop}%
  \bibitem [{\citenamefont {Dahlström}\ \emph {et~al.}(2013)\citenamefont
    {Dahlström}, \citenamefont {Guénot}, \citenamefont {Klünder},
    \citenamefont {Gisselbrecht}, \citenamefont {Mauritsson}, \citenamefont
    {L’Huillier}, \citenamefont {Maquet},\ and\ \citenamefont
    {Taïeb}}]{DAHLSTROM201353}%
    \BibitemOpen
    \bibfield  {author} {\bibinfo {author} {\bibfnamefont {J.}~\bibnamefont
    {Dahlström}}, \bibinfo {author} {\bibfnamefont {D.}~\bibnamefont {Guénot}},
    \bibinfo {author} {\bibfnamefont {K.}~\bibnamefont {Klünder}}, \bibinfo
    {author} {\bibfnamefont {M.}~\bibnamefont {Gisselbrecht}}, \bibinfo {author}
    {\bibfnamefont {J.}~\bibnamefont {Mauritsson}}, \bibinfo {author}
    {\bibfnamefont {A.}~\bibnamefont {L’Huillier}}, \bibinfo {author}
    {\bibfnamefont {A.}~\bibnamefont {Maquet}}, \ and\ \bibinfo {author}
    {\bibfnamefont {R.}~\bibnamefont {Taïeb}},\ }\href {\doibase
    https://doi.org/10.1016/j.chemphys.2012.01.017} {\bibfield  {journal}
    {\bibinfo  {journal} {Chemical Physics}\ }\textbf {\bibinfo {volume} {414}},\
    \bibinfo {pages} {53} (\bibinfo {year} {2013})},\ \bibinfo {note} {attosecond
    spectroscopy}\BibitemShut {NoStop}%
  \bibitem [{\citenamefont {Toma}\ and\ \citenamefont
    {Muller}(2002)}]{E_S_Toma_2002}%
    \BibitemOpen
    \bibfield  {author} {\bibinfo {author} {\bibfnamefont {E.~S.}\ \bibnamefont
    {Toma}}\ and\ \bibinfo {author} {\bibfnamefont {H.~G.}\ \bibnamefont
    {Muller}},\ }\href {\doibase 10.1088/0953-4075/35/16/306} {\bibfield
    {journal} {\bibinfo  {journal} {J. Phys. B}\ }\textbf {\bibinfo {volume} {35}},\ \bibinfo {pages} {3435}
    (\bibinfo {year} {2002})}\BibitemShut {NoStop}%
  \bibitem [{\citenamefont {V\'eniard}\ \emph {et~al.}(1996)\citenamefont
    {V\'eniard}, \citenamefont {Ta\"{\i}eb},\ and\ \citenamefont
    {Maquet}}]{PhysRevA.54.721}%
    \BibitemOpen
    \bibfield  {author} {\bibinfo {author} {\bibfnamefont {V.}~\bibnamefont
    {V\'eniard}}, \bibinfo {author} {\bibfnamefont {R.}~\bibnamefont
    {Ta\"{\i}eb}}, \ and\ \bibinfo {author} {\bibfnamefont {A.}~\bibnamefont
    {Maquet}},\ }\href {\doibase 10.1103/PhysRevA.54.721} {\bibfield  {journal}
    {\bibinfo  {journal} {Phys. Rev. A}\ }\textbf {\bibinfo {volume} {54}},\
    \bibinfo {pages} {721} (\bibinfo {year} {1996})}\BibitemShut {NoStop}%
  \bibitem [{\citenamefont {G{\'e}neaux}\ \emph {et~al.}(2016)\citenamefont
    {G{\'e}neaux}, \citenamefont {Camper}, \citenamefont {Auguste}, \citenamefont
    {Gobert}, \citenamefont {Caillat}, \citenamefont {Ta{\"i}eb},\ and\
    \citenamefont {Ruchon}}]{Géneaux2016}%
    \BibitemOpen
    \bibfield  {author} {\bibinfo {author} {\bibfnamefont {R.}~\bibnamefont
    {G{\'e}neaux}}, \bibinfo {author} {\bibfnamefont {A.}~\bibnamefont {Camper}},
    \bibinfo {author} {\bibfnamefont {T.}~\bibnamefont {Auguste}}, \bibinfo
    {author} {\bibfnamefont {O.}~\bibnamefont {Gobert}}, \bibinfo {author}
    {\bibfnamefont {J.}~\bibnamefont {Caillat}}, \bibinfo {author} {\bibfnamefont
    {R.}~\bibnamefont {Ta{\"i}eb}}, \ and\ \bibinfo {author} {\bibfnamefont
    {T.}~\bibnamefont {Ruchon}},\ }\href {\doibase 10.1038/ncomms12583}
    {\bibfield  {journal} {\bibinfo  {journal} {Nature Communications}\ }\textbf
    {\bibinfo {volume} {7}},\ \bibinfo {pages} {12583} (\bibinfo {year}
    {2016})}\BibitemShut {NoStop}%
  \bibitem [{\citenamefont {Gruson}\ \emph {et~al.}(2016)\citenamefont {Gruson},
    \citenamefont {Barreau}, \citenamefont {Jim{\'e}nez-Galan}, \citenamefont
    {Risoud}, \citenamefont {Caillat}, \citenamefont {Maquet}, \citenamefont
    {Carr{\'e}}, \citenamefont {Lepetit}, \citenamefont {Hergott}, \citenamefont
    {Ruchon}, \citenamefont {Argenti}, \citenamefont {Ta{\"\i}eb}, \citenamefont
    {Mart{\'\i}n},\ and\ \citenamefont {Sali{\`e}res}}]{Gruson734}%
    \BibitemOpen
    \bibfield  {author} {\bibinfo {author} {\bibfnamefont {V.}~\bibnamefont
    {Gruson}}, \bibinfo {author} {\bibfnamefont {L.}~\bibnamefont {Barreau}},
    \bibinfo {author} {\bibfnamefont {{\'A}.}~\bibnamefont {Jim{\'e}nez-Galan}},
    \bibinfo {author} {\bibfnamefont {F.}~\bibnamefont {Risoud}}, \bibinfo
    {author} {\bibfnamefont {J.}~\bibnamefont {Caillat}}, \bibinfo {author}
    {\bibfnamefont {A.}~\bibnamefont {Maquet}}, \bibinfo {author} {\bibfnamefont
    {B.}~\bibnamefont {Carr{\'e}}}, \bibinfo {author} {\bibfnamefont
    {F.}~\bibnamefont {Lepetit}}, \bibinfo {author} {\bibfnamefont {J.-F.}\
    \bibnamefont {Hergott}}, \bibinfo {author} {\bibfnamefont {T.}~\bibnamefont
    {Ruchon}}, \bibinfo {author} {\bibfnamefont {L.}~\bibnamefont {Argenti}},
    \bibinfo {author} {\bibfnamefont {R.}~\bibnamefont {Ta{\"\i}eb}}, \bibinfo
    {author} {\bibfnamefont {F.}~\bibnamefont {Mart{\'\i}n}}, \ and\ \bibinfo
    {author} {\bibfnamefont {P.}~\bibnamefont {Sali{\`e}res}},\ }\href {\doibase
    10.1126/science.aah5188} {\bibfield  {journal} {\bibinfo  {journal}
    {Science}\ }\textbf {\bibinfo {volume} {354}},\ \bibinfo {pages} {734}
    (\bibinfo {year} {2016})}\BibitemShut {NoStop}%
  \bibitem [{\citenamefont {Zhong}\ \emph {et~al.}(2020)\citenamefont {Zhong},
    \citenamefont {Vinbladh}, \citenamefont {Busto}, \citenamefont {Squibb},
    \citenamefont {Isinger}, \citenamefont {Neori{\v{c}}i{\'{c}}}, \citenamefont
    {Laurell}, \citenamefont {Weissenbilder}, \citenamefont {Arnold},
    \citenamefont {Feifel}, \citenamefont {Dahlstr{\"o}m}, \citenamefont
    {Wendin}, \citenamefont {Gisselbrecht}, \citenamefont {Lindroth},\ and\
    \citenamefont {L'Huillier}}]{Zhong2020}%
    \BibitemOpen
    \bibfield  {author} {\bibinfo {author} {\bibfnamefont {S.}~\bibnamefont
    {Zhong}}, \bibinfo {author} {\bibfnamefont {J.}~\bibnamefont {Vinbladh}},
    \bibinfo {author} {\bibfnamefont {D.}~\bibnamefont {Busto}}, \bibinfo
    {author} {\bibfnamefont {R.~J.}\ \bibnamefont {Squibb}}, \bibinfo {author}
    {\bibfnamefont {M.}~\bibnamefont {Isinger}}, \bibinfo {author} {\bibfnamefont
    {L.}~\bibnamefont {Neori{\v{c}}i{\'{c}}}}, \bibinfo {author} {\bibfnamefont
    {H.}~\bibnamefont {Laurell}}, \bibinfo {author} {\bibfnamefont
    {R.}~\bibnamefont {Weissenbilder}}, \bibinfo {author} {\bibfnamefont {C.~L.}\
    \bibnamefont {Arnold}}, \bibinfo {author} {\bibfnamefont {R.}~\bibnamefont
    {Feifel}}, \bibinfo {author} {\bibfnamefont {J.~M.}\ \bibnamefont
    {Dahlstr{\"o}m}}, \bibinfo {author} {\bibfnamefont {G.}~\bibnamefont
    {Wendin}}, \bibinfo {author} {\bibfnamefont {M.}~\bibnamefont
    {Gisselbrecht}}, \bibinfo {author} {\bibfnamefont {E.}~\bibnamefont
    {Lindroth}}, \ and\ \bibinfo {author} {\bibfnamefont {A.}~\bibnamefont
    {L'Huillier}},\ }\href {\doibase 10.1038/s41467-020-18847-1} {\bibfield
    {journal} {\bibinfo  {journal} {Nature Communications}\ }\textbf {\bibinfo
    {volume} {11}},\ \bibinfo {pages} {5042} (\bibinfo {year}
    {2020})}\BibitemShut {NoStop}%
  \bibitem [{\citenamefont {Biswas}\ \emph {et~al.}(2020)\citenamefont {Biswas},
    \citenamefont {F{\"o}rg}, \citenamefont {Ortmann}, \citenamefont
    {Sch{\"o}tz}, \citenamefont {Schweinberger}, \citenamefont {Zimmermann},
    \citenamefont {Pi}, \citenamefont {Baykusheva}, \citenamefont {Masood},
    \citenamefont {Liontos}, \citenamefont {Kamal}, \citenamefont {Kling},
    \citenamefont {Alharbi}, \citenamefont {Alharbi}, \citenamefont {Azzeer},
    \citenamefont {Hartmann}, \citenamefont {W{\"o}rner}, \citenamefont
    {Landsman},\ and\ \citenamefont {Kling}}]{Biswas2020}%
    \BibitemOpen
    \bibfield  {author} {\bibinfo {author} {\bibfnamefont {S.}~\bibnamefont
    {Biswas}}, \bibinfo {author} {\bibfnamefont {B.}~\bibnamefont {F{\"o}rg}},
    \bibinfo {author} {\bibfnamefont {L.}~\bibnamefont {Ortmann}}, \bibinfo
    {author} {\bibfnamefont {J.}~\bibnamefont {Sch{\"o}tz}}, \bibinfo {author}
    {\bibfnamefont {W.}~\bibnamefont {Schweinberger}}, \bibinfo {author}
    {\bibfnamefont {T.}~\bibnamefont {Zimmermann}}, \bibinfo {author}
    {\bibfnamefont {L.}~\bibnamefont {Pi}}, \bibinfo {author} {\bibfnamefont
    {D.}~\bibnamefont {Baykusheva}}, \bibinfo {author} {\bibfnamefont {H.~A.}\
    \bibnamefont {Masood}}, \bibinfo {author} {\bibfnamefont {I.}~\bibnamefont
    {Liontos}}, \bibinfo {author} {\bibfnamefont {A.~M.}\ \bibnamefont {Kamal}},
    \bibinfo {author} {\bibfnamefont {N.~G.}\ \bibnamefont {Kling}}, \bibinfo
    {author} {\bibfnamefont {A.~F.}\ \bibnamefont {Alharbi}}, \bibinfo {author}
    {\bibfnamefont {M.}~\bibnamefont {Alharbi}}, \bibinfo {author} {\bibfnamefont
    {A.~M.}\ \bibnamefont {Azzeer}}, \bibinfo {author} {\bibfnamefont
    {G.}~\bibnamefont {Hartmann}}, \bibinfo {author} {\bibfnamefont {H.~J.}\
    \bibnamefont {W{\"o}rner}}, \bibinfo {author} {\bibfnamefont {A.~S.}\
    \bibnamefont {Landsman}}, \ and\ \bibinfo {author} {\bibfnamefont {M.~F.}\
    \bibnamefont {Kling}},\ }\href {\doibase 10.1038/s41567-020-0887-8}
    {\bibfield  {journal} {\bibinfo  {journal} {Nature Physics}\ }\textbf
    {\bibinfo {volume} {16}},\ \bibinfo {pages} {778} (\bibinfo {year}
    {2020})}\BibitemShut {NoStop}%
  \bibitem [{\citenamefont {Gong}\ \emph {et~al.}(2022)\citenamefont {Gong},
    \citenamefont {Jiang}, \citenamefont {Tong}, \citenamefont {Qiang},
    \citenamefont {Lu}, \citenamefont {Ni}, \citenamefont {Lucchese},
    \citenamefont {Ueda},\ and\ \citenamefont {Wu}}]{PhysRevX.12.011002}%
    \BibitemOpen
    \bibfield  {author} {\bibinfo {author} {\bibfnamefont {X.}~\bibnamefont
    {Gong}}, \bibinfo {author} {\bibfnamefont {W.}~\bibnamefont {Jiang}},
    \bibinfo {author} {\bibfnamefont {J.}~\bibnamefont {Tong}}, \bibinfo {author}
    {\bibfnamefont {J.}~\bibnamefont {Qiang}}, \bibinfo {author} {\bibfnamefont
    {P.}~\bibnamefont {Lu}}, \bibinfo {author} {\bibfnamefont {H.}~\bibnamefont
    {Ni}}, \bibinfo {author} {\bibfnamefont {R.}~\bibnamefont {Lucchese}},
    \bibinfo {author} {\bibfnamefont {K.}~\bibnamefont {Ueda}}, \ and\ \bibinfo
    {author} {\bibfnamefont {J.}~\bibnamefont {Wu}},\ }\href {\doibase
    10.1103/PhysRevX.12.011002} {\bibfield  {journal} {\bibinfo  {journal} {Phys.
    Rev. X}\ }\textbf {\bibinfo {volume} {12}},\ \bibinfo {pages} {011002}
    (\bibinfo {year} {2022})}\BibitemShut {NoStop}%
  \bibitem [{\citenamefont {Swoboda}\ \emph {et~al.}(2010)\citenamefont
    {Swoboda}, \citenamefont {Fordell}, \citenamefont {Kl\"under}, \citenamefont
    {Dahlstr\"om}, \citenamefont {Miranda}, \citenamefont {Buth}, \citenamefont
    {Schafer}, \citenamefont {Mauritsson}, \citenamefont {L'Huillier},\ and\
    \citenamefont {Gisselbrecht}}]{PhysRevLett.104.103003}%
    \BibitemOpen
    \bibfield  {author} {\bibinfo {author} {\bibfnamefont {M.}~\bibnamefont
    {Swoboda}}, \bibinfo {author} {\bibfnamefont {T.}~\bibnamefont {Fordell}},
    \bibinfo {author} {\bibfnamefont {K.}~\bibnamefont {Kl\"under}}, \bibinfo
    {author} {\bibfnamefont {J.~M.}\ \bibnamefont {Dahlstr\"om}}, \bibinfo
    {author} {\bibfnamefont {M.}~\bibnamefont {Miranda}}, \bibinfo {author}
    {\bibfnamefont {C.}~\bibnamefont {Buth}}, \bibinfo {author} {\bibfnamefont
    {K.~J.}\ \bibnamefont {Schafer}}, \bibinfo {author} {\bibfnamefont
    {J.}~\bibnamefont {Mauritsson}}, \bibinfo {author} {\bibfnamefont
    {A.}~\bibnamefont {L'Huillier}}, \ and\ \bibinfo {author} {\bibfnamefont
    {M.}~\bibnamefont {Gisselbrecht}},\ }\href {\doibase
    10.1103/PhysRevLett.104.103003} {\bibfield  {journal} {\bibinfo  {journal}
    {Phys. Rev. Lett.}\ }\textbf {\bibinfo {volume} {104}},\ \bibinfo {pages}
    {103003} (\bibinfo {year} {2010})}\BibitemShut {NoStop}%
  \bibitem [{\citenamefont {Kheifets}(2021)}]{PhysRevA.104.L021103}%
    \BibitemOpen
    \bibfield  {author} {\bibinfo {author} {\bibfnamefont {A.~S.}\ \bibnamefont
    {Kheifets}},\ }\href {\doibase 10.1103/PhysRevA.104.L021103} {\bibfield
    {journal} {\bibinfo  {journal} {Phys. Rev. A}\ }\textbf {\bibinfo {volume}
    {104}},\ \bibinfo {pages} {L021103} (\bibinfo {year} {2021})}\BibitemShut
    {NoStop}%
  \bibitem [{\citenamefont {Autuori}\ \emph {et~al.}(2022)\citenamefont
    {Autuori}, \citenamefont {Platzer}, \citenamefont {Lejman}, \citenamefont
    {Gallician}, \citenamefont {Maëder}, \citenamefont {Covolo}, \citenamefont
    {Bosse}, \citenamefont {Dalui}, \citenamefont {Bresteau}, \citenamefont
    {Hergott}, \citenamefont {Tcherbakoff}, \citenamefont {Marroux},
    \citenamefont {Loriot}, \citenamefont {Lépine}, \citenamefont {Poisson},
    \citenamefont {Taïeb}, \citenamefont {Caillat},\ and\ \citenamefont
    {Salières}}]{doi:10.1126/sciadv.abl7594}%
    \BibitemOpen
    \bibfield  {author} {\bibinfo {author} {\bibfnamefont {A.}~\bibnamefont
    {Autuori}}, \bibinfo {author} {\bibfnamefont {D.}~\bibnamefont {Platzer}},
    \bibinfo {author} {\bibfnamefont {M.}~\bibnamefont {Lejman}}, \bibinfo
    {author} {\bibfnamefont {G.}~\bibnamefont {Gallician}}, \bibinfo {author}
    {\bibfnamefont {L.}~\bibnamefont {Maëder}}, \bibinfo {author} {\bibfnamefont
    {A.}~\bibnamefont {Covolo}}, \bibinfo {author} {\bibfnamefont
    {L.}~\bibnamefont {Bosse}}, \bibinfo {author} {\bibfnamefont
    {M.}~\bibnamefont {Dalui}}, \bibinfo {author} {\bibfnamefont
    {D.}~\bibnamefont {Bresteau}}, \bibinfo {author} {\bibfnamefont {J.-F.}\
    \bibnamefont {Hergott}}, \bibinfo {author} {\bibfnamefont {O.}~\bibnamefont
    {Tcherbakoff}}, \bibinfo {author} {\bibfnamefont {H.~J.~B.}\ \bibnamefont
    {Marroux}}, \bibinfo {author} {\bibfnamefont {V.}~\bibnamefont {Loriot}},
    \bibinfo {author} {\bibfnamefont {F.}~\bibnamefont {Lépine}}, \bibinfo
    {author} {\bibfnamefont {L.}~\bibnamefont {Poisson}}, \bibinfo {author}
    {\bibfnamefont {R.}~\bibnamefont {Taïeb}}, \bibinfo {author} {\bibfnamefont
    {J.}~\bibnamefont {Caillat}}, \ and\ \bibinfo {author} {\bibfnamefont
    {P.}~\bibnamefont {Salières}},\ }\href {\doibase 10.1126/sciadv.abl7594}
    {\bibfield  {journal} {\bibinfo  {journal} {Science Advances}\ }\textbf
    {\bibinfo {volume} {8}},\ \bibinfo {pages} {eabl7594} (\bibinfo {year}
    {2022})}\BibitemShut {NoStop}%
  \bibitem [{\citenamefont {Heuser}\ \emph {et~al.}(2016)\citenamefont {Heuser},
    \citenamefont {Jim\'enez~Gal\'an}, \citenamefont {Cirelli}, \citenamefont
    {Marante}, \citenamefont {Sabbar}, \citenamefont {Boge}, \citenamefont
    {Lucchini}, \citenamefont {Gallmann}, \citenamefont {Ivanov}, \citenamefont
    {Kheifets}, \citenamefont {Dahlstr\"om}, \citenamefont {Lindroth},
    \citenamefont {Argenti}, \citenamefont {Mart\'{\i}n},\ and\ \citenamefont
    {Keller}}]{PhysRevA.94.063409}%
    \BibitemOpen
    \bibfield  {author} {\bibinfo {author} {\bibfnamefont {S.}~\bibnamefont
    {Heuser}}, \bibinfo {author} {\bibfnamefont {A.}~\bibnamefont
    {Jim\'enez~Gal\'an}}, \bibinfo {author} {\bibfnamefont {C.}~\bibnamefont
    {Cirelli}}, \bibinfo {author} {\bibfnamefont {C.}~\bibnamefont {Marante}},
    \bibinfo {author} {\bibfnamefont {M.}~\bibnamefont {Sabbar}}, \bibinfo
    {author} {\bibfnamefont {R.}~\bibnamefont {Boge}}, \bibinfo {author}
    {\bibfnamefont {M.}~\bibnamefont {Lucchini}}, \bibinfo {author}
    {\bibfnamefont {L.}~\bibnamefont {Gallmann}}, \bibinfo {author}
    {\bibfnamefont {I.}~\bibnamefont {Ivanov}}, \bibinfo {author} {\bibfnamefont
    {A.~S.}\ \bibnamefont {Kheifets}}, \bibinfo {author} {\bibfnamefont {J.~M.}\
    \bibnamefont {Dahlstr\"om}}, \bibinfo {author} {\bibfnamefont
    {E.}~\bibnamefont {Lindroth}}, \bibinfo {author} {\bibfnamefont
    {L.}~\bibnamefont {Argenti}}, \bibinfo {author} {\bibfnamefont
    {F.}~\bibnamefont {Mart\'{\i}n}}, \ and\ \bibinfo {author} {\bibfnamefont
    {U.}~\bibnamefont {Keller}},\ }\href {\doibase 10.1103/PhysRevA.94.063409}
    {\bibfield  {journal} {\bibinfo  {journal} {Phys. Rev. A}\ }\textbf {\bibinfo
    {volume} {94}},\ \bibinfo {pages} {063409} (\bibinfo {year}
    {2016})}\BibitemShut {NoStop}%
  \bibitem [{\citenamefont {Baggesen}\ and\ \citenamefont
    {Madsen}(2010)}]{PhysRevLett.104.043602}%
    \BibitemOpen
    \bibfield  {author} {\bibinfo {author} {\bibfnamefont {J.~C.}\ \bibnamefont
    {Baggesen}}\ and\ \bibinfo {author} {\bibfnamefont {L.~B.}\ \bibnamefont
    {Madsen}},\ }\href {\doibase 10.1103/PhysRevLett.104.043602} {\bibfield
    {journal} {\bibinfo  {journal} {Phys. Rev. Lett.}\ }\textbf {\bibinfo
    {volume} {104}},\ \bibinfo {pages} {043602} (\bibinfo {year}
    {2010})}\BibitemShut {NoStop}%
  \bibitem [{\citenamefont {Pazourek}\ \emph {et~al.}(2012)\citenamefont
    {Pazourek}, \citenamefont {Feist}, \citenamefont {Nagele},\ and\
    \citenamefont {Burgd\"orfer}}]{PhysRevLett.108.163001}%
    \BibitemOpen
    \bibfield  {author} {\bibinfo {author} {\bibfnamefont {R.}~\bibnamefont
    {Pazourek}}, \bibinfo {author} {\bibfnamefont {J.}~\bibnamefont {Feist}},
    \bibinfo {author} {\bibfnamefont {S.}~\bibnamefont {Nagele}}, \ and\ \bibinfo
    {author} {\bibfnamefont {J.}~\bibnamefont {Burgd\"orfer}},\ }\href {\doibase
    10.1103/PhysRevLett.108.163001} {\bibfield  {journal} {\bibinfo  {journal}
    {Phys. Rev. Lett.}\ }\textbf {\bibinfo {volume} {108}},\ \bibinfo {pages}
    {163001} (\bibinfo {year} {2012})}\BibitemShut {NoStop}%
  \bibitem [{\citenamefont {Ossiander}\ \emph {et~al.}(2017)\citenamefont
    {Ossiander}, \citenamefont {Siegrist}, \citenamefont {Shirvanyan},
    \citenamefont {Pazourek}, \citenamefont {Sommer}, \citenamefont {Latka},
    \citenamefont {Guggenmos}, \citenamefont {Nagele}, \citenamefont {Feist},
    \citenamefont {Burgd{\"o}rfer}, \citenamefont {Kienberger},\ and\
    \citenamefont {Schultze}}]{Ossiander2017}%
    \BibitemOpen
    \bibfield  {author} {\bibinfo {author} {\bibfnamefont {M.}~\bibnamefont
    {Ossiander}}, \bibinfo {author} {\bibfnamefont {F.}~\bibnamefont {Siegrist}},
    \bibinfo {author} {\bibfnamefont {V.}~\bibnamefont {Shirvanyan}}, \bibinfo
    {author} {\bibfnamefont {R.}~\bibnamefont {Pazourek}}, \bibinfo {author}
    {\bibfnamefont {A.}~\bibnamefont {Sommer}}, \bibinfo {author} {\bibfnamefont
    {T.}~\bibnamefont {Latka}}, \bibinfo {author} {\bibfnamefont
    {A.}~\bibnamefont {Guggenmos}}, \bibinfo {author} {\bibfnamefont
    {S.}~\bibnamefont {Nagele}}, \bibinfo {author} {\bibfnamefont
    {J.}~\bibnamefont {Feist}}, \bibinfo {author} {\bibfnamefont
    {J.}~\bibnamefont {Burgd{\"o}rfer}}, \bibinfo {author} {\bibfnamefont
    {R.}~\bibnamefont {Kienberger}}, \ and\ \bibinfo {author} {\bibfnamefont
    {M.}~\bibnamefont {Schultze}},\ }\href {\doibase 10.1038/nphys3941}
    {\bibfield  {journal} {\bibinfo  {journal} {Nature Physics}\ }\textbf
    {\bibinfo {volume} {13}},\ \bibinfo {pages} {280} (\bibinfo {year}
    {2017})}\BibitemShut {NoStop}%
  \bibitem [{\citenamefont {Donsa}\ \emph {et~al.}(2020)\citenamefont {Donsa},
    \citenamefont {Ederer}, \citenamefont {Pazourek}, \citenamefont
    {Burgd\"orfer},\ and\ \citenamefont {B\ifmmode~\check{r}\else
    \v{r}\fi{}ezinov\'a}}]{PhysRevA.102.033112}%
    \BibitemOpen
    \bibfield  {author} {\bibinfo {author} {\bibfnamefont {S.}~\bibnamefont
    {Donsa}}, \bibinfo {author} {\bibfnamefont {M.}~\bibnamefont {Ederer}},
    \bibinfo {author} {\bibfnamefont {R.}~\bibnamefont {Pazourek}}, \bibinfo
    {author} {\bibfnamefont {J.}~\bibnamefont {Burgd\"orfer}}, \ and\ \bibinfo
    {author} {\bibfnamefont {I.}~\bibnamefont {B\ifmmode~\check{r}\else
    \v{r}\fi{}ezinov\'a}},\ }\href {\doibase 10.1103/PhysRevA.102.033112}
    {\bibfield  {journal} {\bibinfo  {journal} {Phys. Rev. A}\ }\textbf {\bibinfo
    {volume} {102}},\ \bibinfo {pages} {033112} (\bibinfo {year}
    {2020})}\BibitemShut {NoStop}%
  \bibitem [{\citenamefont {Sussman}(2011)}]{10.1119/1.3553018}%
    \BibitemOpen
    \bibfield  {author} {\bibinfo {author} {\bibfnamefont {B.~J.}\ \bibnamefont
    {Sussman}},\ }\href {\doibase 10.1119/1.3553018} {\bibfield  {journal}
    {\bibinfo  {journal} {American Journal of Physics}\ }\textbf {\bibinfo
    {volume} {79}},\ \bibinfo {pages} {477} (\bibinfo {year} {2011})}\BibitemShut
    {NoStop}%
  \bibitem [{\citenamefont {Kfir}\ \emph {et~al.}(2015)\citenamefont {Kfir},
    \citenamefont {Grychtol}, \citenamefont {Turgut}, \citenamefont {Knut},
    \citenamefont {Zusin}, \citenamefont {Popmintchev}, \citenamefont
    {Popmintchev}, \citenamefont {Nembach}, \citenamefont {Shaw}, \citenamefont
    {Fleischer}, \citenamefont {Kapteyn}, \citenamefont {Murnane},\ and\
    \citenamefont {Cohen}}]{Kfir2015}%
    \BibitemOpen
    \bibfield  {author} {\bibinfo {author} {\bibfnamefont {O.}~\bibnamefont
    {Kfir}}, \bibinfo {author} {\bibfnamefont {P.}~\bibnamefont {Grychtol}},
    \bibinfo {author} {\bibfnamefont {E.}~\bibnamefont {Turgut}}, \bibinfo
    {author} {\bibfnamefont {R.}~\bibnamefont {Knut}}, \bibinfo {author}
    {\bibfnamefont {D.}~\bibnamefont {Zusin}}, \bibinfo {author} {\bibfnamefont
    {D.}~\bibnamefont {Popmintchev}}, \bibinfo {author} {\bibfnamefont
    {T.}~\bibnamefont {Popmintchev}}, \bibinfo {author} {\bibfnamefont
    {H.}~\bibnamefont {Nembach}}, \bibinfo {author} {\bibfnamefont {J.~M.}\
    \bibnamefont {Shaw}}, \bibinfo {author} {\bibfnamefont {A.}~\bibnamefont
    {Fleischer}}, \bibinfo {author} {\bibfnamefont {H.}~\bibnamefont {Kapteyn}},
    \bibinfo {author} {\bibfnamefont {M.}~\bibnamefont {Murnane}}, \ and\
    \bibinfo {author} {\bibfnamefont {O.}~\bibnamefont {Cohen}},\ }\href
    {\doibase 10.1038/nphoton.2014.293} {\bibfield  {journal} {\bibinfo
    {journal} {Nature Photonics}\ }\textbf {\bibinfo {volume} {9}},\ \bibinfo
    {pages} {99} (\bibinfo {year} {2015})}\BibitemShut {NoStop}%
  \bibitem [{\citenamefont {Donsa}\ \emph {et~al.}(2019)\citenamefont {Donsa},
    \citenamefont {Douguet}, \citenamefont {Burgd\"orfer}, \citenamefont
    {B\ifmmode~\check{r}\else \v{r}\fi{}ezinov\'a},\ and\ \citenamefont
    {Argenti}}]{PhysRevLett.123.133203}%
    \BibitemOpen
    \bibfield  {author} {\bibinfo {author} {\bibfnamefont {S.}~\bibnamefont
    {Donsa}}, \bibinfo {author} {\bibfnamefont {N.}~\bibnamefont {Douguet}},
    \bibinfo {author} {\bibfnamefont {J.}~\bibnamefont {Burgd\"orfer}}, \bibinfo
    {author} {\bibfnamefont {I.}~\bibnamefont {B\ifmmode~\check{r}\else
    \v{r}\fi{}ezinov\'a}}, \ and\ \bibinfo {author} {\bibfnamefont
    {L.}~\bibnamefont {Argenti}},\ }\href {\doibase
    10.1103/PhysRevLett.123.133203} {\bibfield  {journal} {\bibinfo  {journal}
    {Phys. Rev. Lett.}\ }\textbf {\bibinfo {volume} {123}},\ \bibinfo {pages}
    {133203} (\bibinfo {year} {2019})}\BibitemShut {NoStop}%
  \bibitem [{\citenamefont {Sörngård}\ \emph {et~al.}(2020)\citenamefont
    {Sörngård}, \citenamefont {Dahlström},\ and\ \citenamefont
    {Lindroth}}]{Sorngard_2020}%
    \BibitemOpen
    \bibfield  {author} {\bibinfo {author} {\bibfnamefont {J.}~\bibnamefont
    {Sörngård}}, \bibinfo {author} {\bibfnamefont {J.~M.}\ \bibnamefont
    {Dahlström}}, \ and\ \bibinfo {author} {\bibfnamefont {E.}~\bibnamefont
    {Lindroth}},\ }\href {\doibase 10.1088/1361-6455/ab84c6} {\bibfield
    {journal} {\bibinfo  {journal} {J. Phys. B}\ }\textbf {\bibinfo {volume} {53}},\ \bibinfo {pages}
    {134003} (\bibinfo {year} {2020})}\BibitemShut {NoStop}%
  \bibitem [{\citenamefont {Han}\ \emph {et~al.}(2023)\citenamefont {Han},
    \citenamefont {Ji}, \citenamefont {Bal{\v{c}}i{\={u}}nas}, \citenamefont
    {Ueda},\ and\ \citenamefont {W{\"o}rner}}]{Han2023}%
    \BibitemOpen
    \bibfield  {author} {\bibinfo {author} {\bibfnamefont {M.}~\bibnamefont
    {Han}}, \bibinfo {author} {\bibfnamefont {J.-B.}\ \bibnamefont {Ji}},
    \bibinfo {author} {\bibfnamefont {T.}~\bibnamefont {Bal{\v{c}}i{\={u}}nas}},
    \bibinfo {author} {\bibfnamefont {K.}~\bibnamefont {Ueda}}, \ and\ \bibinfo
    {author} {\bibfnamefont {H.~J.}\ \bibnamefont {W{\"o}rner}},\ }\href
    {\doibase 10.1038/s41567-022-01832-4} {\bibfield  {journal} {\bibinfo
    {journal} {Nature Physics}\ }\textbf {\bibinfo {volume} {19}},\ \bibinfo
    {pages} {230} (\bibinfo {year} {2023})}\BibitemShut {NoStop}%
  \bibitem [{\citenamefont {Sarsa}\ \emph {et~al.}(2004)\citenamefont {Sarsa},
    \citenamefont {Gálvez},\ and\ \citenamefont {Buendía}}]{SARSA2004163}%
    \BibitemOpen
    \bibfield  {author} {\bibinfo {author} {\bibfnamefont {A.}~\bibnamefont
    {Sarsa}}, \bibinfo {author} {\bibfnamefont {F.}~\bibnamefont {Gálvez}}, \
    and\ \bibinfo {author} {\bibfnamefont {E.}~\bibnamefont {Buendía}},\ }\href
    {\doibase https://doi.org/10.1016/j.adt.2004.07.003} {\bibfield  {journal}
    {\bibinfo  {journal} {Atomic Data and Nuclear Data Tables}\ }\textbf
    {\bibinfo {volume} {88}},\ \bibinfo {pages} {163} (\bibinfo {year}
    {2004})}\BibitemShut {NoStop}%
  \bibitem [{\citenamefont {Kramida}\ \emph {et~al.}(2022)\citenamefont
    {Kramida}, \citenamefont {{Yu.~Ralchenko}}, \citenamefont {Reader},\ and\
    \citenamefont {{and NIST ASD Team}}}]{NIST_ASD}%
    \BibitemOpen
    \bibfield  {author} {\bibinfo {author} {\bibfnamefont {A.}~\bibnamefont
    {Kramida}}, \bibinfo {author} {\bibnamefont {{Yu.~Ralchenko}}}, \bibinfo
    {author} {\bibfnamefont {J.}~\bibnamefont {Reader}}, \ and\ \bibinfo {author}
    {\bibnamefont {{and NIST ASD Team}}},\ }\href@noop {} {}\bibinfo
    {howpublished} {{NIST Atomic Spectra Database (ver. 5.10), [Online].
    Available: {\tt{https://physics.nist.gov/asd}} [2023, September 4]. National
    Institute of Standards and Technology, Gaithersburg, MD.}} (\bibinfo {year}
    {2022})\BibitemShut {NoStop}%
  \bibitem [{\citenamefont {Rescigno}\ and\ \citenamefont
    {McCurdy}(2000)}]{PhysRevA.62.032706}%
    \BibitemOpen
    \bibfield  {author} {\bibinfo {author} {\bibfnamefont {T.~N.}\ \bibnamefont
    {Rescigno}}\ and\ \bibinfo {author} {\bibfnamefont {C.~W.}\ \bibnamefont
    {McCurdy}},\ }\href {\doibase 10.1103/PhysRevA.62.032706} {\bibfield
    {journal} {\bibinfo  {journal} {Phys. Rev. A}\ }\textbf {\bibinfo {volume}
    {62}},\ \bibinfo {pages} {032706} (\bibinfo {year} {2000})}\BibitemShut
    {NoStop}%
  \bibitem [{\citenamefont {Jiang}\ and\ \citenamefont {Tian}(2017)}]{Jiang:17}%
    \BibitemOpen
    \bibfield  {author} {\bibinfo {author} {\bibfnamefont {W.-C.}\ \bibnamefont
    {Jiang}}\ and\ \bibinfo {author} {\bibfnamefont {X.-Q.}\ \bibnamefont
    {Tian}},\ }\href {\doibase 10.1364/OE.25.026832} {\bibfield  {journal}
    {\bibinfo  {journal} {Opt. Express}\ }\textbf {\bibinfo {volume} {25}},\
    \bibinfo {pages} {26832} (\bibinfo {year} {2017})}\BibitemShut {NoStop}%
  \bibitem [{\citenamefont {Liang}\ \emph {et~al.}(2022)\citenamefont {Liang},
    \citenamefont {Zhou}, \citenamefont {Liao}, \citenamefont {Jiang},
    \citenamefont {Li},\ and\ \citenamefont {Lu}}]{doi:10.34133/2022/9842716}%
    \BibitemOpen
    \bibfield  {author} {\bibinfo {author} {\bibfnamefont {J.}~\bibnamefont
    {Liang}}, \bibinfo {author} {\bibfnamefont {Y.}~\bibnamefont {Zhou}},
    \bibinfo {author} {\bibfnamefont {Y.}~\bibnamefont {Liao}}, \bibinfo {author}
    {\bibfnamefont {W.-C.}\ \bibnamefont {Jiang}}, \bibinfo {author}
    {\bibfnamefont {M.}~\bibnamefont {Li}}, \ and\ \bibinfo {author}
    {\bibfnamefont {P.}~\bibnamefont {Lu}},\ }\href {\doibase
    10.34133/2022/9842716} {\bibfield  {journal} {\bibinfo  {journal} {Ultrafast
    Science}\ }\textbf {\bibinfo {volume} {2022}} (\bibinfo {year} {2022}),\
    10.34133/2022/9842716}\BibitemShut {NoStop}%
  \bibitem [{\citenamefont {Arb\'o}\ \emph {et~al.}(2008)\citenamefont {Arb\'o},
    \citenamefont {Miraglia}, \citenamefont {Gravielle}, \citenamefont
    {Schiessl}, \citenamefont {Persson},\ and\ \citenamefont
    {Burgd\"orfer}}]{PhysRevA.77.013401}%
    \BibitemOpen
    \bibfield  {author} {\bibinfo {author} {\bibfnamefont {D.~G.}\ \bibnamefont
    {Arb\'o}}, \bibinfo {author} {\bibfnamefont {J.~E.}\ \bibnamefont
    {Miraglia}}, \bibinfo {author} {\bibfnamefont {M.~S.}\ \bibnamefont
    {Gravielle}}, \bibinfo {author} {\bibfnamefont {K.}~\bibnamefont {Schiessl}},
    \bibinfo {author} {\bibfnamefont {E.}~\bibnamefont {Persson}}, \ and\
    \bibinfo {author} {\bibfnamefont {J.}~\bibnamefont {Burgd\"orfer}},\ }\href
    {\doibase 10.1103/PhysRevA.77.013401} {\bibfield  {journal} {\bibinfo
    {journal} {Phys. Rev. A}\ }\textbf {\bibinfo {volume} {77}},\ \bibinfo
    {pages} {013401} (\bibinfo {year} {2008})}\BibitemShut {NoStop}%
  \bibitem [{\citenamefont {Joachain}\ \emph {et~al.}(2011)\citenamefont
    {Joachain}, \citenamefont {Kylstra},\ and\ \citenamefont
    {Potvliege}}]{joachain_kylstra_potvliege_2011}%
    \BibitemOpen
    \bibfield  {author} {\bibinfo {author} {\bibfnamefont {C.~J.}\ \bibnamefont
    {Joachain}}, \bibinfo {author} {\bibfnamefont {N.~J.}\ \bibnamefont
    {Kylstra}}, \ and\ \bibinfo {author} {\bibfnamefont {R.~M.}\ \bibnamefont
    {Potvliege}},\ }\href {\doibase 10.1017/CBO9780511993459} {\emph {\bibinfo
    {title} {Atoms in Intense Laser Fields}}}\ (\bibinfo  {publisher} {Cambridge
    University Press},\ \bibinfo {year} {2011})\BibitemShut {NoStop}%
  \bibitem [{\citenamefont {Kl\"under}\ \emph {et~al.}(2011)\citenamefont
    {Kl\"under}, \citenamefont {Dahlstr\"om}, \citenamefont {Gisselbrecht},
    \citenamefont {Fordell}, \citenamefont {Swoboda}, \citenamefont {Gu\'enot},
    \citenamefont {Johnsson}, \citenamefont {Caillat}, \citenamefont
    {Mauritsson}, \citenamefont {Maquet}, \citenamefont {Ta\"{\i}eb},\ and\
    \citenamefont {L'Huillier}}]{PhysRevLett.106.143002}%
    \BibitemOpen
    \bibfield  {author} {\bibinfo {author} {\bibfnamefont {K.}~\bibnamefont
    {Kl\"under}}, \bibinfo {author} {\bibfnamefont {J.~M.}\ \bibnamefont
    {Dahlstr\"om}}, \bibinfo {author} {\bibfnamefont {M.}~\bibnamefont
    {Gisselbrecht}}, \bibinfo {author} {\bibfnamefont {T.}~\bibnamefont
    {Fordell}}, \bibinfo {author} {\bibfnamefont {M.}~\bibnamefont {Swoboda}},
    \bibinfo {author} {\bibfnamefont {D.}~\bibnamefont {Gu\'enot}}, \bibinfo
    {author} {\bibfnamefont {P.}~\bibnamefont {Johnsson}}, \bibinfo {author}
    {\bibfnamefont {J.}~\bibnamefont {Caillat}}, \bibinfo {author} {\bibfnamefont
    {J.}~\bibnamefont {Mauritsson}}, \bibinfo {author} {\bibfnamefont
    {A.}~\bibnamefont {Maquet}}, \bibinfo {author} {\bibfnamefont
    {R.}~\bibnamefont {Ta\"{\i}eb}}, \ and\ \bibinfo {author} {\bibfnamefont
    {A.}~\bibnamefont {L'Huillier}},\ }\href {\doibase
    10.1103/PhysRevLett.106.143002} {\bibfield  {journal} {\bibinfo  {journal}
    {Phys. Rev. Lett.}\ }\textbf {\bibinfo {volume} {106}},\ \bibinfo {pages}
    {143002} (\bibinfo {year} {2011})}\BibitemShut {NoStop}%
  \bibitem [{\citenamefont {Breit}\ and\ \citenamefont
    {Bethe}(1954)}]{PhysRev.93.888}%
    \BibitemOpen
    \bibfield  {author} {\bibinfo {author} {\bibfnamefont {G.}~\bibnamefont
    {Breit}}\ and\ \bibinfo {author} {\bibfnamefont {H.~A.}\ \bibnamefont
    {Bethe}},\ }\href {\doibase 10.1103/PhysRev.93.888} {\bibfield  {journal}
    {\bibinfo  {journal} {Phys. Rev.}\ }\textbf {\bibinfo {volume} {93}},\
    \bibinfo {pages} {888} (\bibinfo {year} {1954})}\BibitemShut {NoStop}%
  \bibitem [{\citenamefont {Altshuler}(1956)}]{Altshuler1956}%
    \BibitemOpen
    \bibfield  {author} {\bibinfo {author} {\bibfnamefont {S.}~\bibnamefont
    {Altshuler}},\ }\href {\doibase 10.1007/BF02745414} {\bibfield  {journal}
    {\bibinfo  {journal} {Il Nuovo Cimento (1955-1965)}\ }\textbf {\bibinfo
    {volume} {3}},\ \bibinfo {pages} {246} (\bibinfo {year} {1956})}\BibitemShut
    {NoStop}%
  \bibitem [{\citenamefont {{Starace}}(1982)}]{staraceTheo}%
    \BibitemOpen
    \bibfield  {author} {\bibinfo {author} {\bibfnamefont {A.~F.}\ \bibnamefont
    {{Starace}}},\ }\href {\doibase 10.1007/978-3-642-46453-9_1} {\enquote
    {\bibinfo {title} {{Theory of Atomic Photoionization}},}\ } (\bibinfo {year}
    {1982})\BibitemShut {NoStop}%
  \bibitem [{Note1()}]{Note1}%
    \BibitemOpen
    \bibinfo {note} {Alternatively, absorbing an XUV photon from the XUV field
    $\protect \textbf {E}_{\protect \rm XUV}(t+\tau )$ in advance of the IR field
    $\protect \textbf {E}_{\protect \rm IR}(t)$ corresponds to an interaction
    phase of $e^{-i\Omega _{\protect \rm 2q+1}\tau }$ with $\Omega _{\protect \rm
    2q+1}>0$. The equivalence of these two perspectives is established upon the
    energy-preserving condition during the ionization process.}\BibitemShut
    {Stop}%
  \bibitem [{Note2()}]{Note2}%
    \BibitemOpen
    \bibinfo {note} {Note that the in the coordinate system used here, the $xOy$
    plane is defined as the polarization plane of the circularly polarized
    fields; and the $z$ axis is defined as the propagation direction of the
    circularly polarized fields and the polarization axis of the linearly
    polarized fields. This definition of the coordinate system for the circularly
    polarized fields differs from that used in our TDSE calculations
    [Eqs.\protect \,(\ref {IR_A}) and (\ref {xuv})], however, the observed
    physical quantities are unchanged under the rotation [$\protect \mathcal
    {C}:(\protect \hat {e}_x,\protect \hat {e}_y,\protect \hat {e}_z)\rightarrow
    \protect \mathcal {C}^\prime :(\protect \hat {e}_z,\protect \hat
    {e}_x,\protect \hat {e}_y)$] of the coordinate system fixed with an
    observer.}\BibitemShut {Stop}%
  \bibitem [{Note3()}]{Note3}%
    \BibitemOpen
    \bibinfo {note} {According to the dipole selection rules, the involved
    ionization channels are different in the cases of linearly and circularly
    polarized fields and thus the retrieved RABBIT phases are
    inequivalent.}\BibitemShut {Stop}%
  \bibitem [{\citenamefont {Arfken}\ and\ \citenamefont
    {Weber}(1999)}]{arfken1999mathematical}%
    \BibitemOpen
    \bibfield  {author} {\bibinfo {author} {\bibfnamefont {G.~B.}\ \bibnamefont
    {Arfken}}\ and\ \bibinfo {author} {\bibfnamefont {H.~J.}\ \bibnamefont
    {Weber}},\ }\href@noop {} {\enquote {\bibinfo {title} {Mathematical methods
    for physicists},}\ } (\bibinfo {year} {1999})\BibitemShut {NoStop}%
  \bibitem [{\citenamefont {Dalgarno}\ \emph {et~al.}(1955)\citenamefont
    {Dalgarno}, \citenamefont {Lewis},\ and\ \citenamefont
    {Bates}}]{doi:10.1098/rspa.1955.0246}%
    \BibitemOpen
    \bibfield  {author} {\bibinfo {author} {\bibfnamefont {A.}~\bibnamefont
    {Dalgarno}}, \bibinfo {author} {\bibfnamefont {J.~T.}\ \bibnamefont {Lewis}},
    \ and\ \bibinfo {author} {\bibfnamefont {D.~R.}\ \bibnamefont {Bates}},\
    }\href {\doibase 10.1098/rspa.1955.0246} {\bibfield  {journal} {\bibinfo
    {journal} {Proc. R. Soc. A}\ }\textbf {\bibinfo {volume} {233}},\ \bibinfo {pages}
    {70} (\bibinfo {year} {1955})}\BibitemShut {NoStop}%
  \bibitem [{\citenamefont {Pi}\ and\ \citenamefont
    {Starace}(2014)}]{PhysRevA.90.023403}%
    \BibitemOpen
    \bibfield  {author} {\bibinfo {author} {\bibfnamefont {L.-W.}\ \bibnamefont
    {Pi}}\ and\ \bibinfo {author} {\bibfnamefont {A.~F.}\ \bibnamefont
    {Starace}},\ }\href {\doibase 10.1103/PhysRevA.90.023403} {\bibfield
    {journal} {\bibinfo  {journal} {Phys. Rev. A}\ }\textbf {\bibinfo {volume}
    {90}},\ \bibinfo {pages} {023403} (\bibinfo {year} {2014})}\BibitemShut
    {NoStop}%
  \bibitem [{\citenamefont {Aymar}\ and\ \citenamefont {Crance}(1981)}]{1981}%
    \BibitemOpen
    \bibfield  {author} {\bibinfo {author} {\bibfnamefont {M.}~\bibnamefont
    {Aymar}}\ and\ \bibinfo {author} {\bibfnamefont {M.}~\bibnamefont {Crance}},\
    }\href {\doibase 10.1088/0022-3700/14/19/011} {\bibfield  {journal} {\bibinfo
     {journal} {J. Phys. B}\ }\textbf {\bibinfo {volume} {14}},\ \bibinfo {pages}
    {3585} (\bibinfo {year} {1981})}\BibitemShut {NoStop}%
  \bibitem [{\citenamefont {Gao}\ and\ \citenamefont
    {Starace}(1987)}]{doi:10.1063/1.4903436}%
    \BibitemOpen
    \bibfield  {author} {\bibinfo {author} {\bibfnamefont {B.}~\bibnamefont
    {Gao}}\ and\ \bibinfo {author} {\bibfnamefont {A.~F.}\ \bibnamefont
    {Starace}},\ }\href {\doibase 10.1063/1.4903436} {\bibfield  {journal}
    {\bibinfo  {journal} {Computers in Physics}\ }\textbf {\bibinfo {volume}
    {1}},\ \bibinfo {pages} {70} (\bibinfo {year} {1987})}\BibitemShut {NoStop}%
  \bibitem [{\citenamefont {Cormier}\ and\ \citenamefont
    {Lambropoulos}(1995)}]{E_Cormier_1995}%
    \BibitemOpen
    \bibfield  {author} {\bibinfo {author} {\bibfnamefont {E.}~\bibnamefont
    {Cormier}}\ and\ \bibinfo {author} {\bibfnamefont {P.}~\bibnamefont
    {Lambropoulos}},\ }\href {\doibase 10.1088/0953-4075/28/23/013} {\bibfield
    {journal} {\bibinfo  {journal} {J. Phys. B}\ }\textbf {\bibinfo {volume} {28}},\ \bibinfo {pages} {5043}
    (\bibinfo {year} {1995})}\BibitemShut {NoStop}%
  \bibitem [{\citenamefont {Kobe}\ and\ \citenamefont
    {Smirl}(1978)}]{10.1119/1.11264}%
    \BibitemOpen
    \bibfield  {author} {\bibinfo {author} {\bibfnamefont {D.~H.}\ \bibnamefont
    {Kobe}}\ and\ \bibinfo {author} {\bibfnamefont {A.~L.}\ \bibnamefont
    {Smirl}},\ }\href {\doibase 10.1119/1.11264} {\bibfield  {journal} {\bibinfo
    {journal} {American Journal of Physics}\ }\textbf {\bibinfo {volume} {46}},\
    \bibinfo {pages} {624} (\bibinfo {year} {1978})}\BibitemShut {NoStop}%
  \bibitem [{\citenamefont {V\'abek}\ \emph {et~al.}(2022)\citenamefont
    {V\'abek}, \citenamefont {Bachau},\ and\ \citenamefont
    {Catoire}}]{PhysRevA.106.053115}%
    \BibitemOpen
    \bibfield  {author} {\bibinfo {author} {\bibfnamefont {J.}~\bibnamefont
    {V\'abek}}, \bibinfo {author} {\bibfnamefont {H.}~\bibnamefont {Bachau}}, \
    and\ \bibinfo {author} {\bibfnamefont {F.}~\bibnamefont {Catoire}},\ }\href
    {\doibase 10.1103/PhysRevA.106.053115} {\bibfield  {journal} {\bibinfo
    {journal} {Phys. Rev. A}\ }\textbf {\bibinfo {volume} {106}},\ \bibinfo
    {pages} {053115} (\bibinfo {year} {2022})}\BibitemShut {NoStop}%
  \bibitem [{\citenamefont {Busto}\ \emph {et~al.}(2019)\citenamefont {Busto},
    \citenamefont {Vinbladh}, \citenamefont {Zhong}, \citenamefont {Isinger},
    \citenamefont {Nandi}, \citenamefont {Maclot}, \citenamefont {Johnsson},
    \citenamefont {Gisselbrecht}, \citenamefont {L'Huillier}, \citenamefont
    {Lindroth},\ and\ \citenamefont {Dahlstr\"om}}]{PhysRevLett.123.133201}%
    \BibitemOpen
    \bibfield  {author} {\bibinfo {author} {\bibfnamefont {D.}~\bibnamefont
    {Busto}}, \bibinfo {author} {\bibfnamefont {J.}~\bibnamefont {Vinbladh}},
    \bibinfo {author} {\bibfnamefont {S.}~\bibnamefont {Zhong}}, \bibinfo
    {author} {\bibfnamefont {M.}~\bibnamefont {Isinger}}, \bibinfo {author}
    {\bibfnamefont {S.}~\bibnamefont {Nandi}}, \bibinfo {author} {\bibfnamefont
    {S.}~\bibnamefont {Maclot}}, \bibinfo {author} {\bibfnamefont
    {P.}~\bibnamefont {Johnsson}}, \bibinfo {author} {\bibfnamefont
    {M.}~\bibnamefont {Gisselbrecht}}, \bibinfo {author} {\bibfnamefont
    {A.}~\bibnamefont {L'Huillier}}, \bibinfo {author} {\bibfnamefont
    {E.}~\bibnamefont {Lindroth}}, \ and\ \bibinfo {author} {\bibfnamefont
    {J.~M.}\ \bibnamefont {Dahlstr\"om}},\ }\href {\doibase
    10.1103/PhysRevLett.123.133201} {\bibfield  {journal} {\bibinfo  {journal}
    {Phys. Rev. Lett.}\ }\textbf {\bibinfo {volume} {123}},\ \bibinfo {pages}
    {133201} (\bibinfo {year} {2019})}\BibitemShut {NoStop}%
  \bibitem [{\citenamefont {Bertolino}\ \emph {et~al.}(2020)\citenamefont
    {Bertolino}, \citenamefont {Busto}, \citenamefont {Zapata},\ and\
    \citenamefont {Dahlström}}]{Bertolino_2020}%
    \BibitemOpen
    \bibfield  {author} {\bibinfo {author} {\bibfnamefont {M.}~\bibnamefont
    {Bertolino}}, \bibinfo {author} {\bibfnamefont {D.}~\bibnamefont {Busto}},
    \bibinfo {author} {\bibfnamefont {F.}~\bibnamefont {Zapata}}, \ and\ \bibinfo
    {author} {\bibfnamefont {J.~M.}\ \bibnamefont {Dahlström}},\ }\href
    {\doibase 10.1088/1361-6455/ab84c4} {\bibfield  {journal} {\bibinfo
    {journal} {J. Phys. B}\
    }\textbf {\bibinfo {volume} {53}},\ \bibinfo {pages} {144002} (\bibinfo
    {year} {2020})}\BibitemShut {NoStop}%
  \bibitem [{\citenamefont {Bertolino}\ and\ \citenamefont
    {Dahlstr\"om}(2021)}]{PhysRevResearch.3.013270}%
    \BibitemOpen
    \bibfield  {author} {\bibinfo {author} {\bibfnamefont {M.}~\bibnamefont
    {Bertolino}}\ and\ \bibinfo {author} {\bibfnamefont {J.~M.}\ \bibnamefont
    {Dahlstr\"om}},\ }\href {\doibase 10.1103/PhysRevResearch.3.013270}
    {\bibfield  {journal} {\bibinfo  {journal} {Phys. Rev. Res.}\ }\textbf
    {\bibinfo {volume} {3}},\ \bibinfo {pages} {013270} (\bibinfo {year}
    {2021})}\BibitemShut {NoStop}%
  \bibitem [{\citenamefont {Fano}(1985)}]{PhysRevA.32.617}%
    \BibitemOpen
    \bibfield  {author} {\bibinfo {author} {\bibfnamefont {U.}~\bibnamefont
    {Fano}},\ }\href {\doibase 10.1103/PhysRevA.32.617} {\bibfield  {journal}
    {\bibinfo  {journal} {Phys. Rev. A}\ }\textbf {\bibinfo {volume} {32}},\
    \bibinfo {pages} {617} (\bibinfo {year} {1985})}\BibitemShut {NoStop}%
  \bibitem [{\citenamefont {Jim\'enez-Gal\'an}\ \emph {et~al.}(2014)\citenamefont
    {Jim\'enez-Gal\'an}, \citenamefont {Argenti},\ and\ \citenamefont
    {Mart\'{\i}n}}]{PhysRevLett.113.263001}%
    \BibitemOpen
    \bibfield  {author} {\bibinfo {author} {\bibfnamefont {A.}~\bibnamefont
    {Jim\'enez-Gal\'an}}, \bibinfo {author} {\bibfnamefont {L.}~\bibnamefont
    {Argenti}}, \ and\ \bibinfo {author} {\bibfnamefont {F.}~\bibnamefont
    {Mart\'{\i}n}},\ }\href {\doibase 10.1103/PhysRevLett.113.263001} {\bibfield
    {journal} {\bibinfo  {journal} {Phys. Rev. Lett.}\ }\textbf {\bibinfo
    {volume} {113}},\ \bibinfo {pages} {263001} (\bibinfo {year}
    {2014})}\BibitemShut {NoStop}%
  \bibitem [{\citenamefont {Jim\'enez-Gal\'an}\ \emph {et~al.}(2016)\citenamefont
    {Jim\'enez-Gal\'an}, \citenamefont {Mart\'{\i}n},\ and\ \citenamefont
    {Argenti}}]{PhysRevA.93.023429}%
    \BibitemOpen
    \bibfield  {author} {\bibinfo {author} {\bibfnamefont {A.}~\bibnamefont
    {Jim\'enez-Gal\'an}}, \bibinfo {author} {\bibfnamefont {F.}~\bibnamefont
    {Mart\'{\i}n}}, \ and\ \bibinfo {author} {\bibfnamefont {L.}~\bibnamefont
    {Argenti}},\ }\href {\doibase 10.1103/PhysRevA.93.023429} {\bibfield
    {journal} {\bibinfo  {journal} {Phys. Rev. A}\ }\textbf {\bibinfo {volume}
    {93}},\ \bibinfo {pages} {023429} (\bibinfo {year} {2016})}\BibitemShut
    {NoStop}%
  \bibitem [{\citenamefont {Zhao}\ and\ \citenamefont
    {Lin}(2005)}]{PhysRevA.71.060702}%
    \BibitemOpen
    \bibfield  {author} {\bibinfo {author} {\bibfnamefont {Z.~X.}\ \bibnamefont
    {Zhao}}\ and\ \bibinfo {author} {\bibfnamefont {C.~D.}\ \bibnamefont {Lin}},\
    }\href {\doibase 10.1103/PhysRevA.71.060702} {\bibfield  {journal} {\bibinfo
    {journal} {Phys. Rev. A}\ }\textbf {\bibinfo {volume} {71}},\ \bibinfo
    {pages} {060702} (\bibinfo {year} {2005})}\BibitemShut {NoStop}%
  \bibitem [{\citenamefont {Mao}\ \emph {et~al.}(2023)\citenamefont {Mao},
    \citenamefont {Yao}, \citenamefont {He}, \citenamefont {Zhang}, \citenamefont
    {Li},\ and\ \citenamefont {He}}]{PhysRevA.108.053117}%
    \BibitemOpen
    \bibfield  {author} {\bibinfo {author} {\bibfnamefont {Y.-J.}\ \bibnamefont
    {Mao}}, \bibinfo {author} {\bibfnamefont {H.-B.}\ \bibnamefont {Yao}},
    \bibinfo {author} {\bibfnamefont {M.}~\bibnamefont {He}}, \bibinfo {author}
    {\bibfnamefont {Z.-H.}\ \bibnamefont {Zhang}}, \bibinfo {author}
    {\bibfnamefont {Y.}~\bibnamefont {Li}}, \ and\ \bibinfo {author}
    {\bibfnamefont {F.}~\bibnamefont {He}},\ }\href {\doibase
    10.1103/PhysRevA.108.053117} {\bibfield  {journal} {\bibinfo  {journal}
    {Phys. Rev. A}\ }\textbf {\bibinfo {volume} {108}},\ \bibinfo {pages}
    {053117} (\bibinfo {year} {2023})}\BibitemShut {NoStop}%
  \bibitem [{\citenamefont {Pan}\ \emph {et~al.}(2023)\citenamefont {Pan},
    \citenamefont {Hu}, \citenamefont {Zhang}, \citenamefont {Zhang},
    \citenamefont {Zhou}, \citenamefont {Lu}, \citenamefont {Lu}, \citenamefont
    {Ni}, \citenamefont {Wu},\ and\ \citenamefont {He}}]{Pan2023}%
    \BibitemOpen
    \bibfield  {author} {\bibinfo {author} {\bibfnamefont {S.}~\bibnamefont
    {Pan}}, \bibinfo {author} {\bibfnamefont {C.}~\bibnamefont {Hu}}, \bibinfo
    {author} {\bibfnamefont {W.}~\bibnamefont {Zhang}}, \bibinfo {author}
    {\bibfnamefont {Z.}~\bibnamefont {Zhang}}, \bibinfo {author} {\bibfnamefont
    {L.}~\bibnamefont {Zhou}}, \bibinfo {author} {\bibfnamefont {C.}~\bibnamefont
    {Lu}}, \bibinfo {author} {\bibfnamefont {P.}~\bibnamefont {Lu}}, \bibinfo
    {author} {\bibfnamefont {H.}~\bibnamefont {Ni}}, \bibinfo {author}
    {\bibfnamefont {J.}~\bibnamefont {Wu}}, \ and\ \bibinfo {author}
    {\bibfnamefont {F.}~\bibnamefont {He}},\ }\href {\doibase
    10.1038/s41377-023-01075-9} {\bibfield  {journal} {\bibinfo  {journal}
    {Light: Science {\&} Applications}\ }\textbf {\bibinfo {volume} {12}},\
    \bibinfo {pages} {35} (\bibinfo {year} {2023})}\BibitemShut {NoStop}%
  \bibitem [{\citenamefont {Sun}\ and\ \citenamefont
    {Lou}(2003)}]{PhysRevLett.91.023002}%
    \BibitemOpen
    \bibfield  {author} {\bibinfo {author} {\bibfnamefont {Z.}~\bibnamefont
    {Sun}}\ and\ \bibinfo {author} {\bibfnamefont {N.}~\bibnamefont {Lou}},\
    }\href {\doibase 10.1103/PhysRevLett.91.023002} {\bibfield  {journal}
    {\bibinfo  {journal} {Phys. Rev. Lett.}\ }\textbf {\bibinfo {volume} {91}},\
    \bibinfo {pages} {023002} (\bibinfo {year} {2003})}\BibitemShut {NoStop}%
  \bibitem [{\citenamefont {Fischer}\ \emph {et~al.}(1998)\citenamefont
    {Fischer}, \citenamefont {Madison}, \citenamefont {Niu},\ and\ \citenamefont
    {Raizen}}]{PhysRevA.58.R2648}%
    \BibitemOpen
    \bibfield  {author} {\bibinfo {author} {\bibfnamefont {M.~C.}\ \bibnamefont
    {Fischer}}, \bibinfo {author} {\bibfnamefont {K.~W.}\ \bibnamefont
    {Madison}}, \bibinfo {author} {\bibfnamefont {Q.}~\bibnamefont {Niu}}, \ and\
    \bibinfo {author} {\bibfnamefont {M.~G.}\ \bibnamefont {Raizen}},\ }\href
    {\doibase 10.1103/PhysRevA.58.R2648} {\bibfield  {journal} {\bibinfo
    {journal} {Phys. Rev. A}\ }\textbf {\bibinfo {volume} {58}},\ \bibinfo
    {pages} {R2648} (\bibinfo {year} {1998})}\BibitemShut {NoStop}%
  \bibitem [{\citenamefont {Silva}\ and\ \citenamefont
    {Jim\'enez-Gal\'an}(2022)}]{PhysRevA.106.053103}%
    \BibitemOpen
    \bibfield  {author} {\bibinfo {author} {\bibfnamefont {R.~E.~F.}\
    \bibnamefont {Silva}}\ and\ \bibinfo {author} {\bibfnamefont
    {A.}~\bibnamefont {Jim\'enez-Gal\'an}},\ }\href {\doibase
    10.1103/PhysRevA.106.053103} {\bibfield  {journal} {\bibinfo  {journal}
    {Phys. Rev. A}\ }\textbf {\bibinfo {volume} {106}},\ \bibinfo {pages}
    {053103} (\bibinfo {year} {2022})}\BibitemShut {NoStop}%
  \bibitem [{\citenamefont {Wu}(1996)}]{PhysRevA.54.1586}%
    \BibitemOpen
    \bibfield  {author} {\bibinfo {author} {\bibfnamefont {Y.}~\bibnamefont
    {Wu}},\ }\href {\doibase 10.1103/PhysRevA.54.1586} {\bibfield  {journal}
    {\bibinfo  {journal} {Phys. Rev. A}\ }\textbf {\bibinfo {volume} {54}},\
    \bibinfo {pages} {1586} (\bibinfo {year} {1996})}\BibitemShut {NoStop}%
  \bibitem [{\citenamefont {Sargent}\ and\ \citenamefont
    {Horwitz}(1976)}]{PhysRevA.13.1962}%
    \BibitemOpen
    \bibfield  {author} {\bibinfo {author} {\bibfnamefont {M.}~\bibnamefont
    {Sargent}}\ and\ \bibinfo {author} {\bibfnamefont {P.}~\bibnamefont
    {Horwitz}},\ }\href {\doibase 10.1103/PhysRevA.13.1962} {\bibfield  {journal}
    {\bibinfo  {journal} {Phys. Rev. A}\ }\textbf {\bibinfo {volume} {13}},\
    \bibinfo {pages} {1962} (\bibinfo {year} {1976})}\BibitemShut {NoStop}%
  \bibitem [{\citenamefont {Song}\ \emph {et~al.}(2016)\citenamefont {Song},
    \citenamefont {Ai}, \citenamefont {Qiu},\ and\ \citenamefont
    {Deng}}]{PhysRevA.93.052324}%
    \BibitemOpen
    \bibfield  {author} {\bibinfo {author} {\bibfnamefont {X.-K.}\ \bibnamefont
    {Song}}, \bibinfo {author} {\bibfnamefont {Q.}~\bibnamefont {Ai}}, \bibinfo
    {author} {\bibfnamefont {J.}~\bibnamefont {Qiu}}, \ and\ \bibinfo {author}
    {\bibfnamefont {F.-G.}\ \bibnamefont {Deng}},\ }\href {\doibase
    10.1103/PhysRevA.93.052324} {\bibfield  {journal} {\bibinfo  {journal} {Phys.
    Rev. A}\ }\textbf {\bibinfo {volume} {93}},\ \bibinfo {pages} {052324}
    (\bibinfo {year} {2016})}\BibitemShut {NoStop}%
  \bibitem [{\citenamefont {Laurent}\ \emph {et~al.}(2012)\citenamefont
    {Laurent}, \citenamefont {Cao}, \citenamefont {Li}, \citenamefont {Wang},
    \citenamefont {Ben-Itzhak},\ and\ \citenamefont
    {Cocke}}]{PhysRevLett.109.083001}%
    \BibitemOpen
    \bibfield  {author} {\bibinfo {author} {\bibfnamefont {G.}~\bibnamefont
    {Laurent}}, \bibinfo {author} {\bibfnamefont {W.}~\bibnamefont {Cao}},
    \bibinfo {author} {\bibfnamefont {H.}~\bibnamefont {Li}}, \bibinfo {author}
    {\bibfnamefont {Z.}~\bibnamefont {Wang}}, \bibinfo {author} {\bibfnamefont
    {I.}~\bibnamefont {Ben-Itzhak}}, \ and\ \bibinfo {author} {\bibfnamefont
    {C.~L.}\ \bibnamefont {Cocke}},\ }\href {\doibase
    10.1103/PhysRevLett.109.083001} {\bibfield  {journal} {\bibinfo  {journal}
    {Phys. Rev. Lett.}\ }\textbf {\bibinfo {volume} {109}},\ \bibinfo {pages}
    {083001} (\bibinfo {year} {2012})}\BibitemShut {NoStop}%
  \bibitem [{\citenamefont {Bharti}\ \emph {et~al.}(2021)\citenamefont {Bharti},
    \citenamefont {Atri-Schuller}, \citenamefont {Menning}, \citenamefont
    {Hamilton}, \citenamefont {Moshammer}, \citenamefont {Pfeifer}, \citenamefont
    {Douguet}, \citenamefont {Bartschat},\ and\ \citenamefont
    {Harth}}]{PhysRevA.103.022834}%
    \BibitemOpen
    \bibfield  {author} {\bibinfo {author} {\bibfnamefont {D.}~\bibnamefont
    {Bharti}}, \bibinfo {author} {\bibfnamefont {D.}~\bibnamefont
    {Atri-Schuller}}, \bibinfo {author} {\bibfnamefont {G.}~\bibnamefont
    {Menning}}, \bibinfo {author} {\bibfnamefont {K.~R.}\ \bibnamefont
    {Hamilton}}, \bibinfo {author} {\bibfnamefont {R.}~\bibnamefont {Moshammer}},
    \bibinfo {author} {\bibfnamefont {T.}~\bibnamefont {Pfeifer}}, \bibinfo
    {author} {\bibfnamefont {N.}~\bibnamefont {Douguet}}, \bibinfo {author}
    {\bibfnamefont {K.}~\bibnamefont {Bartschat}}, \ and\ \bibinfo {author}
    {\bibfnamefont {A.}~\bibnamefont {Harth}},\ }\href {\doibase
    10.1103/PhysRevA.103.022834} {\bibfield  {journal} {\bibinfo  {journal}
    {Phys. Rev. A}\ }\textbf {\bibinfo {volume} {103}},\ \bibinfo {pages}
    {022834} (\bibinfo {year} {2021})}\BibitemShut {NoStop}%
  \bibitem [{\citenamefont {Maroju}\ \emph {et~al.}(2023)\citenamefont {Maroju},
    \citenamefont {Di~Fraia}, \citenamefont {Plekan}, \citenamefont {Bonanomi},
    \citenamefont {Merzuk}, \citenamefont {Busto}, \citenamefont {Makos},
    \citenamefont {Schmoll}, \citenamefont {Shah}, \citenamefont {Ribi{\v{c}}},
    \citenamefont {Giannessi}, \citenamefont {De~Ninno}, \citenamefont
    {Spezzani}, \citenamefont {Penco}, \citenamefont {Demidovich}, \citenamefont
    {Danailov}, \citenamefont {Coreno}, \citenamefont {Zangrando}, \citenamefont
    {Simoncig}, \citenamefont {Manfredda}, \citenamefont {Squibb}, \citenamefont
    {Feifel}, \citenamefont {Bengtsson}, \citenamefont {Simpson}, \citenamefont
    {Csizmadia}, \citenamefont {Dumergue}, \citenamefont {K{\"u}hn},
    \citenamefont {Ueda}, \citenamefont {Li}, \citenamefont {Schafer},
    \citenamefont {Frassetto}, \citenamefont {Poletto}, \citenamefont {Prince},
    \citenamefont {Mauritsson}, \citenamefont {Callegari},\ and\ \citenamefont
    {Sansone}}]{Maroju2023}%
    \BibitemOpen
    \bibfield  {author} {\bibinfo {author} {\bibfnamefont {P.~K.}\ \bibnamefont
    {Maroju}}, \bibinfo {author} {\bibfnamefont {M.}~\bibnamefont {Di~Fraia}},
    \bibinfo {author} {\bibfnamefont {O.}~\bibnamefont {Plekan}}, \bibinfo
    {author} {\bibfnamefont {M.}~\bibnamefont {Bonanomi}}, \bibinfo {author}
    {\bibfnamefont {B.}~\bibnamefont {Merzuk}}, \bibinfo {author} {\bibfnamefont
    {D.}~\bibnamefont {Busto}}, \bibinfo {author} {\bibfnamefont
    {I.}~\bibnamefont {Makos}}, \bibinfo {author} {\bibfnamefont
    {M.}~\bibnamefont {Schmoll}}, \bibinfo {author} {\bibfnamefont
    {R.}~\bibnamefont {Shah}}, \bibinfo {author} {\bibfnamefont {P.~R.}\
    \bibnamefont {Ribi{\v{c}}}}, \bibinfo {author} {\bibfnamefont
    {L.}~\bibnamefont {Giannessi}}, \bibinfo {author} {\bibfnamefont
    {G.}~\bibnamefont {De~Ninno}}, \bibinfo {author} {\bibfnamefont
    {C.}~\bibnamefont {Spezzani}}, \bibinfo {author} {\bibfnamefont
    {G.}~\bibnamefont {Penco}}, \bibinfo {author} {\bibfnamefont
    {A.}~\bibnamefont {Demidovich}}, \bibinfo {author} {\bibfnamefont
    {M.}~\bibnamefont {Danailov}}, \bibinfo {author} {\bibfnamefont
    {M.}~\bibnamefont {Coreno}}, \bibinfo {author} {\bibfnamefont
    {M.}~\bibnamefont {Zangrando}}, \bibinfo {author} {\bibfnamefont
    {A.}~\bibnamefont {Simoncig}}, \bibinfo {author} {\bibfnamefont
    {M.}~\bibnamefont {Manfredda}}, \bibinfo {author} {\bibfnamefont {R.~J.}\
    \bibnamefont {Squibb}}, \bibinfo {author} {\bibfnamefont {R.}~\bibnamefont
    {Feifel}}, \bibinfo {author} {\bibfnamefont {S.}~\bibnamefont {Bengtsson}},
    \bibinfo {author} {\bibfnamefont {E.~R.}\ \bibnamefont {Simpson}}, \bibinfo
    {author} {\bibfnamefont {T.}~\bibnamefont {Csizmadia}}, \bibinfo {author}
    {\bibfnamefont {M.}~\bibnamefont {Dumergue}}, \bibinfo {author}
    {\bibfnamefont {S.}~\bibnamefont {K{\"u}hn}}, \bibinfo {author}
    {\bibfnamefont {K.}~\bibnamefont {Ueda}}, \bibinfo {author} {\bibfnamefont
    {J.}~\bibnamefont {Li}}, \bibinfo {author} {\bibfnamefont {K.~J.}\
    \bibnamefont {Schafer}}, \bibinfo {author} {\bibfnamefont {F.}~\bibnamefont
    {Frassetto}}, \bibinfo {author} {\bibfnamefont {L.}~\bibnamefont {Poletto}},
    \bibinfo {author} {\bibfnamefont {K.~C.}\ \bibnamefont {Prince}}, \bibinfo
    {author} {\bibfnamefont {J.}~\bibnamefont {Mauritsson}}, \bibinfo {author}
    {\bibfnamefont {C.}~\bibnamefont {Callegari}}, \ and\ \bibinfo {author}
    {\bibfnamefont {G.}~\bibnamefont {Sansone}},\ }\href {\doibase
    10.1038/s41566-022-01127-3} {\bibfield  {journal} {\bibinfo  {journal}
    {Nature Photonics}\ }\textbf {\bibinfo {volume} {17}},\ \bibinfo {pages}
    {200} (\bibinfo {year} {2023})}\BibitemShut {NoStop}%
  \bibitem [{\citenamefont {Rabi}\ \emph {et~al.}(1954)\citenamefont {Rabi},
    \citenamefont {Ramsey},\ and\ \citenamefont {Schwinger}}]{RevModPhys.26.167}%
    \BibitemOpen
    \bibfield  {author} {\bibinfo {author} {\bibfnamefont {I.~I.}\ \bibnamefont
    {Rabi}}, \bibinfo {author} {\bibfnamefont {N.~F.}\ \bibnamefont {Ramsey}}, \
    and\ \bibinfo {author} {\bibfnamefont {J.}~\bibnamefont {Schwinger}},\ }\href
    {\doibase 10.1103/RevModPhys.26.167} {\bibfield  {journal} {\bibinfo
    {journal} {Rev. Mod. Phys.}\ }\textbf {\bibinfo {volume} {26}},\ \bibinfo
    {pages} {167} (\bibinfo {year} {1954})}\BibitemShut {NoStop}%
  \bibitem [{\citenamefont {{Cohen-Tannoudji}}\ \emph {et~al.}(1986)\citenamefont
    {{Cohen-Tannoudji}}, \citenamefont {{Diu}},\ and\ \citenamefont
    {{Laloe}}}]{1986qmv1.book.....C}%
    \BibitemOpen
    \bibfield  {author} {\bibinfo {author} {\bibfnamefont {C.}~\bibnamefont
    {{Cohen-Tannoudji}}}, \bibinfo {author} {\bibfnamefont {B.}~\bibnamefont
    {{Diu}}}, \ and\ \bibinfo {author} {\bibfnamefont {F.}~\bibnamefont
    {{Laloe}}},\ }\href@noop {} {\emph {\bibinfo {title} {{Quantum Mechanics,
    Volume 1}}}},\ Vol.~\bibinfo {volume} {1}\ (\bibinfo {year}
    {1986})\BibitemShut {NoStop}%
  \bibitem [{\citenamefont {Blanes}\ \emph {et~al.}(2009)\citenamefont {Blanes},
    \citenamefont {Casas}, \citenamefont {Oteo},\ and\ \citenamefont
    {Ros}}]{BLANES2009151}%
    \BibitemOpen
    \bibfield  {author} {\bibinfo {author} {\bibfnamefont {S.}~\bibnamefont
    {Blanes}}, \bibinfo {author} {\bibfnamefont {F.}~\bibnamefont {Casas}},
    \bibinfo {author} {\bibfnamefont {J.}~\bibnamefont {Oteo}}, \ and\ \bibinfo
    {author} {\bibfnamefont {J.}~\bibnamefont {Ros}},\ }\href {\doibase
    https://doi.org/10.1016/j.physrep.2008.11.001} {\bibfield  {journal}
    {\bibinfo  {journal} {Physics Reports}\ }\textbf {\bibinfo {volume} {470}},\
    \bibinfo {pages} {151} (\bibinfo {year} {2009})}\BibitemShut {NoStop}%
  \bibitem [{\citenamefont {Harutyunyan}\ \emph {et~al.}(2023)\citenamefont
    {Harutyunyan}, \citenamefont {Holweck}, \citenamefont {Sugny},\ and\
    \citenamefont {Gu\'erin}}]{PhysRevLett.131.200801}%
    \BibitemOpen
    \bibfield  {author} {\bibinfo {author} {\bibfnamefont {M.}~\bibnamefont
    {Harutyunyan}}, \bibinfo {author} {\bibfnamefont {F.}~\bibnamefont
    {Holweck}}, \bibinfo {author} {\bibfnamefont {D.}~\bibnamefont {Sugny}}, \
    and\ \bibinfo {author} {\bibfnamefont {S.}~\bibnamefont {Gu\'erin}},\ }\href
    {\doibase 10.1103/PhysRevLett.131.200801} {\bibfield  {journal} {\bibinfo
    {journal} {Phys. Rev. Lett.}\ }\textbf {\bibinfo {volume} {131}},\ \bibinfo
    {pages} {200801} (\bibinfo {year} {2023})}\BibitemShut {NoStop}%
  \bibitem [{\citenamefont {Torosov}\ and\ \citenamefont
    {Vitanov}(2019)}]{PhysRevA.99.013402}%
    \BibitemOpen
    \bibfield  {author} {\bibinfo {author} {\bibfnamefont {B.~T.}\ \bibnamefont
    {Torosov}}\ and\ \bibinfo {author} {\bibfnamefont {N.~V.}\ \bibnamefont
    {Vitanov}},\ }\href {\doibase 10.1103/PhysRevA.99.013402} {\bibfield
    {journal} {\bibinfo  {journal} {Phys. Rev. A}\ }\textbf {\bibinfo {volume}
    {99}},\ \bibinfo {pages} {013402} (\bibinfo {year} {2019})}\BibitemShut
    {NoStop}%
  \bibitem [{\citenamefont {Eberly}(1998)}]{Eberly:98}%
    \BibitemOpen
    \bibfield  {author} {\bibinfo {author} {\bibfnamefont {J.}~\bibnamefont
    {Eberly}},\ }\href {\doibase 10.1364/OE.2.000173} {\bibfield  {journal}
    {\bibinfo  {journal} {Opt. Express}\ }\textbf {\bibinfo {volume} {2}},\
    \bibinfo {pages} {173} (\bibinfo {year} {1998})}\BibitemShut {NoStop}%
  \bibitem [{\citenamefont {Milošević}(2013)}]{10.1063/1.4797476}%
    \BibitemOpen
    \bibfield  {author} {\bibinfo {author} {\bibfnamefont {D.~B.}\ \bibnamefont
    {Milošević}},\ }\href {\doibase 10.1063/1.4797476} {\bibfield  {journal}
    {\bibinfo  {journal} {Journal of Mathematical Physics}\ }\textbf {\bibinfo
    {volume} {54}},\ \bibinfo {pages} {042101} (\bibinfo {year}
    {2013})}\BibitemShut {NoStop}%
  \bibitem [{\citenamefont {Patchkovskii}\ \emph {et~al.}(2023)\citenamefont
    {Patchkovskii}, \citenamefont {Benda}, \citenamefont {Ertel},\ and\
    \citenamefont {Busto}}]{PhysRevA.107.043105}%
    \BibitemOpen
    \bibfield  {author} {\bibinfo {author} {\bibfnamefont {S.}~\bibnamefont
    {Patchkovskii}}, \bibinfo {author} {\bibfnamefont {J.}~\bibnamefont {Benda}},
    \bibinfo {author} {\bibfnamefont {D.}~\bibnamefont {Ertel}}, \ and\ \bibinfo
    {author} {\bibfnamefont {D.}~\bibnamefont {Busto}},\ }\href {\doibase
    10.1103/PhysRevA.107.043105} {\bibfield  {journal} {\bibinfo  {journal}
    {Phys. Rev. A}\ }\textbf {\bibinfo {volume} {107}},\ \bibinfo {pages}
    {043105} (\bibinfo {year} {2023})}\BibitemShut {NoStop}%
  \end{thebibliography}
%

\end{document}